\documentstyle [12pt,epsfig]{article} 
\makeatletter
\@addtoreset{equation}{section}
\makeatother

\newcommand{\ba}{\begin{eqnarray}}
\newcommand{\ea}{\end{eqnarray}}

\begin{document}
\newcommand{\BS}{\bigskip}
\newcommand{\SECTION}[1]{\BS{\large\section{\bf #1}}}
\newcommand{\SUBSECTION}[1]{\BS{\large\subsection{\bf #1}}}
\newcommand{\SUBSUBSECTION}[1]{\BS{\large\subsubsection{\bf #1}}}

\newcommand {\mv}  {M_{\rm{V}}}
\newcommand {\mQ}  {m_{\rm{Q}}}
\newcommand {\mg}  {m_{\rm{g}}}
\newcommand {\Mg}  {M_{\rm{g}}}
\newcommand {\pst}  {\tilde{\psi}(p)}
\newcommand {\Jp}  {\rm{J/\psi}}
\newcommand {\Up}  {\Upsilon}
\newcommand {\rbf}  {\overline{r}_f}
\newcommand {\vbf}  {\overline{v}_f}
\newcommand {\abf}  {\overline{a}_f}
\newcommand {\sbf}  {\overline{s}_f}
\newcommand {\sbq}  {\overline{s}_q}
\newcommand {\vbq}  {\overline{v}_q}
\newcommand {\abq}  {\overline{a}_q}
\newcommand {\rbQ}  {\overline{r}_Q}
\newcommand {\vbQ}  {\overline{v}_Q}
\newcommand {\abQ}  {\overline{a}_Q}
\newcommand {\rbl}  {\overline{r}_l}
\newcommand {\vbl}  {\overline{v}_l}
\newcommand {\abl}  {\overline{a}_l}
\newcommand {\sbl}  {\overline{s}_l}
\newcommand {\rbc}  {\overline{r}_c}
\newcommand {\vbc}  {\overline{v}_c}
\newcommand {\abc}  {\overline{a}_c}
\newcommand {\sbc}  {\overline{s}_c}
\newcommand {\rbb}  {\overline{r}_b}
\newcommand {\vbb}  {\overline{v}_b}
\newcommand {\abb}  {\overline{a}_b}
\newcommand {\sbb}  {\overline{s}_b}
\newcommand {\gbl}  {\overline{g}_b^L}
\newcommand {\gbr}  {\overline{g}_b^R}
\newcommand {\gcl}  {\overline{g}_c^L}
\newcommand {\gcr}  {\overline{g}_c^R}
\newcommand {\Afbf}  {A_{FB}^{0,f}}
\newcommand {\Afbl}  {A_{FB}^{0,l}}
\newcommand {\Afbc}  {A_{FB}^{0,c}}
\newcommand {\Afbb}  {A_{FB}^{0,b}}
\newcommand {\tpol}  {$\tau$-polarisation}
\newcommand {\alps}  {\alpha_s(M_Z)}
\newcommand {\alsm}  {\alpha_s(\mu)}
\newcommand {\alsQ}  {\alpha_s(m_Q)}
\newcommand {\alsc}  {\alpha_s(m_c)}
\newcommand {\alsb}  {\alpha_s(m_b)}
\begin{titlepage}
\hspace*{8cm} {UGVA-DPNC 2001/1-185 January 2001 }
\newline
\hspace*{11cm}
hep-ph/0101158
\begin{center}
\vspace*{2cm}
{\large \bf A Phenomenological Analysis of Gluon Mass Effects in
Inclusive  Radiative Decays of the $\rm{J/\psi}$ and $\Upsilon$ }
\vspace*{1.5cm}
\end{center}
\begin{center}
{\bf J.H.Field$^*$ }
\end{center}
\begin{center}
{ 
D\'{e}partement de Physique Nucl\'{e}aire et Corpusculaire
 Universit\'{e} de Gen\`{e}ve . 24, quai Ernest-Ansermet
 CH-1211 Gen\`{e}ve 4.
}
\end{center}
\vspace*{2cm}
\begin{abstract}
The shapes of the inclusive photon spectra in the processes $\Jp \rightarrow \gamma X$
 and $\Up \rightarrow \gamma X$ have been analysed using all available experimental data.
 Relativistic, higher order QCD and gluon mass corrections were taken into account in
 the fitted functions. Only on including the gluon mass corrections, were consistent and acceptable fits
 obtained. Values of $0.721^{+0.016}_{-0.068}$ GeV and $1.18^{+0.09}_{-0.29}$ GeV were
 found for the effective gluon masses (corresponding to Born level diagrams) for
 the $\Jp$ and $\Up$ respectively. The width ratios 
$\Gamma(V \rightarrow {\rm hadrons})/\Gamma(V \rightarrow \gamma+ {\rm hadrons})~V=\Jp,~\Up$
 were used to determine $\alpha_s(1.5 {\rm GeV})$ and $\alpha_s(4.9 {\rm GeV})$. Values
 consistent with the current world average $\alpha_s$ were obtained only when gluon
 mass correction factors, calculated using the fitted values of the effective gluon mass,
 were applied. A gluon mass $\simeq 1$ GeV, as suggested with these results, is consistent
 with previous analytical theoretical calculations and independent phenomenological
 estimates, as well as with a recent, more accurate, lattice calculation of the gluon
 propagator in the infra-red region.          
\end{abstract}
\vspace*{1cm}
\newline 
{\it Keywords ;} Radiative decays of heavy quarkonia, Relativistic Corrections, Quantum Chromodynamics,
 Gluon mass effects.
\newline
\newline
$*$ e-mail address: john.field@cern.ch
\newline
\newline
 Second Revised Version May 2002
\end{titlepage}
\SECTION{\bf{Introduction}}
 As suggested by the inventors of QCD~\cite{FGML,SW}, the colour symmetry of the theory 
 is, conventionally, assumed to be unbroken, so that, theoretically~\cite{PDG}, the
 gluon is supposed to have a vanishing mass. It was also conjectured, by the same authors,
 that the resulting infra-red divergences of the theory at large distances
 (`infra-red slavery') might explain the confinement of quarks. As is also
 well known, in the contrary case that gluons are massive, there is a possible
 breakdown of renormalisability as well as violation of unitarity at high 
 energy by certain tree level amplitudes. These problems are common to all
 non-abelian gauge theories with massive vector mesons~\cite{CHLS,CLT}.
 \par These problems may be solved, as in the Standard Electroweak Model,
  by the introduction, also for the strong interaction, of spontaneous symmetry breaking
 and the Higgs
  mechanism~\cite{BCQCD}. Since, however, there is no experimental evidence for
 the existence of a Higgs boson for the strong interaction, or for 
 electrically charged gluons, which are also predicted by some of
 these `broken colour' theories, it is still generally supposed, in spite
 of the infra-red divergent nature of such a theory, that the QCD colour
 symmetry remains unbroken.
\par A possible way out of this dilemma (infra-red divergences if the 
 gluon mass is zero, breakdown of renormalisabilty and unitarity if it 
 is not) was suggested by Cornwall~\cite{Cornwall,Cornwall1} who suggested
 that non-vanishing gluon mass might be dynamically generated in a theory
 in which the colour 
 gauge symmetry remained unbroken. Other authors~\cite{CF,CdE} pointed out
 that a gauge invariant, renormalisable and infra-red finite, version of QCD
 with massive gluons is possible, provided that a suitable four-vertex
 Faddeev-Popov ghost field is introduced into the theory.
\par The aim of the present paper is not to pursue further these
 theoretical considerations\footnote{ The interested reader is referred to
 Reference~\cite{FPP} for recent developments, and citations of
 the related literature}, but rather to seek {\it direct experimental
 evidence} on the mass of the gluon. The processes considered, the radiative 
 decays of ground state vector heavy quarkonia into a photon and light hadrons,
 are particularly well adapted to such a study, as the observed final state
 results from the hadronisation of a pure two gluon final state at the
 lowest order in perturbative QCD (pQCD). These are the `golden' physical
 processes for the determination of the gluon mass, that may be compared
 to the neutral kaon system for the study of CP violation or Tritium $\beta$-decay
 for the direct determination of the mass of the electron antineutrino.
 Indeed, the analogy between the process $\Jp \rightarrow \gamma X$ and
 Tritium $\beta$-decay is a very close one. In both cases it is the study of
 the end-point region of a spectrum ( that of the electron for Tritium $\beta$-decay,
 of the photon for the radiative $\Jp$ decay) that give the mass limits 
 on the $\overline{\nu}_e$ or the gluon mass. The $\Jp$, being the lightest
 quarkonia is the most sensitive to the gluon mass, just as the Tritium $\beta$-decay,
 with a very low energy release, gives the best direct limit on the $\overline{\nu}_e$
 mass. Indeed, as will be shown below, the suppression of the spectrum end point
 due to gluon mass effects, is much more severe in the case of $\Jp$ radiative
 decays than for the heavier $\Up$ state.
\par Already in 1980, Parisi and Petronzio~\cite{PP}(PP) had suggested a mass of
 $\simeq$ 800 MeV for the gluon, on the basis of the strong suppression of the
 end point of the photon spectrum in radiative $\Jp$ decays, as measured by the
 MARK II collaboration~\cite{Mark II}. In order to relate, however, in a precise
 way, the shape of the photon spectrum to the gluon mass, two other important
 physical effects, which also soften the shape of the photon spectrum, must
 also be properly accounted for. These are: (i) relativistic corrections and
 (ii) higher order QCD corrections. Because of the only recently available complete
 next-to-leading-order (NLO) pQCD calculation of the photon spectrum in the
 decays $\Up \rightarrow \gamma X$~\cite{Kramer} and a much improved 
 understanding of the phenomenology of relativistic corrections based on
 several recent and independent potential model calculations, the 
 analysis presented below is the first to take fully into account
 the important effects (i) and (ii) and so confirm the conclusion of
 PP that the gluon mass is $\simeq$ 1 GeV.
 At the time of writing, no calculation
 yet exists in which the effects (i) and (ii), as well as that of the 
 gluon mass are taken into account in a unified way, so the present analysis
 is inevitably a phenomenological one where the three different types of
 corrections are assumed, loosely speaking, to `factorise'. Since, however,
 it is clear that the gluon mass effects are, by far, the most important,
 no large uncertainity on the results obtained are expected to result from
 this approximation.
\par The results presented below also confirm the conclusions of two
 previous, closely related, papers written by M.Consoli and the present author
~\cite{mcjhf1,mcjhf2}(MCJHF1,MCJHF2). Some brief comments
 are made here
 on these papers: some more detailed remarks are made in Section 8
 below.
\par In MCJHF1 effective gluon masses, $\mg$, determined from fits
 to $\Jp \rightarrow \gamma X$ and $\Up \rightarrow \gamma X$ were used,
 in conjunction with gluon mass correction factors calculated by PP
 (or re-calculated using pure phase-space considerations) to derive 
 a large number of $\alpha_s$ values from different charmonium and
 bottomonium branching ratios. Agreement with the expected pQCD
 evolution of $\alpha_s$  from the scale $M_{\Jp}/2$ to $M_{\Up}/2$
 was only obtained when the gluon mass corrections were applied. Also,
 only in this case, was good agreement found between the derived values
 of $\alpha_s$ and those obtained from deep inlastic scattering 
 experiments. In this paper only the Photiadis~\cite{Photiadis} higher order (HO) QCD
 correction (that is only applicable in the end-point region of the
 photon spectrum, and does not include real gluon radiation effects)
 was used, and relativistic corrections were completely neglected.
\par The second paper, MCJHF2, made essential use of the
 recently proposed non-relativistic quantum chromodynamics (NRQCD)
 formalism of Bodwin, Braaten and Lepage (BBL)~\cite{NRQCD}, in
 which both non-relativistic and HO QCD corrections (but not gluon
 mass effects) were treated in a rigorous way, order-by-order in perturbation
 theory, using an Operator Product Expansion. As suggested by BBL,
 the values of $\alpha_s$, and the leading relativistic correction
 parameter $r \simeq \langle v^2 \rangle$ were treated as free
 parameters in fits to various charmonium and bottomonium
 decay widths. Similar fits were also performed to the
 inclusive photon spectra in $\Jp$ and $\Up$ decays. No 
 consistent values of $r$ and $\alpha_s$ were found in the
 absence of gluon mass corrections. When the latter were
 included, consistent values of $\alpha_s$ similar to those
 found in MCJHF1 were obtained. However, in this case, the
 values of $r$ were found to be much smaller in absolute
 value than the expectations from potential model
 calculations, and even (as discussed further in Section 8
 below) of the wrong sign. The conclusion concerning the
 inability of the NRQCD formalism to describe the 
 experimental data, in the absence of gluon mass 
 corrections, was not however affected by the incorrect 
 treatment of relativistic corrections. 
\par The structure of this paper is as follows. Sections
 2, 3 and 4 are devoted to descriptions of the implementation
 of relativistic, HO QCD and gluon mass corrections respectively.
 Fits to the experimental data on  $\Jp \rightarrow \gamma X$
 and $\Up \rightarrow \gamma X$ to obtain, in each case, the corresponding
 effective gluon mass $\mg$, are described in Sections 5 and 6. Section 7
describes the determination of $\alsc$ and $\alsb$ from the experimental
 branching ratios
 $\Gamma(V \rightarrow {\rm hadrons})/\Gamma(V \rightarrow \gamma+{\rm hadrons})$,
 $V = \Jp, \Up$. The values of $\alpha_s$ obtained in this way are unaffected
 by relativistic corrections. In Section 8, the effective gluon mass values 
 obtained in this paper are compared with other estimates of the gluon mass in
 the literature. Finally Section 9 contains a brief summary and outlook.
 Details of the method used to simulate the effects of experimental 
 resolution on the inclusive photon spectrum are given in an Appendix.

\SECTION{\bf{Relativistic Corrections}}
\begin{table}
\begin{center}
\begin{tabular}{|c||c|c|} \hline 
  Reference &   $\rm{J/\psi}$ & $\Up$ \\
\hline
\hline
Bradley~\cite{Bradley} & 0.44  &  0.069 \\
\hline
Eichten {\it el al.}~\cite{Eichten} & 0.20  &  0.096 \\
\hline
Bergstr\"{o}m {\it el al.} I~\cite{Bergs1} &    &    \\
$V(r)=0.2r-\frac{0.25}{r}$ & 0.47  & ---  \\
$V(r)= \frac{1}{2}\mu \omega_0^2 r^2$ & 0.21  & ---  \\
$V(r)=  Tr-\frac{\alpha_s(r)}{r}$ & 0.34  & ---  \\
       &   &    \\
\hline
Bergstr\"{o}m {\it el al.} II~\cite{Bergs2} &    &    \\
$V(r)=0.2r-\frac{4 \alpha_s}{3r}$ & 0.47  & ---  \\
$V(r)=0.163r-\frac{4 \alpha_s}{3r}$ & ---  & 0.47  \\
     &   &    \\ 
\hline
Beyer {\it el al.}~\cite{Beyer} & 0.21  &  0.18 \\
\hline
Chiang {\it el al.}~\cite{Chiang} & 0.21  &  0.078 \\
\hline
Chao {\it el al.}~\cite{Chao} & 0.26  &  0.13 \\
\hline
Schuler~\cite{Schuler} &    &   \\
$V(r)= \lambda r^{\nu}$ &   & \\
$\nu = -0.1$ & 0.36  & 0.075  \\
$\nu = 0.0$ & 0.32  & 0.066  \\
$\nu = 0.3$ & 0.25  & 0.048  \\
\hline
\end{tabular}
\caption[]{{\it Estimations of $<v^2>$ for the $\rm{J/\psi}$ and the $\Upsilon$}}
\end{center}
\end{table}
\begin{figure} 
 \begin{center}
   \includegraphics[width=10cm]{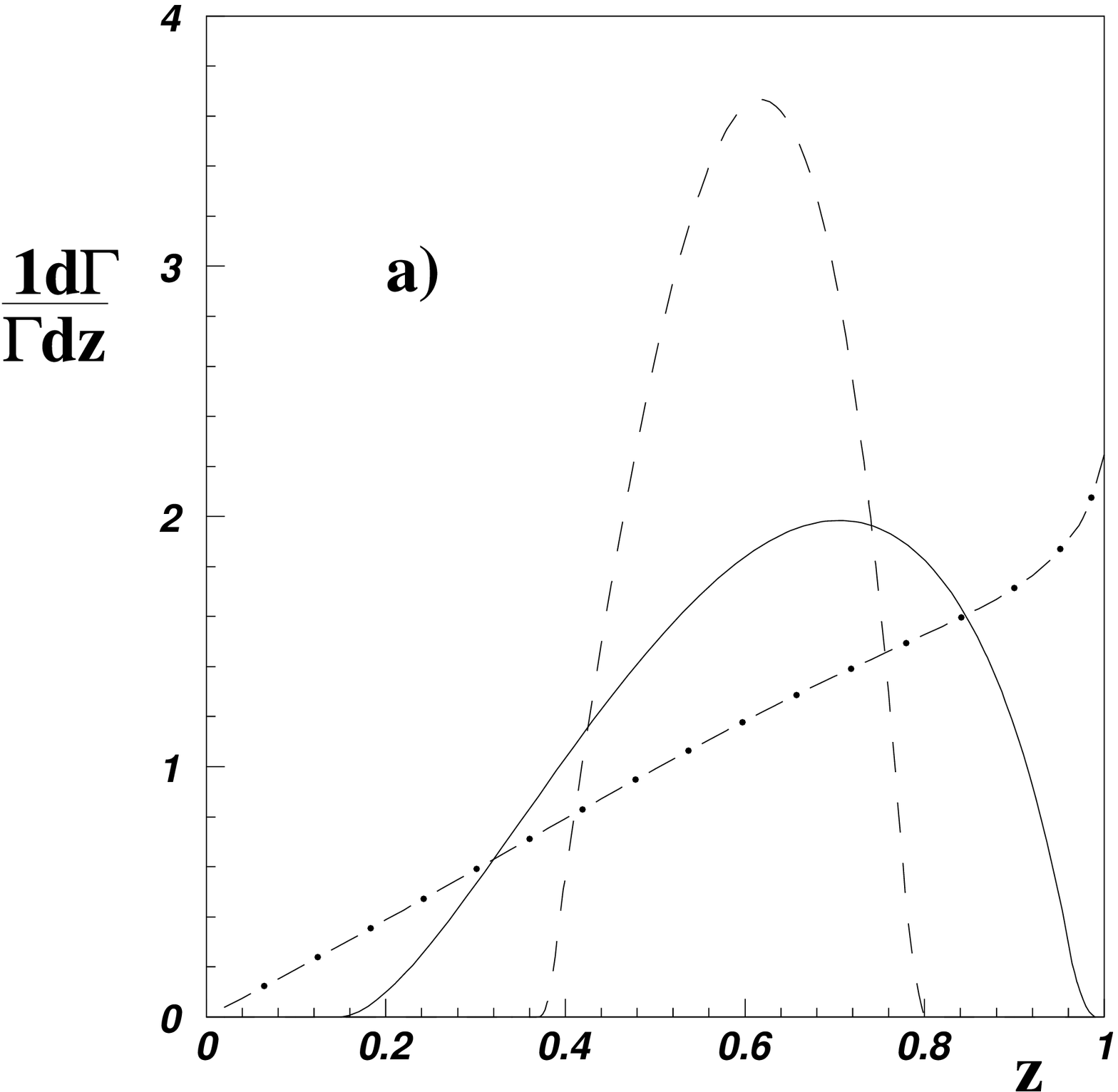}
   \includegraphics[width=10cm]{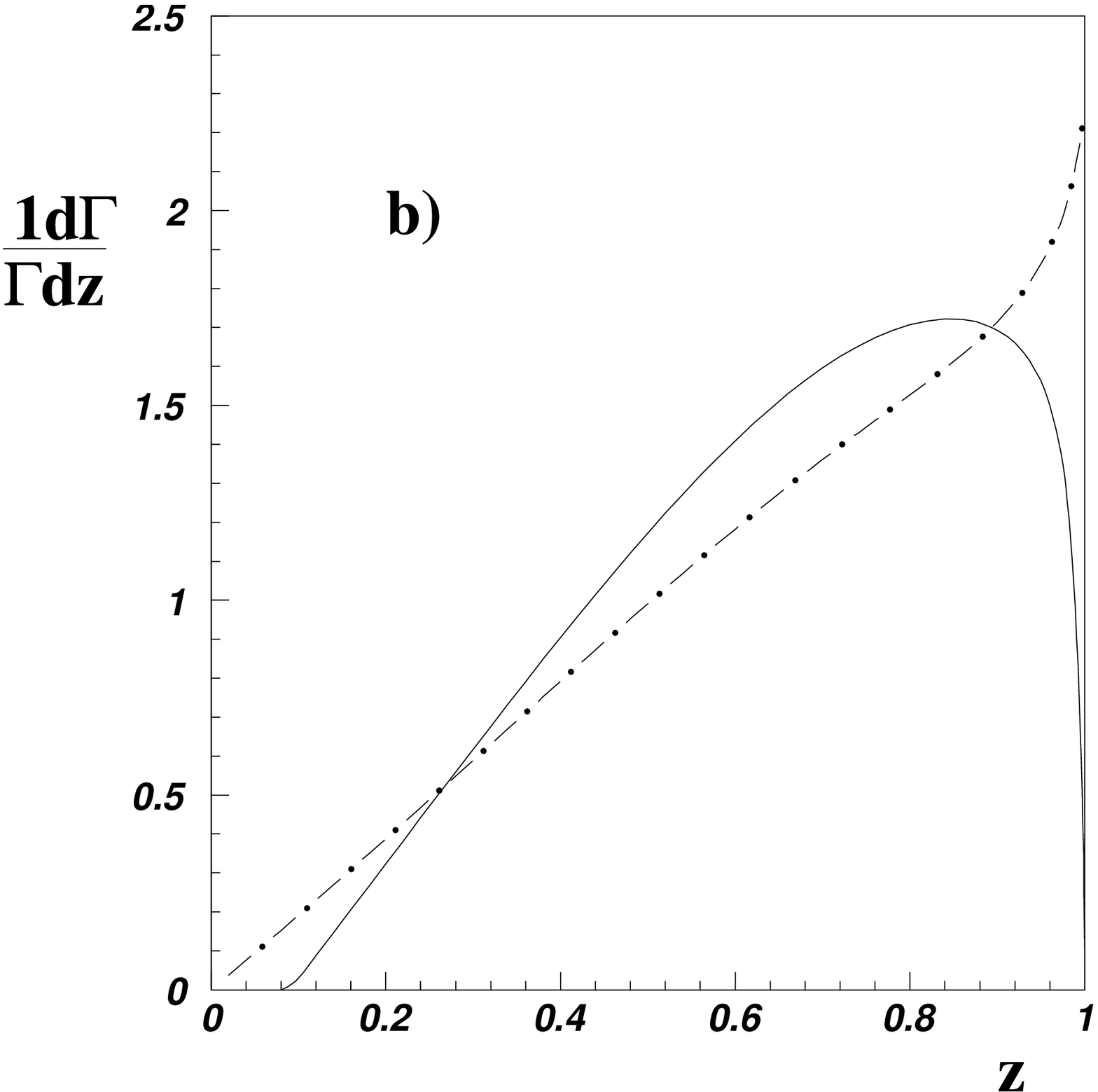}
 \end{center}
 \caption{{\it Theoretical predictions for relativistic corrections to the
 inclusive photon spectrum
 in  a) $\Jp$, and b) $\Up$, radiative decays. In a), the LO QCD prediction~\cite{OrePow}
 is shown as the dot-dashed line. The solid line gives the prediction 
 of Eqn(2.8) for $<v^2> = 0.28$. The dashed line shows the prediction
 of Eqn(2.8) neglecting the order $<v^2>^2$ term. In this case only
 the positive part of the prediction is shown. In b), the LO QCD prediction~\cite{OrePow}
 is again shown as the dot-dashed line. The solid line gives the prediction 
 of Eqn(2.8) for $<v^2> = 0.09$. In both a) and b) the peak of the function
 in Eqn(2.8) at small values of $z$ has been suppressed}} 
\label{relcor}
\end{figure}
 Relativistic corrections to the van Royen-Weisskopf formula~\cite{vRWeiss} for the 
 decay rate of a vector meson V into a charged lepton pair:
\begin{equation}
 \Gamma({\rm{V} \rightarrow \ell^+\ell^-}) = \frac{16 \pi \alpha(\mv)^2 e_{\rm{Q}}^2}{\mv^2}
 |\psi(0)|^2
\end{equation}
were calculated by Bergstr\"{o}m {\it {\it et al.}}~\cite{Bergs}. A relativistic correction factor,
$f_{RC}$, to the leptonic decay width
 $\Gamma({\rm{V} \rightarrow \ell^+\ell^-})$ was found with the general form:
\begin{equation}
f_{RC}({\rm{V} \rightarrow \ell^+\ell^-}) = (1-\frac{1}{3}r)^2
\end{equation}
where 
\begin{equation}
 r = \int \frac{d^3p}{(2 \pi)^3}\frac{(E(p)-\mQ)}{E(p)}\frac{\pst}{\psi(0)}
\end{equation}
and $E$, $p$ and $\mQ$ are the energy, momentum and mass of the bound heavy quark.
$\pst$ is the wavefunction in relative momentum space, related to the spatial wavefunction
 at the
origin, $\psi(0)$, by the expression:
\begin{equation}
 \psi(0) = \int \frac{d^3p}{(2 \pi)^3} \pst.
\end{equation}
In the approximation where the valence quarks of the vector meson are considered to 
be symmetrically bound in the meson rest frame, so that $E(p)=\mv/2$, it follows that
\begin{eqnarray}
\frac{E(p)-\mQ}{E(p)} = \frac{\epsilon}{\mv} &  = & \frac{\sqrt{p^2+\mQ^2}-\mQ}
{\frac{\mv}{2}} \nonumber \\
 & = & \frac{p^2}{\mQ\mv}+O(p^4) \nonumber \\
 & = &  \frac{p^2}{2\mQ^2}+O(p^4) \nonumber \\
 & = &  \frac{v^2}{2}+O(v^4).
\end{eqnarray}
Here $\epsilon$ is the `binding energy' $\mv-2\mQ$ and $v$ is the 
velocity
 of the heavy quark\footnote{In units with $c=1$}. Using Eqn(2.5), (2.2) may be written as:
\begin{equation}
f_{RC}({\rm{V} \rightarrow \ell^+\ell^-}) = (1-\frac{1}{6}<v^2>)^2 + O(v^6) 
\end{equation}
where $<v^2>$ is the mean value of the squared velocity, that 
depends on the bound state potential. Similar relativistic corrections
were calculated for several decay processes of heavy quarkonia by
Keung and Muzinich (KM)~\cite{KM}. The calculation was based on a non-relativistic
reduction of the Bethe-Salpeter equation~\cite{BetSal} for the relativistic
quark-antiquark bound state problem. The results of KM were presented as
$O(v^2)$ corrections to the decay rate rather than to the decay amplitude,
as in Eqn(2.6) above. In the present paper all relativistic corrections
are applied at the amplitude level so that additional $O(v^4)$  terms are 
added to the results quoted by KM to `complete the square' and obtain a
positive definite decay rate. This correction is important only for
charmonium decays where, because of the relatively large value of $<v^2>$,
the corrected decay rate becomes negative for both small and large values
of $z \equiv 2 E_{\gamma}/\mv$, if only the  $O(v^2)$ correction terms are 
retained. KM confirm the relativistic correction factor for
 $\rm{V} \rightarrow \ell^+\ell^-$ given in Eqn(2.6) and find also:
\begin{equation}
f_{RC}({\rm{V} \rightarrow ggg}) = f_{RC}({\rm{V} \rightarrow \gamma gg}) =  (1-2.16<v^2>)^2 + O(v^6). 
\end{equation} 
Of particular importance for the present study, KM also give, in their Eqn(3.5), the
relativistic correction to the inclusive photon spectrum in ${\rm{V} \rightarrow \gamma gg}$.
`Completing the square' to obtain a positive definite differential decay rate yields
the spectrum: 
\begin{eqnarray}
\frac{1}{\Gamma}\frac{d \Gamma}{dz} & = & \frac{1}{C_N}\left[\sqrt{f_0(z)}+\frac{g(z)<v^2>}
{2\sqrt{f_0(z)}} \right]^2   \nonumber \\
 & = & \frac{1}{C_N}\left[f_0(z)+g(z)<v^2>+\frac{(g(z)<v^2>)^2}
{4 f_0(z)} \right],  
\end{eqnarray}
where~\cite{GSrev}:
\begin{equation} 
C_N=(\pi^2-9)
\left[1+<v^2>\left({{5}\over{3}}-{{1}\over{4}}{{(9\pi^2-68)}\over{(\pi^2-9)}}\right)\right]\sim
(\pi^2-9)(1-4.32<v^2>), 
\end{equation}
and 
\begin{equation}
 g(z) \equiv  \frac{5 f_0(z)}{3}-\frac{f_1(z)}{12}.
\end{equation}
 The functions $f_0(z)$ and $f_1(z)$  are reported in Eqns(15) and (16) of
 Reference
 \cite{mcjhf2}. In the approximation used here, relativistic corrections
 are completely specified by the single parameter $<v^2>$. Although
 one may hope, in the future, to determine this non-perturbative parameter
 by lattice QCD methods~\cite{lattcal}, the only existing estimates are 
derived from potential models of the quarkonium bound state. Some of
 the estimates of $<v^2>$ for the $\rm{J/\psi}$ and the $\Upsilon$,
 that have been given in the literature, are presented in Table 1.
 Usually in these papers the relativistic correction factor for the decay 
 $\rm{V} \rightarrow \ell^+\ell^-$ is quoted. For the entries
 in Table 1, this is converted into a value of $<v^2>$ using
 Eqn(2.6). In the case of Chiang {\it et al.}~\cite{Chiang} Eqn(2.7)
 is used, and for Chao {\it et al.}~\cite{Chao} the ratio
 $f_{RC}({\rm{V} \rightarrow ggg})/f_{RC}(\rm{V} \rightarrow
 \ell^+\ell^-)$. Schuler~\cite{Schuler} calculated directly
  values of $<v^2>$ for a series of different charmonium
  and bottomonium states as a function of the prameter $\nu$ in
  a power-like potential of the form $V(r)=\lambda r^{\nu}$. 
 The range of different values of $<v^2>$ presented in
 Table 1 is very wide: 0.20-0.47 for the $\rm{J/\psi}$ and 0.048-0.47
 for the $\Upsilon$. Apart from the estimates of Bergstr\"{o}m
 {\it et al.}~\cite{Bergs2} and Beyer {\it et al.}~\cite{Beyer}
 the value of $<v^2>$ is found to be significantly larger for
 the  $\rm{J/\psi}$ than for the $\Upsilon$, as intuitively expected, 
 given the smaller mass of the charm quark. The near equality of 
 the values of $<v^2>$ for the  $\rm{J/\psi}$ and the $\Upsilon$
 and the very large value found for the $\Upsilon$ in 
 References~\cite{Bergs2} and~\cite{Beyer} may be a consequence of
 an extreme choice
 of the parameters of the potential in the case of the $\Upsilon$.
 \par In the present paper, more weight is given to the more recent
 results of Chiang {\it et al},  Chao {\it et al} and Schuler which
 are roughly consistent with each other. In the following, the values
 taken are: $<v^2> = 0.28$ for the  $\rm{J/\psi}$ and $<v^2> = 0.09$
 for the $\Upsilon$, which lie near the middle of the range of values
obtained by these last three authors. As it will be seen that the 
 effects of relativistic corrections on the shape of the fitted 
 photon spectra are, after the inclusion of gluon mass effects,
 small (as already conjectured in
 Refs~\cite{mcjhf1,mcjhf2}) the conclusions of the present work
 are not sensitive to the precise values assumed for $<v^2>$.
 The relativistically corrected inclusive photon spectra for the 
 $\rm{J/\psi}$ and the $\Upsilon$ calculated using Eqn(2.8) are shown,
 in comparison with the lowest order (LO) QCD prediction~\cite{OrePow}, in
 Figs.1a and 1b respectively. Also shown in Fig.1a is the curve
 given by truncating the correction to the decay rate above
 $O(v^2)$. In this case the spectrum is set to zero if the
 prediction is negative. It may be remarked that Eqn(2.8) shows
 singular behaviour as $z \rightarrow 0$\footnote{
 The corresponding peaks near $z =0.0$ are suppressed in Fig.(1)}.
 However, this does not affect
 any of the fits presented below, as no experimental
 measurements exist for $z < 0.2$.

\SECTION{\bf{Higher Order QCD Corrections}}
\begin{figure}[htbp]
\begin{center}\hspace*{-0.5cm}\mbox{
\epsfysize10.0cm\epsffile{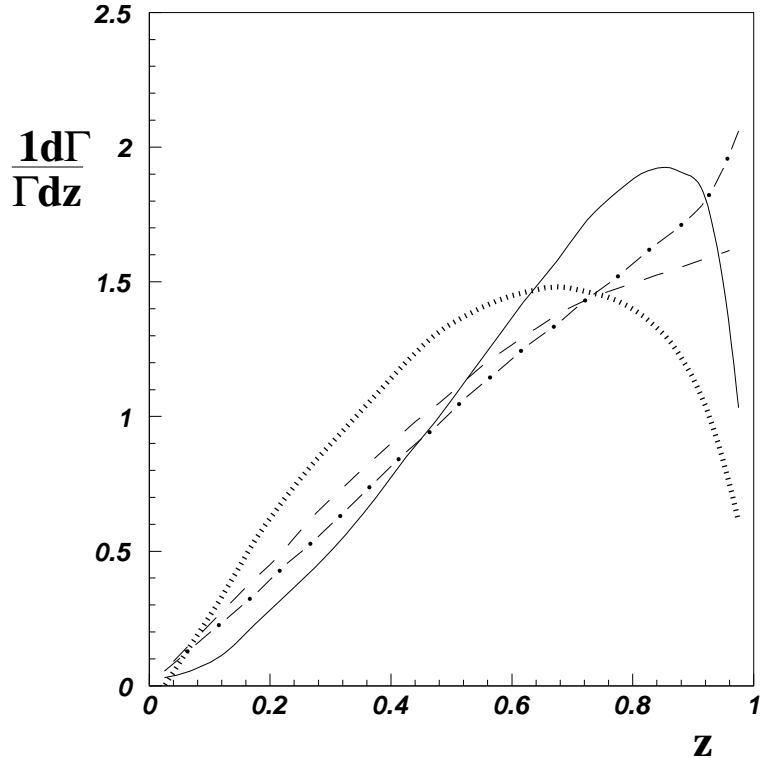}}
\caption{{\it QCD predictions for the inclusive photon spectrum in
 radiative $\Up$ decays.
 Dot-dashed: LO~\cite{OrePow}, dotted: RDF~\cite{RDF},  dashed:
 Photiadis~\cite{Photiadis} and solid curve: Kr\"{a}mer(NLO)~\cite{Kramer}.}}
\label{fig-fig1}
\end{center}
 \end{figure}
 To lowest order (LO) in perturbative QCD, the inclusive photon spectrum is
 described by the process ${\rm{V} \rightarrow \gamma gg}$. Assuming massless gluons
 and neglecting relativistic corrections, the shape of the photon spectrum is the same
 as in orthopositronium decay~\cite{OrePow}:
\begin{eqnarray}
\frac{1}{\Gamma}\frac{d \Gamma}{dz} & = & \frac{1}{(\pi^2-9)}\left[\frac{4(1-z)\ln(1-z)}
{z^2} \right. \nonumber \\
 &   &  \left. -\frac{4(1-z)^2\ln(1-z)}{(2-z)^2}+\frac{2z(2-z)}{z^2}+\frac{2z(1-z)}{(2-z)^2}\right].
\end{eqnarray}
The first estimation of higher order QCD corrections to the spectrum was made by
R.D.Field (RDF)~\cite{RDF}. These QCD effects were calculated using a parton shower
Monte Carlo program in which the process ${\rm g \rightarrow gg}$ was iterated.
 The invariant mass of the cascading virtual gluons was cut off at the scale
 $\mu_c = 0.45$ GeV and a value $\Lambda = 0.2$ GeV was used for the QCD scale 
 parameter in the parton shower. For $\Up$ decays the average, perturbatively generated, 
`effective gluon mass', i.e. the mass of the virtual gluon initiating the parton
 cascade, was 1.6 GeV. Because of the low value of the cut-off scale, the shapes of the
 photon spectra for $\Jp$ and $\Up$ decays were predicted to be similar. In
 both cases the average value of $z$ was found to be 0.57, and even in the
 case of the decay of a hypothetical state with a mass of 60 GeV, the
 average $z$ increased only to 0.59. The RDF spectrum for $\Up$ decays is shown in 
 Fig.2 as the dotted line. 
  The parton cascade used by RDF does not take into account
 QCD coherence effects in gluon radiation~\cite{QCDcoh} usually implemented
 in parton shower Monte Carlo event generators by an `angular ordering'
 ansatz~\cite{angord}. The effect of this coherence, which is the QCD 
 analogue of the `Chudakov Effect'~\cite{Chudakov} in QED, is to suppress
 corrections due to real gluon radiation for kinematical configurations that
 yield photons close to the kinematical end point. In this case the two
 primary recoiling gluons are almost collinear, forming an effectively
 colourless current from which the radiation of large angle
 secondary gluons is strongly suppressed by destructive interference~\cite{Hautmn}.
 The radiation of almost collinear gluons is not suppressed, but such
 radiation will hardly modify the shape of the LO spectrum. It is thus to
 be expected that the neglect of QCD coherence in the parton shower used
 to calculate the RDF spectrum will result in a too strong suppression of the
 spectrum in the end point region. The comparison, shown below, with a complete NLO
 perturbative QCD calculation, where such coherence effects are taken
 into account, indicates that this is indeed the case.
  The RDF spectrum 
 gives a good description of four out of five of the experimental measurements
 of the $\Up$ spectrum (see Section 6 below). It will be seen however, that,
 for the case of $\Jp$ decays, the spectrum is much too hard to describe the
 experimental measurements.
\par The second estimate of higher order QCD corrections to the inclusive
 photon spectrum in $\Up$ decays was made by Photiadis~\cite{Photiadis}.
 This calculation, expected to be valid near to $z=1$, neglected completely
 real gluon radiation, which, as discussed above, is strongly
 suppressed in this region, but resummed to all orders in $\alpha_s$ the
 leading logarithmic terms of the form $\ln(1-z)$ resulting from the exchange
 of virtual gluons and quarks between the two recoiling gluons of the LO
 diagram. As shown in Fig.2 (the dashed curve), these effects give only a modest
 suppression of the LO spectrum near $z=1$.
\par The most recent result on higher order QCD corrections to the $\Up$ spectrum
is the complete NLO calculation of Kr\"{a}mer
~\cite{Kramer} that is also shown in Fig.2 as the solid curve. It can be seen
 that strong suppression
 occurs only very near to $z=1$, and is much less marked than in the case of the 
 RDF spectrum. 
\par Since the Photiadis calculation does not include the effects of 
 real gluon radiation, it can be argued that the corrections calculated by both
Kr\"{a}mer and Photiadis should be applied. This will double count virtual 
 corrections of the type shown in Fig.2c of Reference~\cite{Kramer}, but
 should give a better description, particularly away from the end point
 region, than using only the Photiadis correction. 
\par To date, no calculations of the inclusive photon spectrum in heavy quarkonia
 decays have been made taking into account, at the same time, higher order QCD
corrections, relativistic corrections and genuine gluon mass effects
\footnote{That is, including a fixed gluon mass in the calculation of both the
 invariant amplitude and the phase space, and taking into account the 
longitudinal polarisation states of the gluons. This is to be contrasted
with the parton shower model of RDF where an effective gluon mass (actually
 a time-like gluon virtuality) is perturbatively generated from massless
gluons. The distinction between `genuine' and `effective' gluon masses is
 discussed further in Section 4.}. Indeed, since the pioneering paper of PP~\cite{PP},
gluon mass effects have been completely calculated only for the LO processes:
${\rm{V} \rightarrow ggg}$ and ${\rm{V} \rightarrow \gamma gg}$~\cite{LW}
(see Section 4 below).
\par In the present analysis, the higher order QCD calculations of Photiadis
 and Kr\"{a}mer, made for $\Up$ decays are also used, unmodified, for $\Jp$
 decays. In fact it will be seen that the observed end point suppression
 of the photon spectrum of the $\Jp$ is so large, as compared to
 the predicted effect of both relativistic and higher order QCD corrections,
 that these play only a minor role. Indeed, the value of the effective
 gluon mass, $\mg$, needed to describe the experimental spectrum is little affected
 by the inclusion of these corrections. The ansatz used to apply the
 higher order (HO) QCD corrections is to multiply the relativistically
 corrected photon spectrum given by Eqn(2.8) by the QCD correction factor:
\begin{equation}
C_{QCD}=\frac{d\Gamma_{HO}}{dz}/\frac{d\Gamma_{LO}}{dz}.
\end{equation}
\par In the case of fits with $\mg \neq 0$, phase space limitations are
 taken into account by the replacement: $z \rightarrow z/z_{MAX}$ in Eqn(3.2)
 where 
\begin{equation}
z_{MAX} = 1-\frac{4 \mg^2}{\mv^2}.
\end{equation}
\par In view of the large value found for the ratio $\mg/\mv$ for the $\Jp$, 
it is to be expected that phase space suppression effects will be even 
 more important for the HO corrections than for the LO process. This will
 reduce even further the effect of such corrections on the fitted value
 of $\mg$.

\SECTION{\bf{Gluon Mass Effects}}

 The possibility of gluon mass effects in the decay $\Jp \rightarrow \gamma X$ 
 was first considered  by PP~\cite{PP}. They noted that
the very strong suppression of the end point region of the photon spectrum
 measured by the Mark II collaboration~\cite{Mark II} could be explained by 
 introducing a gluon mass of about 0.8 GeV. The comparison of the PP prediction
 with the experimental data did not, however, take into account experimental resolution
 effects which are very large in this case. Also relativistic and higher order
 QCD corrections were not included. The aim of the present paper
 is perform a similar comparison to that made by PP, but using all available
 experimental data on both $\Jp$ and $\Up$ radiative decays, as well as
 including experimental resolution effects, relativistic corrections 
 and higher order QCD corrections. As the latter two effects also 
 suppress the end point region of the photon spectrum, only a complete
 quantitative analysis, including all relevant effects, can show if
 the introduction of a non-vanishing effective gluon mass is required to describe
 the experimental data.   
 \par Introducing an `effective gluon mass'
 in the calculation of the Born diagram has two effects:
\begin{itemize}
\item[(i)] Restriction of the available phase space, i.e. modification of the 
 boundary of the Dalitz plot. 
\item[(ii)] Contributions to the amplitude from longitudinal gluon
 polarisation states.
\end{itemize}
 As will be seen, the effect (i) is, by far, the most important.
\par The gluon mass is `effective' because it
 is defined only at the level of the Born diagram. When such a prediction is fitted to
 the data, which includes QCD corrections to all orders, the value of $\mg$ is expected
 to be different from the value obtained if the gluon mass were correctly included also
 in  higher order diagrams in the prediction. In fact, a `genuine' gluon mass $\Mg$ might
 be operationally defined as the effective mass corresponding to a hypothetical
 all-orders pQCD calculation with massive gluons.
 If phase space limitations are very important,
 as in the
 case of the $\Jp$, the tree level `effective' value is expected to be lower than
 the `genuine'
 value that would be found in a fit that properly includes gluon mass effects
 at all orders. If, on the other hand, the `genuine' gluon mass is small compared to that of
 the decaying state, the effects of the non-vanishing gluon mass will be limited
 to a small region near the boundaries of phase space. In this case the tree level
 `effective' mass is mainly generated perturbatively by splitting into
 gluon and quark pairs (as in the RDF model) and is expected
 to be much larger than the `genuine' value. However, the `genuine' mass $\Mg$ found
  by comparing the prediction of the all-orders pQCD calculation to the data
  is expected to be independent of the mass of the decaying state. The above argument
  also shows that, for some mass of the decaying state, the tree level `effective' mass
  and the `genuine' gluon mass should be equal. It may be conjectured that this is
  almost the case for the $\Up$.

\par The correction curves for the processes $\Jp \rightarrow {\rm ggg}$ and 
 $ \eta_c \rightarrow {\rm gg}$ calculated by PP and shown in their Fig.1
 took into account both the effects (i) and (ii), but no
 explicit formulae were given. In the present paper essential use is made of
 formulae including
 both effects (i) and (ii) obtained by Liu and Wetzel~\cite{LW}.
 For the decay  ${\rm{V} \rightarrow \gamma gg}$, the fully differential
 spectrum is:
\begin{eqnarray}
\frac{1}{\Gamma_0}\frac{d\Gamma}{dzdx_1dx_2} = &  & 
\frac{1}{(\pi^2-9)}\frac{1}{z^2(x'_2)^2(x'_3)^2}\left[   \right. 
 \frac{8}{3} \eta (1-\frac{25}{2}\eta)(1-2\eta)^2 \nonumber \\
   &+&32z(1-2\eta)(1-\frac{\eta}{4})\eta^2+z^2(1-2\eta)(1-6\eta-6\eta^2)
  \nonumber \\  
   &-&2(z^3+(x'_2)^3+(x'_3)^3)(1-\frac{10}{3}\eta-2\eta^2)+z^4(1+\frac{\eta}{2})
 \nonumber \\
   &+&((x'_2)^2+(x'_3)^2)(1-8\eta+22\eta^2+8\eta^3)-(x'_2)^2(x'_3)^2\eta  \nonumber \\
   &+&(z^4+(x'_2)^4+(x'_3)^4)(1+\frac{\eta}{2}) \left] \right.
\end{eqnarray} 

\[ z \equiv 2E_{\gamma}/\mv,~~~\eta\equiv (\mg/\mv)^2,
~~~x_i \equiv 2 E_i/\mv,~~~x'_i \equiv x_i-2\eta,~~~
 E_i = {\rm~gluon~energy}  \] 
 Here ${\Gamma_0}$ is the radiative width uncorrected for gluon mass effects. 
 The allowed phase space region is defined by the conditions\footnote{Note that $\eta$ 
 is defined differently here than in Reference~\cite{mcjhf2}.}:
\begin{eqnarray}
2 & = & z+x_2+x_3    \\
  0 & \le & z  \le ~ 1-4\eta    \\ 
 x_2^{min} & \le & x_2~\le~ x_2^{max}  \\
 x_2^{max}  & = & 1-\frac{z}{2}\left[1-\sqrt{1-\frac{4 \eta}{1-z}}\right] \\
 x_2^{min} & = & 1-\frac{z}{2}\left[1+\sqrt{1-\frac{4 \eta}{1-z}}\right] 
\end{eqnarray}
\par In the gluon mass dependent fits to be presented below, the functions 
 $f_0(z)$ and $f_1(z)$ in the KM formula (2.8) are replaced by the functions:
\begin{eqnarray}
f_0(z,\mg) & = & \int_{x_2^{min}}^{x_2^{max}}dx_2 f_0(z,x_2,\mg)  \\
f_1(z,\mg) & = & \int_{x_2^{min}}^{x_2^{max}}dx_2 f_1(z,x_2)  
\end{eqnarray}
Where $f_0(z,x_2,\mg)$ is derived from Eqn.(4.1) and $f_1(z,x_2)$ from 
 Eqn.(3.5) of KM.
 Thus phase space effects are taken properly into account in both $f_0$ and $f_1$,
 whereas the effect of longitudinal polarisation states are included only in $f_0$.
 As  will be shown below, the latter effect is 
 much smaller than the former, so that the effect on the fit results of the
 uncalculated contribution of gluon longitudinal 
 polarisation states
 on the relativistic correction coefficient $f_1$ is expected to completely
 negligible.

\SECTION{\bf{The Decay $\Jp \rightarrow \gamma X$}}
\begin{figure}
 \begin{center}
   \includegraphics*[width=7cm,]{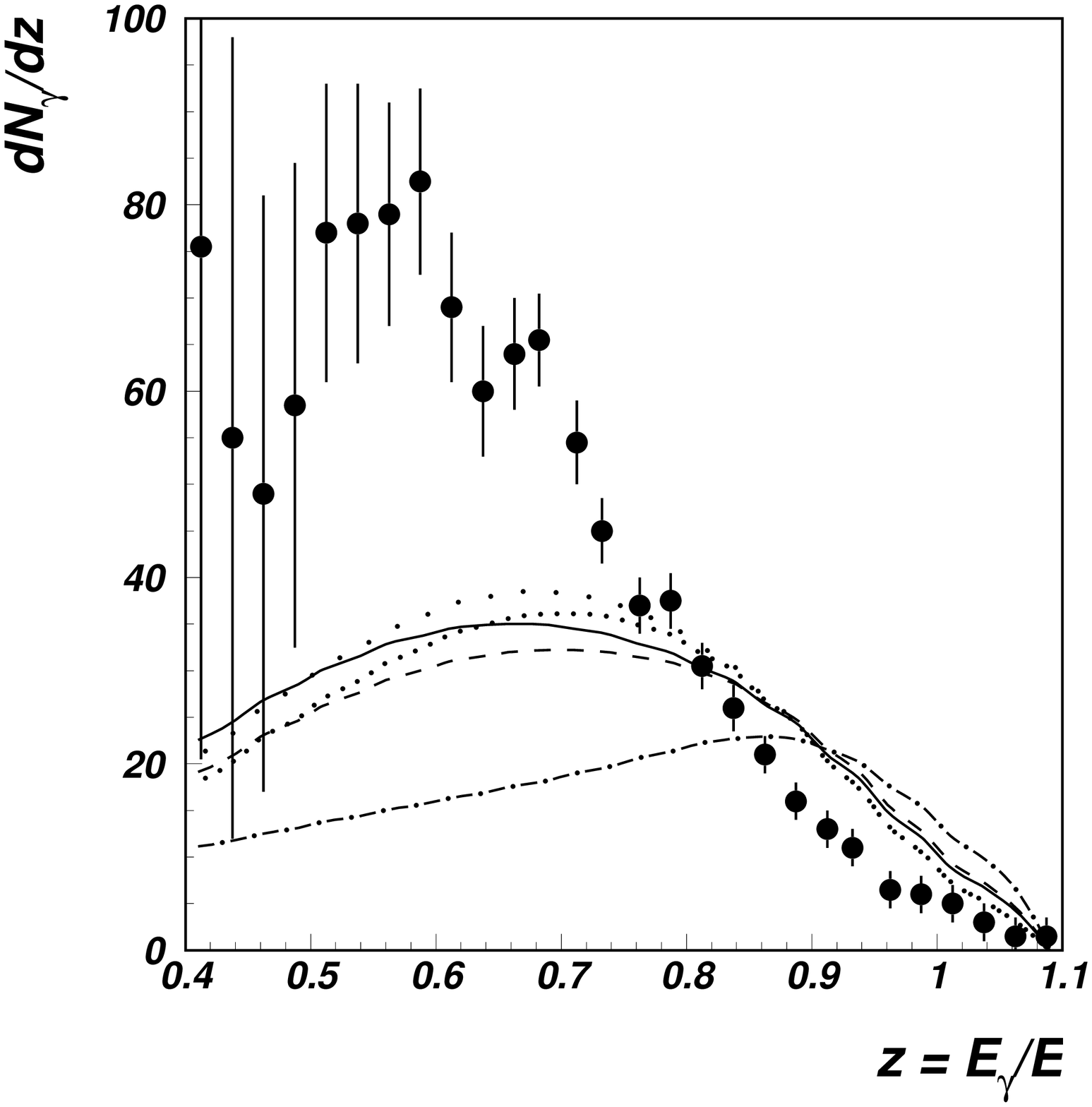} 
   \includegraphics*[width=7cm]{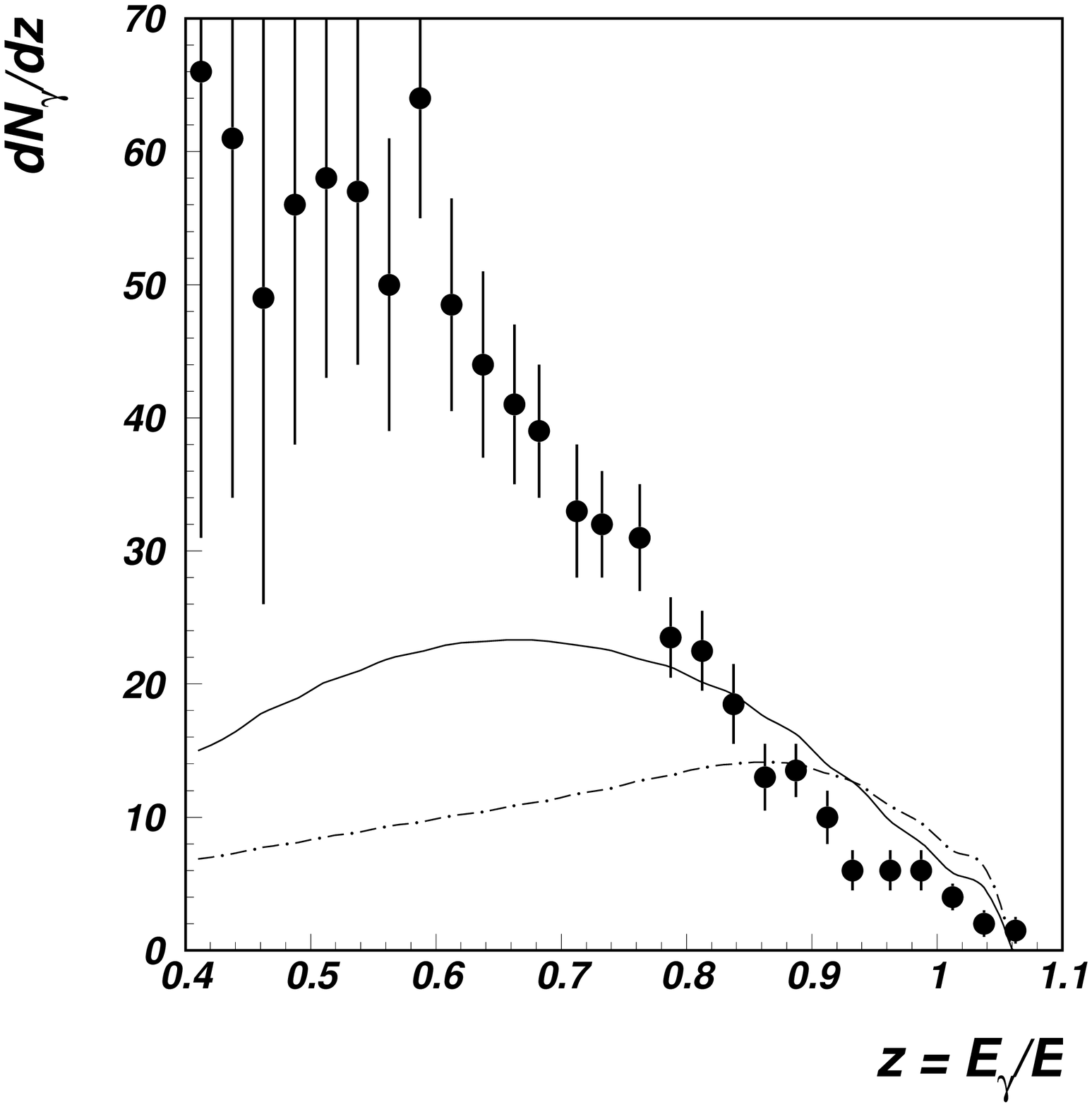}\\
\vspace{1cm}  
   \includegraphics*[width=7cm]{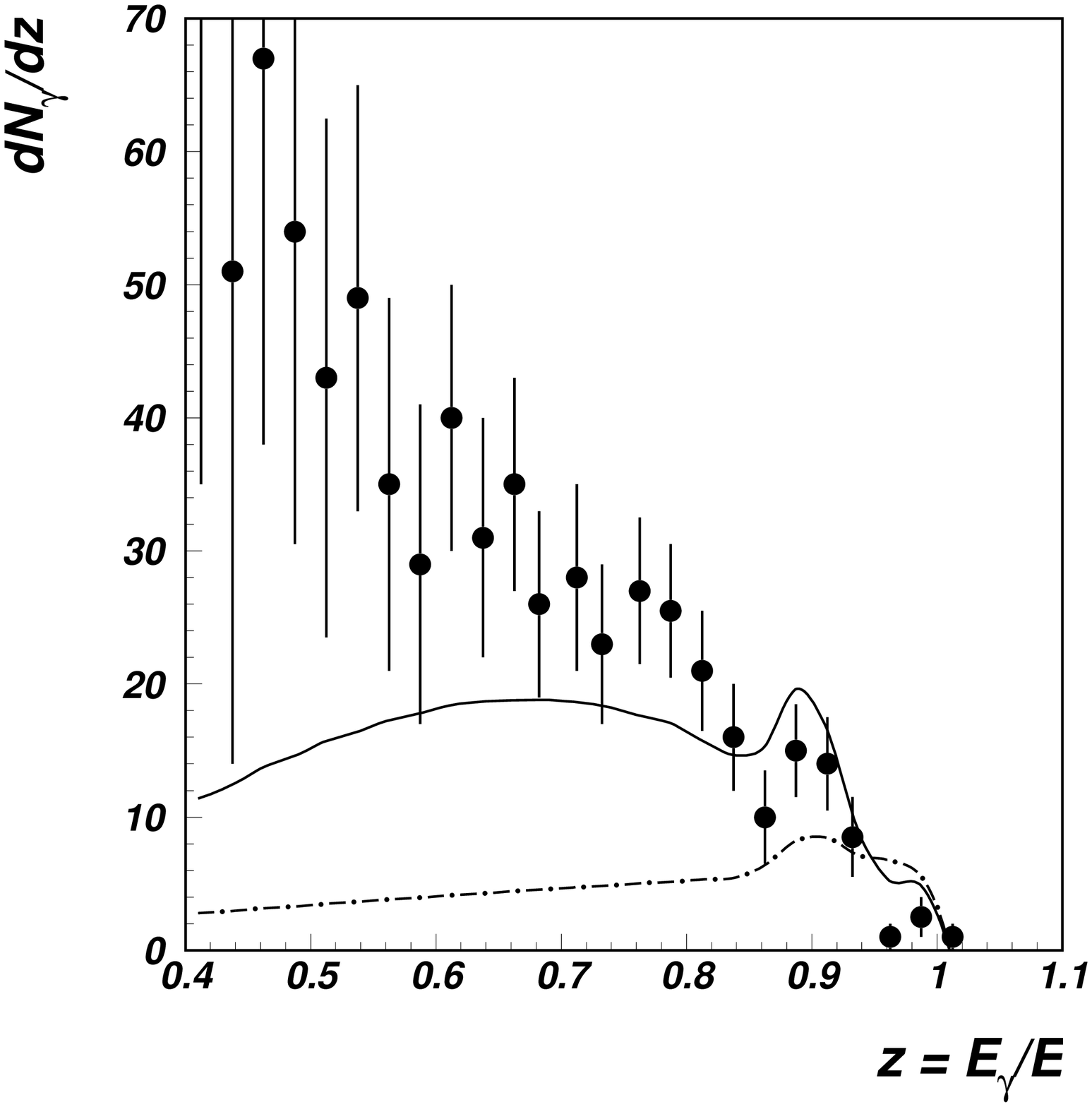}
 \end{center}
\caption{{\it Fits, assuming $\mg=0$,
 to inclusive photon spectra in $\Jp$ decays.  Mark II (top left),
 Mark II cascade (top right),  Mark II conversion (bottom). Dash-dotted line:
 LO QCD prediction~\cite{OrePow}, dashed line: Relativistic Correction (RC) included
~\cite{KM}, solid line: RC and Photiadis~\cite{Photiadis} HO QCD correction, fine
 dotted line: RC and Kr\"{a}mer~\cite{Kramer} HO QCD correction and dotted line:
RC, Photiadis and Kr\"{a}mer corrections}}
\label{jpsimg0}
\end{figure}
\begin{figure}
 \begin{center}
   \includegraphics*[width=7cm,]{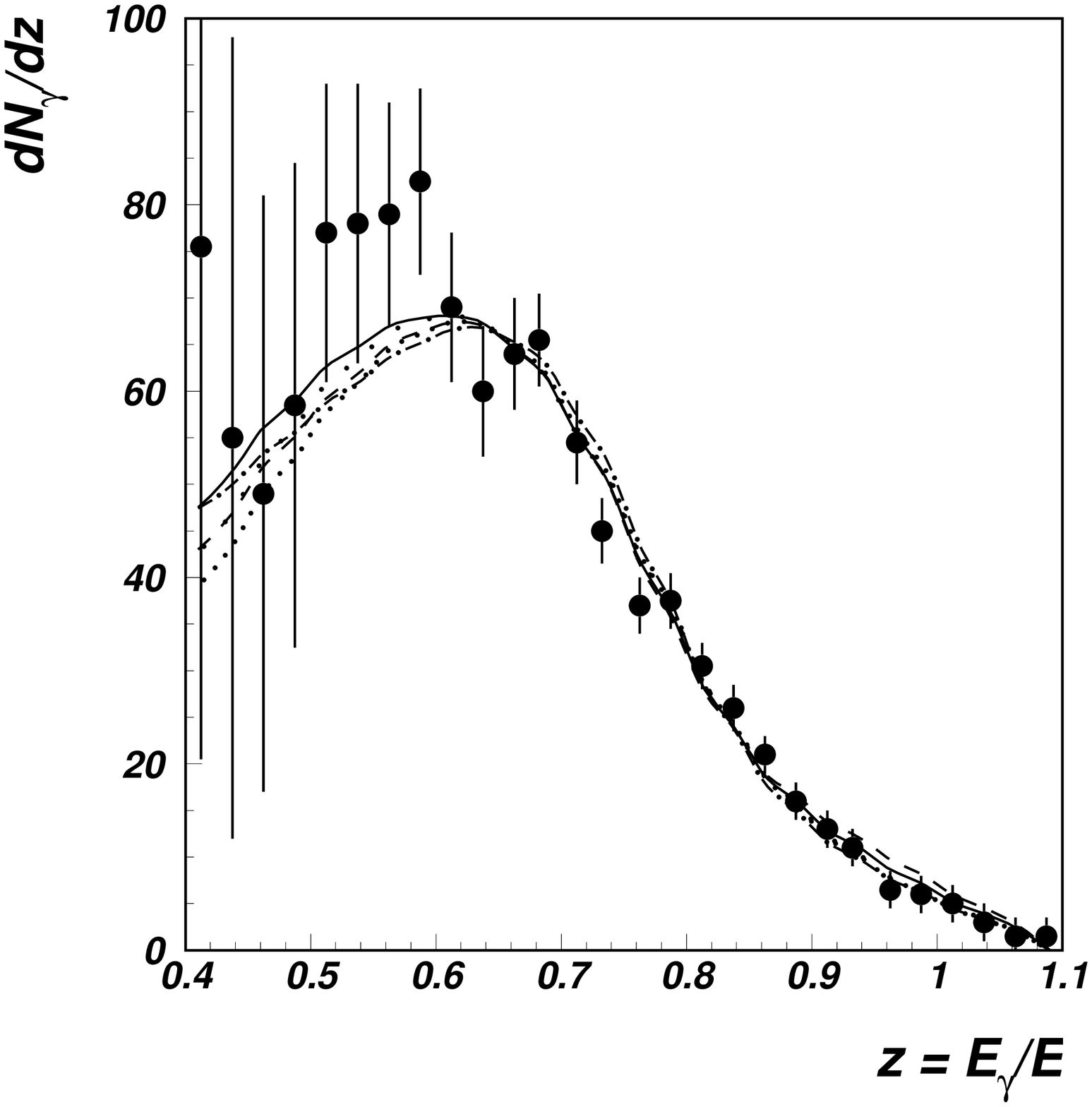} 
   \includegraphics*[width=7cm]{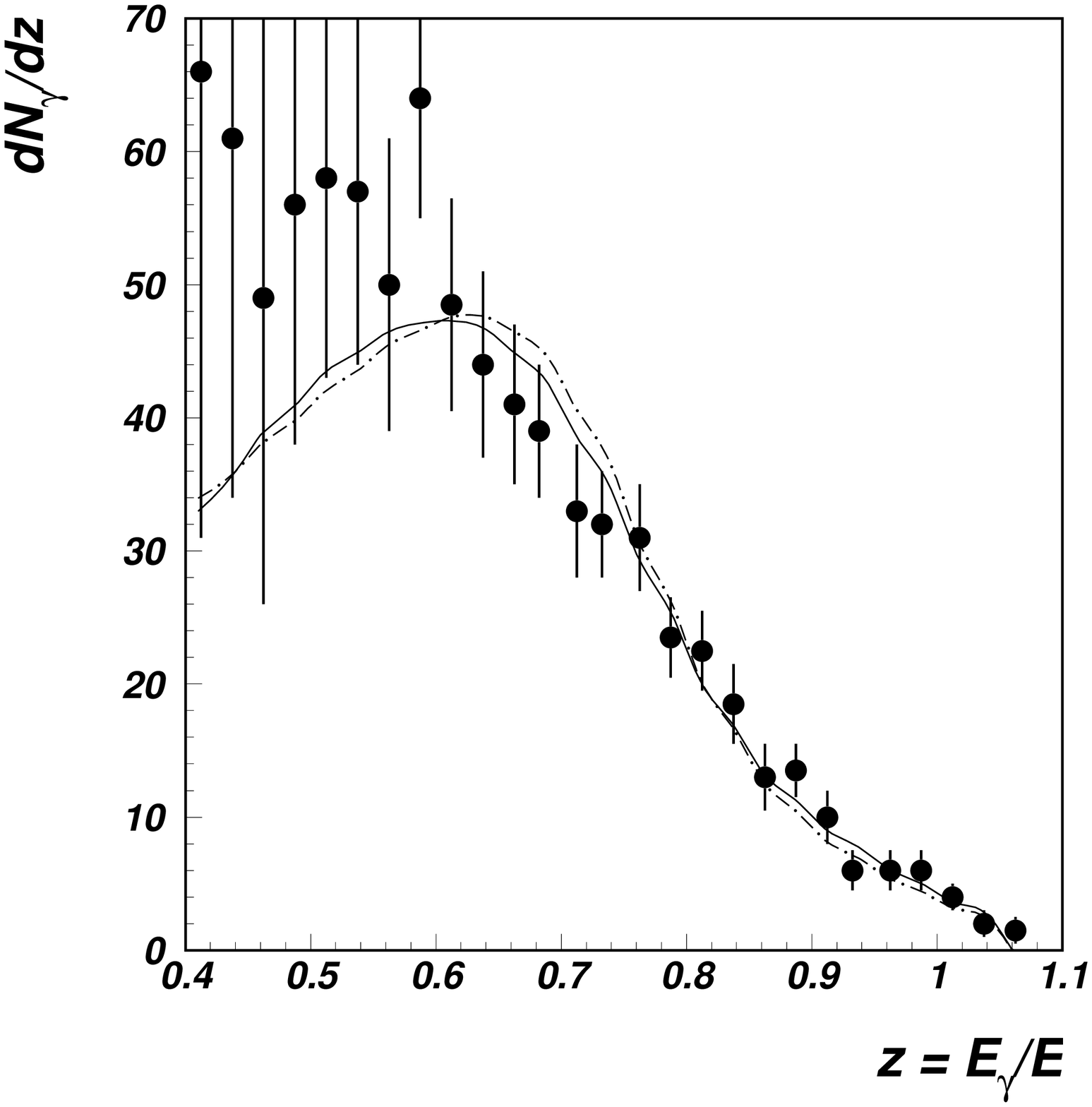}\\
\vspace{1cm}  
   \includegraphics*[width=7cm]{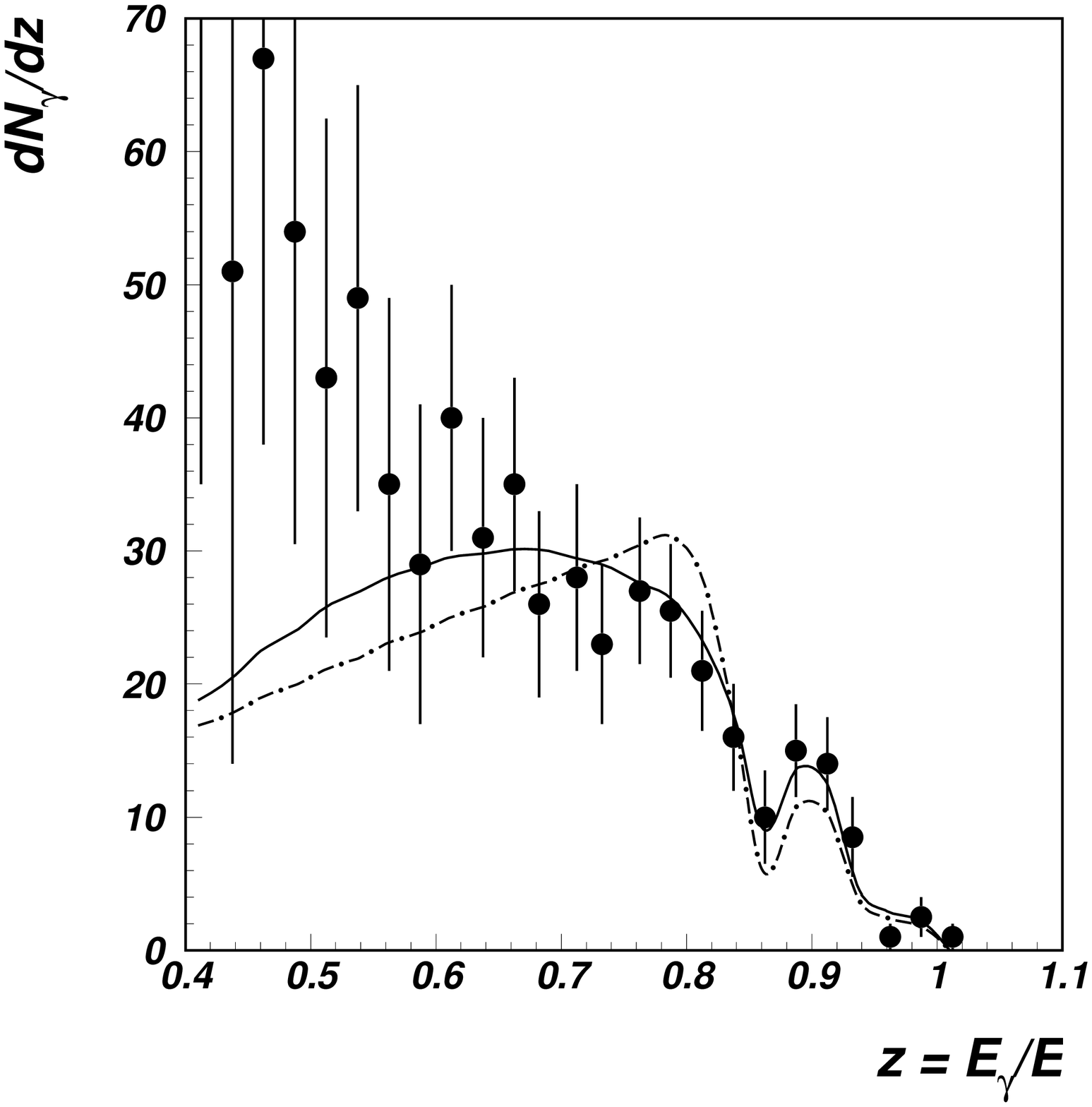}
 \end{center}
\caption{{\it Fits, for the effective gluon mass $\mg$,
 to inclusive photon spectra in $\Jp$ decays. Mark II (top left),
 Mark II cascade (top right),  Mark II conversion (bottom).
 Fit curves are defined as in Fig.3.}}
\label{jpsimg}
\end{figure} 
\begin{table}
\begin{center}
\begin{tabular}{|c|c|c|c|} \hline
Experiment & Decay process &  $ f_{E_{\gamma}} = \sigma_{E_{\gamma}}/E_{\gamma}({\rm GeV})$
  & $f_{E_{\gamma}}$ at $E_{\gamma} = \mv/2$ ($\%$) \\
\hline  
   & $\Jp \rightarrow \gamma X $  & $0.12/\sqrt{E_{\gamma}}$ &  9.6   \\  \cline{2-4}
 Mark II   & $\psi' \rightarrow \pi^+\pi^-(\Jp \rightarrow \gamma X)$ & $0.12/\sqrt{E_{\gamma}}$ &  9.6   \\  \cline{2-4}
   & $\Jp \rightarrow (\gamma  \rightarrow e^+e^-) X$   & $0.022 E_{\gamma}^{0.25}$ &  3.4    \\
\hline
 CUSB  & $\Up \rightarrow \gamma X $ & $0.039/E_{\gamma}^{0.25}$ &  3.2   \\ 
\hline 
 ARGUS  & $\Up \rightarrow \gamma X $ & $\sqrt{0.0052+0.0042/E_{\gamma}}$ &  7.8   \\
\hline 
 Crystal Ball  & $\Up \rightarrow \gamma X $ & $0.027/E_{\gamma}^{0.25}$   &  1.8   \\
\hline 
 CLEO2    & $\Up \rightarrow \gamma X $ & $0.0035/E_{\gamma}^{0.75} +0.019-0.001E_{\gamma}
 $  &  1.5   \\ 
\hline 
\end{tabular}
\caption[]{{\it  Photon energy resolutions of the different heavy quarkonia
 radiative decay experiments.}}
\end{center}
\end{table}

\begin{table}
\begin{center}
\begin{tabular}{|c|c|c|c|} \hline 
  Channel $C$ & z  &  BR$\times$10$^4$  & Relative BR  \\
\hline
\hline
\multicolumn{4}{|c|}{ Resonances }\\
\hline
 $\pi^0$ & 0.998 & 0.39 $\pm$ 0.13 & 0.009 \\
$\eta$ & 0.969 & 8.6 $\pm$ 0.8 & 0.2 \\
$\eta'$ & 0.904 & 43.1 $\pm$ 3.0 & 1.0 \\
$\eta(1440) \rightarrow \rho \rho$  & 0.784 & 17.0 $\pm$ 4.0 & 0.39 \\
$\eta(1440) \rightarrow \rho \gamma$  & 0.784 & 0.64 $\pm$ 0.14 & 0.015 \\
$\eta(1440) \rightarrow K\overline{K}\pi$  & 0.784 & 9.1 $\pm$ 1.8 & 0.21 \\
$\eta(1440) \rightarrow \eta \pi^+\pi^-$  & 0.784 & 3.0 $\pm$ 0.5 & 0.07 \\
$\eta(1760) \rightarrow \rho \rho$  & 0.677 & 1.3 $\pm$ 0.9 & 0.030 \\
$\eta_2(1870) \rightarrow \pi \pi$  & 0.63 & 6.2 $\pm$ 2.4 & 0.144 \\
$\eta(2225)$ & 0.484 & 2.9 $\pm$ 0.6 & 0.067 \\
$f_2(1270)$ & 0.832 & 13.8 $\pm$ 1.4 & 0.32 \\
$f_1(1285)$ & 0.828 & 6.1 $\pm$ 0.9 & 0.14 \\
$f_1(1420) \rightarrow K\overline{K}\pi$  & 0.790 & 8.3 $\pm$ 1.5 & 0.19 \\
$f_0(1500)$ & 0.765 & 5.7 $\pm$ 0.8 & 0.13 \\
$f_1(1510) \rightarrow \eta \pi^+\pi^-$  & 0.762 & 4.5 $\pm$ 1.2 & 0.10 \\
$f'_2(1525)$ & 0.758 & 4.7 $\pm$ 0.6 & 0.11\\
$f_0(1710) \rightarrow K\overline{K} $ & 0.695 & 8.5 $\pm$ 1.0 & 0.20 \\
$f_2(1950) \rightarrow K^*\overline{K}^* $ & 0.603 & 7.0 $\pm$ 2.2 & 0.16 \\
$f_4(2050)$ & 0.561 & 27.0 $\pm$ 7.0 & 0.63 \\
$f_J(2220) \rightarrow \pi \pi$  & 0.486 & 0.8 $\pm$ 0.4 & 0.019 \\
$f_J(2220) \rightarrow K\overline{K} $  & 0.486 & 0.8 $\pm$ 0.3 & 0.019 \\
$f_J(2220) \rightarrow p\overline{p} $  & 0.486 & 0.15 $\pm$ 0.08 & 0.003 \\
\hline
\multicolumn{2}{|c|}{ Total BR (Res.)}& 180 $\pm$ 10 & 4.17 \\
\hline
\multicolumn{4}{|c|}{ Exclusive states }\\
\hline
$\pi^+\pi^-\pi^0\pi^0$ & -- & 83.0 $\pm$ 31.0 & 1.93 \\
$\pi^+\pi^-\pi^+\pi^-$ & -- & 28.0 $\pm$ 5.0  & 0.65 \\
$K^+K^-\pi^+\pi^-$ & -- & 21.0 $\pm$ 6.0 & 0.49 \\
$\eta \pi \pi$ & -- & 61.0 $\pm$ 10.0 & 1.42 \\
$\rho \rho$ & -- & 45.0 $\pm$ 8.0 & 1.05 \\
$\omega \omega$ & -- & 15.9 $\pm$ 3.3 & 0.37 \\
$\phi \phi$ & -- & 4.0 $\pm$ 1.2  & 0.093 \\
$K^*\overline{K}^*$ & -- & 40.0 $\pm$ 13.0 & 0.92 \\
$p\overline{p}$ & -- & 3.8 $\pm$ 1.0 & 0.09 \\
\hline
\multicolumn{2}{|c|}{ Total BR (Excl.)}& 302 $\pm$ 37 & 7.0 \\
\multicolumn{2}{|c|}{ Total BR (Res.$+$Excl.)}& 482 $\pm$ 38 & 11.2 \\
\hline
\end{tabular}
\caption[]{{\it Composition of the hadronic final state in the $\Jp$ radiative decays:
$\Jp \rightarrow \gamma C $~\cite{PDG}.}}
\end{center}
\end{table}

\begin{table}
\begin{center}
\begin{tabular}{|c|c|c|c|c|} \hline 
 $\eta'$ fraction & $\eta$ fraction  & $\mg$  & $\chi^2_{min}$ & CL \\
\hline
\hline
\multicolumn{5}{|c|}{Mark II (25 dof)}\\
\hline
 0.0 & 0.0 & 0.610 $\pm$ 0.015 & 51.7 & 2$\times$10$^{-3}$ \\
 0.074$^{+0.006}_{-0.011}$ & 0.015 & 0.720 $\pm$ 0.012 & 17.5 & 0.86 \\
\hline
\multicolumn{5}{|c|}{Mark II cascade (25 dof)}\\
\hline
 0.078$^{+0.009}_{-0.011}$ & 0.016 & 0.722$^{+0.019}_{-0.017}$& 18.3 & 0.79 \\
\hline
\multicolumn{5}{|c|}{Mark II conversion (22 dof)}\\
\hline
 0.073$^{+0.016}_{-0.010}$ & 0.015 & 0.597 $\pm$ 0.019 & 20.0 & 0.58 \\
\hline

\end{tabular}
\caption[]{{ \it Results of fits to determine the fractions of
 $ \Jp \rightarrow \gamma \eta',~\gamma \eta $ in $\Jp$ radiative decays}}
\end{center}
\end{table}

\begin{table}
\begin{center}
\begin{tabular}{|c|c|c|} \hline 
 Fitted Model  & $\chi^2_{min}$ & CL \\
\hline
\hline
\multicolumn{3}{|c|}{Mark II (27 dof)}\\
\hline
 LO & 666 & $<$ 10$^{-30}$ \\
\hline
Rel. Corr$^n$ & 336 & $<$ 10$^{-30}$ \\
\hline
Rel. Corr$^n$, QCD(P) & 279 & $<$ 10$^{-30}$ \\
\hline
Rel. Corr$^n$, QCD(K) & 253 & $<$ 10$^{-30}$ \\
\hline
Rel. Corr$^n$, QCD(P$\times$K) & 224 & $<$ 10$^{-30}$ \\
\hline
\multicolumn{3}{|c|}{Mark II cascade (26 dof)}\\
\hline
 LO & 348 & $<$ 10$^{-30}$ \\
\hline
Rel. Corr$^n$ & 183 & 1.5$\times$10$^{-25}$ \\
\hline
Rel. Corr$^n$, QCD(P) & 153 & 5.9$\times$10$^{-20}$ \\
\hline
Rel. Corr$^n$, QCD(K) & 136 & 7.3$\times$10$^{-17}$ \\
\hline
Rel. Corr$^n$, QCD(P$\times$K) & 121 & 3.3$\times$10$^{-14}$ \\
\hline
\multicolumn{3}{|c|}{Mark II conversion (24 dof)}\\
\hline
 LO & 198 & $<$ 2.5$\times$10$^{-29}$ \\
\hline
Rel. Corr$^n$ & 80.9 & 4.4$\times$10$^{-25}$ \\
\hline
Rel. Corr$^n$, QCD(P) & 66.3 & 7.8$\times$10$^{-6}$ \\
\hline
Rel. Corr$^n$, QCD(K) & 50.8 & 1.1$\times$10$^{-3}$ \\
\hline
Rel. Corr$^n$, QCD(P$\times$K) & 45.0 & 5.8$\times$10$^{-3}$ \\
\hline
\end{tabular}
\caption[]{{\it Results of fits with $\mg = 0$ to $\Jp$ inclusive photon
spectra. LO denotes the lowest order QCD prediction (Eqn. (3.1)).
Rel. Corr$^n$ includes relativistic corrections calculated according
 to Eqn.(2.8).  
 QCD(P) and QCD(K) denote, respectively the Photiadis~\cite{
Photiadis} and Kr\"{a}mer ~\cite{Kramer} HO QCD corrections. 
QCD(P$\times$K) means that both corrections are applied.}}
\end{center}
\end{table}

\begin{table}
\begin{center}
\begin{tabular}{|c|c|c|c|} \hline 
 Fitted Model & $\mg$ (GeV)  & $\chi^2_{min}$ & CL \\
\hline
\hline
\multicolumn{4}{|c|}{Mark II (26 dof)}\\
\hline
 LW & 0.734 $\pm$ 0.010 & 24.6 & 0.54 \\
\hline
 LW, Rel. Corr$^n$  & 0.740$_{-0.012}^{+0.009}$ & 22.6 & 0.66 \\
\hline
 LW, Rel. Corr$^n$ & & & \\
 QCD(P)   & 0.721$_{-0.009}^{+0.010}$ & 17.5 & 0.89 \\
\hline
 LW, Rel. Corr$^n$ & & & \\
 QCD(K)   & 0.665$_{-0.013}^{+0.014}$ & 19.9 & 0.80 \\
\hline
 LW, Rel. Corr$^n$ & & & \\
 QCD(P$\times$K)   & 0.653$_{-0.015}^{+0.018}$ & 16.9 & 0.91 \\
\hline
\multicolumn{4}{|c|}{Mark II cascade (25 dof)}\\
\hline
 LW & 0.737$_{-0.018}^{+0.015}$  & 24.4 & 0.50 \\
\hline
 LW, Rel. Corr$^n$  & 0.740$\pm$ 0.017  & 21.9\ & 0.64 \\
\hline
 LW, Rel. Corr$^n$ & & & \\
 QCD(P)   & 0.719$\pm$ 0.019 & 18.4 & 0.82 \\
\hline
 LW, Rel. Corr$^n$ & & & \\
 QCD(K)   &0.667$\pm$ 0.021  & 22.7 & 0.60 \\
\hline
 LW, Rel. Corr$^n$ & & & \\
 QCD(P$\times$K)   & 0.655$\pm$ 0.021  & 20.0 & 0.75 \\
\hline
\multicolumn{4}{|c|}{Mark II conversion (23 dof)}\\
\hline
 LW & 0.623$_{-0.016}^{+0.013}$  & 31.6 & 0.11 \\
\hline
 LW, Rel. Corr$^n$  & 0.607$_{-0.017}^{+0.015}$   & 23.9 & 0.41 \\
\hline
 LW, Rel. Corr$^n$ & & & \\
 QCD(P)   & 0.598$_{-0.017}^{+0.018}$ & 19.9 & 0.65 \\
\hline
 LW, Rel. Corr$^n$ & & & \\
 QCD(K)   &0.537$_{-0.030}^{+0.025}$ & 22.5 & 0.49 \\
\hline
 LW, Rel. Corr$^n$ & & & \\
 QCD(P$\times$K)   & 0.526$_{-0.029}^{+0.027}$ & 20.3 & 0.62 \\
\hline 
\end{tabular}
\caption[]{{\it Results of fits with  variable $\mg$ to $\Jp$ inclusive photon
spectra. LW denotes gluon mass corrections calculated according to
the calculations of Liu and Wetzel~\cite{LW}. Relativistic and HO QCD corrections
 are defined as in Table 5. }}
\end{center}
\end{table}

To date, the inclusive photon spectrum in $\Jp$ decays has been measured by only
one experiment, the Mark II collaboration~\cite{Mark II}. Actually, in Reference
~\cite{Mark II}, three independent measurements of the spectrum are given.
The first (referred to simply as `Mark II') uses the process:
\[{\rm e^+e^- \rightarrow \Jp \rightarrow \gamma X} \]
 where the photons are detected
in the electromagnetic calorimeter of the detector. The second, sample, `Mark II(cascade)',
 uses the process:
\[{\rm e^+e^- \rightarrow \psi' \rightarrow  \Jp \pi \pi \rightarrow
 \gamma X \pi \pi} \]
In this case the acceptance and resolution are similar to the `Mark II' measurement. 
 The third data
sample, `Mark II(conversion)', uses $\Jp$ radiative decay events where the photon converts into an
${\rm e^+e^-}$ pair in the beam pipe or the inner flange of the tracking chamber.
Measurement of the momenta of the ${\rm e^+,e^-}$ in the chamber yields a sample
with reduced statistics but much improved photon energy resolution
 The photon energy resolutions for 
these three event samples are 
given in Table 2.

 \par As well as these inclusive measurements, many exclusive
 measurements have been made where a single resonant state or an exclusive
 multihadron final state is produced in association with a hard photon~\cite{PDG}.
 These measurements are summarised in Table 3 where the values of $z$ for each
 exclusive channel with a single particle recoiling against the photon are given.
 As can be seen from Table 3 (
see also Fig.59 of Reference~\cite{KopWerm}) the most striking feature of the 
photon spectrum near the end point is the strong exclusive production
of $\eta$ and $\eta'$ mesons:
\[ \Jp \rightarrow \gamma \eta,~~\gamma \eta' \]
These two channels alone account for 13$\%$ of the total branching ratio
for $z > 0.6$ and completely dominate the end point region $z > 0.85$. 
Because of the large contribution of these two resonances, it was not possible
to obtain acceptable fits to the $\Jp$ spectra using the function of Eqn.(2.8)
even when HO QCD corrections and gluon mass effects were included. The procedure
adopted was then to fix the ratio $\Gamma(\Jp \rightarrow \gamma \eta')/
\Gamma(\Jp \rightarrow \gamma \eta)$ to the measured value 5.0~\cite{PDG},
and perform fits treating the ratio:
\[ R_{\eta'} = \Gamma(\Jp \rightarrow \gamma \eta')/
\Gamma(\Jp \rightarrow \gamma~ {\rm continuum}) \]
as a free parameter, which include the $\eta'$ and $\eta$ contributions
at the appropriate $z$ values of 0.904 and 0.969 respectively.
Here `$\gamma$~continuum' refers to Eqn(2.8), including also gluon mass effects  
and the Photiadis HO QCD correction. The other two parameters in the fit are
an overall normalisation constant and the effective gluon mass $\mg$. The 
method used to fold in the experimental resolution function is described in
the Appendix. As shown in Table 4,
 attempting to fit the spectrum without explicitly introducing
 the $\eta'$ and $\eta$ contributions (i.e. with $R_{\eta'} = 0$),
 leads to an unacceptably low confidence level (CL) of 3 $\times$10$^{-3}$.
 However, including their contributions, good fits are obtained for
 all three
spectra with consistent values of $R_{\eta'}$. Their weighted average is:
 \[R_{\eta'} = 0.0754 \pm 0.0070  \]
The Mark II and Mark II(cascade) spectra yield consistent values of $\mg$
around 720 MeV, but the Mark II(conversion) spectrum gives a significantly
lower (5.8 $\sigma$) value of 597 MeV, indicating some systematic
 difference in the latter measurement. The weighted average value of 
 $R_{\eta'}$ corresponds to :
\[ BR(\Jp \rightarrow \gamma + {\rm continuum} + \eta + \eta') = 0.063 \pm 0.0070 \]
 which may be compared to the summed branching ratio of all channels reported in
 Table 3 of $0.0482 \pm 0.0038$. So it is estimated that 76$\%$ of all $\Jp$ 
 radiative decays to light hadrons are contributed by the exclusive 
 channels listed in Table 3.
\par The results of fits to the three inclusive photon spectra,
 including the exclusive  $\eta'$ and $\eta$ contributions, estimated using
 the weighted average value of $R_{\eta'}$ obtained above, without
 including gluon mass effects, are presented in Table 5 and Fig.3. No
 acceptable fits are obtained, even after the inclusion of of both
 relativistic and HO QCD corrections. The best confidence levels
 obtained for the fits to the Mark II, Mark II(cascade) and Mark II(conversion)
 spectra are: $< 10^{-30}$, $3.3\times10^{-14}$ and $ 5.8\times10^{-3}$
 respectively. Although, as expected from Fig. 1, the shape of the spectrum is
 drastically modified by the relativistic correction, the change in shape,
 though qualitatively in the right direction, is, by far, not enough to 
 explain the observed spectrum shape. As can be seen in Fig. 3a, the
 estimated effects of HO QCD corrections are even smaller than those
 of the relativistic correction.

\par A similar series of fits, but including gluon mass effects, is 
 presented in Table 6 and Fig.4. Good fits are obtained in all cases, and it
 can be seen that the inclusion of relativistic and HO QCD corrections have
 only a minor effect on the fitted values of $\mg$. For example, in the case
 of the largest statistics (Mark II) data sample, introducing the relativistic
 correction increases the fitted value of $\mg$ of 734 MeV by only 6 MeV.
 Further applying either the Photiadis or Kr\"{a}mer HO QCD corrections 
 give further shifts of -19 MeV or -75 MeV respectively in the fitted $\mg$
 value. In fact because of the relatively large value of $\mg$ as compared
 to $M_{\Jp}$, it
 is to be expected, as mentioned previously, that HO QCD corrections will
 be much reduced by by phase space and propagator suppression effects.
 Choosing then the fit with relativistic and the smallest (Photiadis)
 HO QCD correction as best estimate yields the result:
\[ \mg =0.721_{-0.009~-0.068 }^{+0.010~+0.013}~ {\rm GeV}~~(\Jp) \]
 where the first error is statistical (from the fit to the Mark II spectrum)
 and the second is systematic, conservatively estimated from the full
 spread of the different fit results to the Mark II data given in Table 6.
 In view of the
 size of this systematic error in the Mark II value, and the large
 observed systematic shift in $\mg$ obtained with the Mark II(conversion)
 spectrum, no significant improvement in the knowledge of $\mg$
 is expected by combining the results of the fits to the three spectra.
 The less precise values provided by the Mark II(cascade) and
 Mark II(conversion) data should then be considered as consistency checks.

\par It may be noted that the exclusive $\eta'$ signal is clearly seen in the
 Mark II(conversion) spectrum shown in Figs. 3c and 4c. The shape of the
 observed peak is well described in Fig. 4c by the fit including the 
 relativistic and Photiadis HO QCD corrections, and the experimental
 resolution function given in Table 2. This agreement
 gives an important cross check on the method used here (see the 
 Appendix) to fold in the experimental resolution effects.

\par In order to study the relative importance of phase space effects
 and longitudinal gluon contributions in the gluon mass fits, the fit
 to the Mark II data, including both the relativistic and the Photiadis
 HO QCD corrections, is repeated setting $\mg=0$ in the function
 $f_0(z,x_2,\mg)$ of Eqn(4.7), thus removing the longitudinal
  contributions. A very good fit is still obtained (confidence level 
  0.94) with $\mg = 0.682+0.010-0.013$ GeV. This is only 5.4$\%$
  lower than
 the value $\mg = 0.721+0.010-0.009$ GeV obtained including the longitudinal
  contributions. Thus the gluon mass correction is dominated by phase 
  space effects.

\SECTION{\bf{The Decay $\Up \rightarrow \gamma X$}}

\begin{figure}
 \begin{center}
   \includegraphics*[width=7cm,]{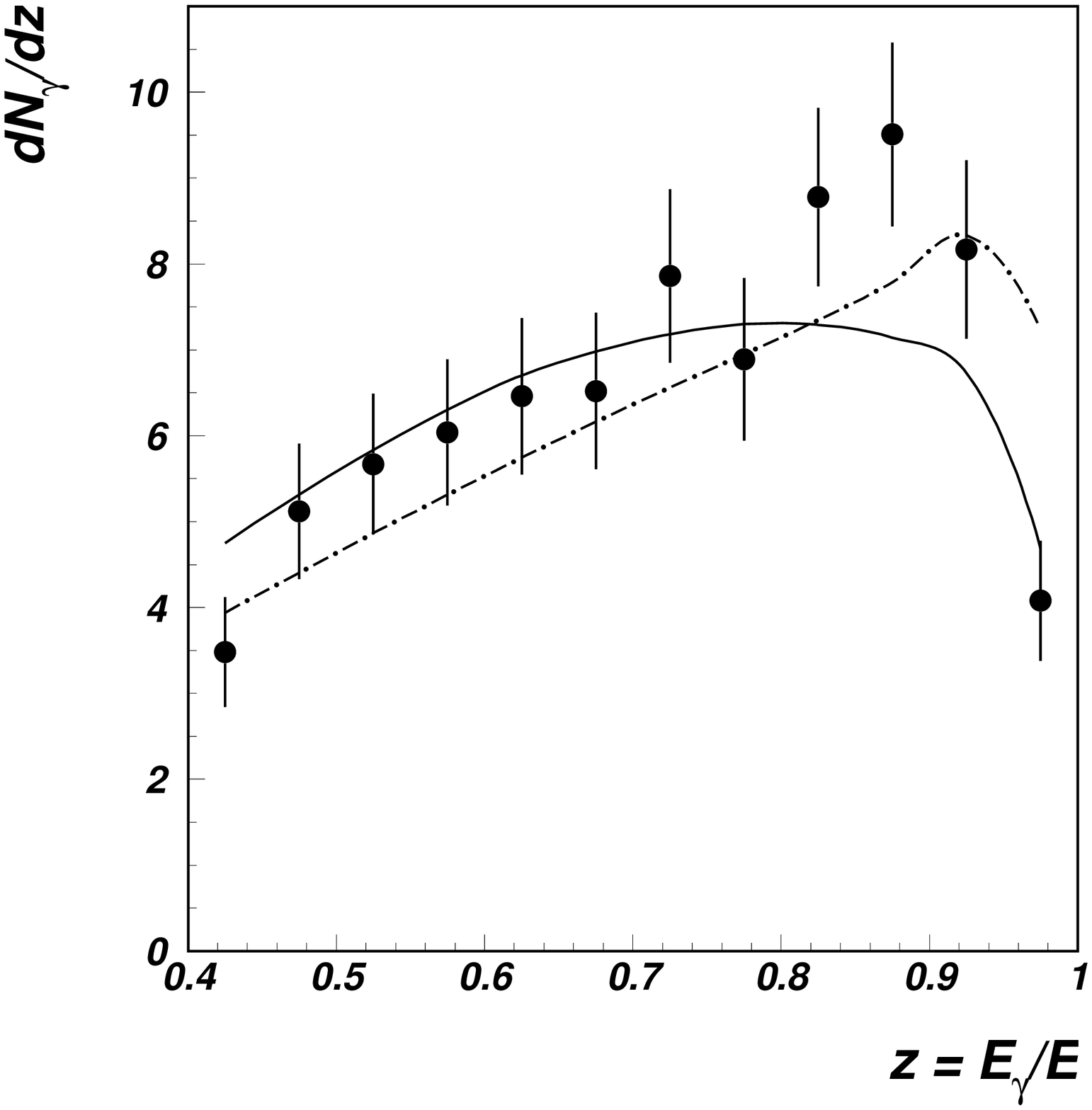} 
   \includegraphics*[width=7cm]{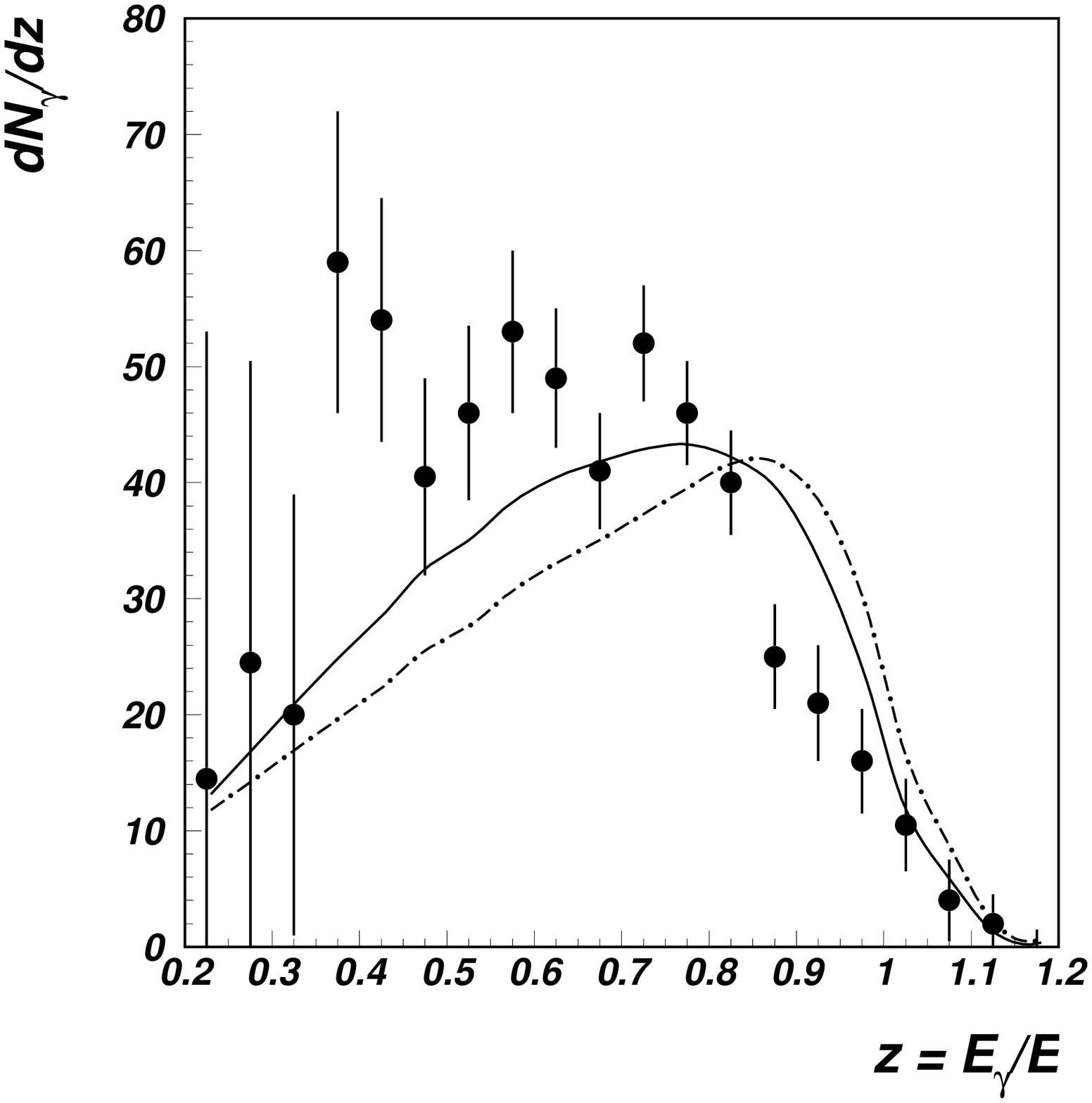}\\
\vspace{1cm}  
   \includegraphics*[width=7cm]{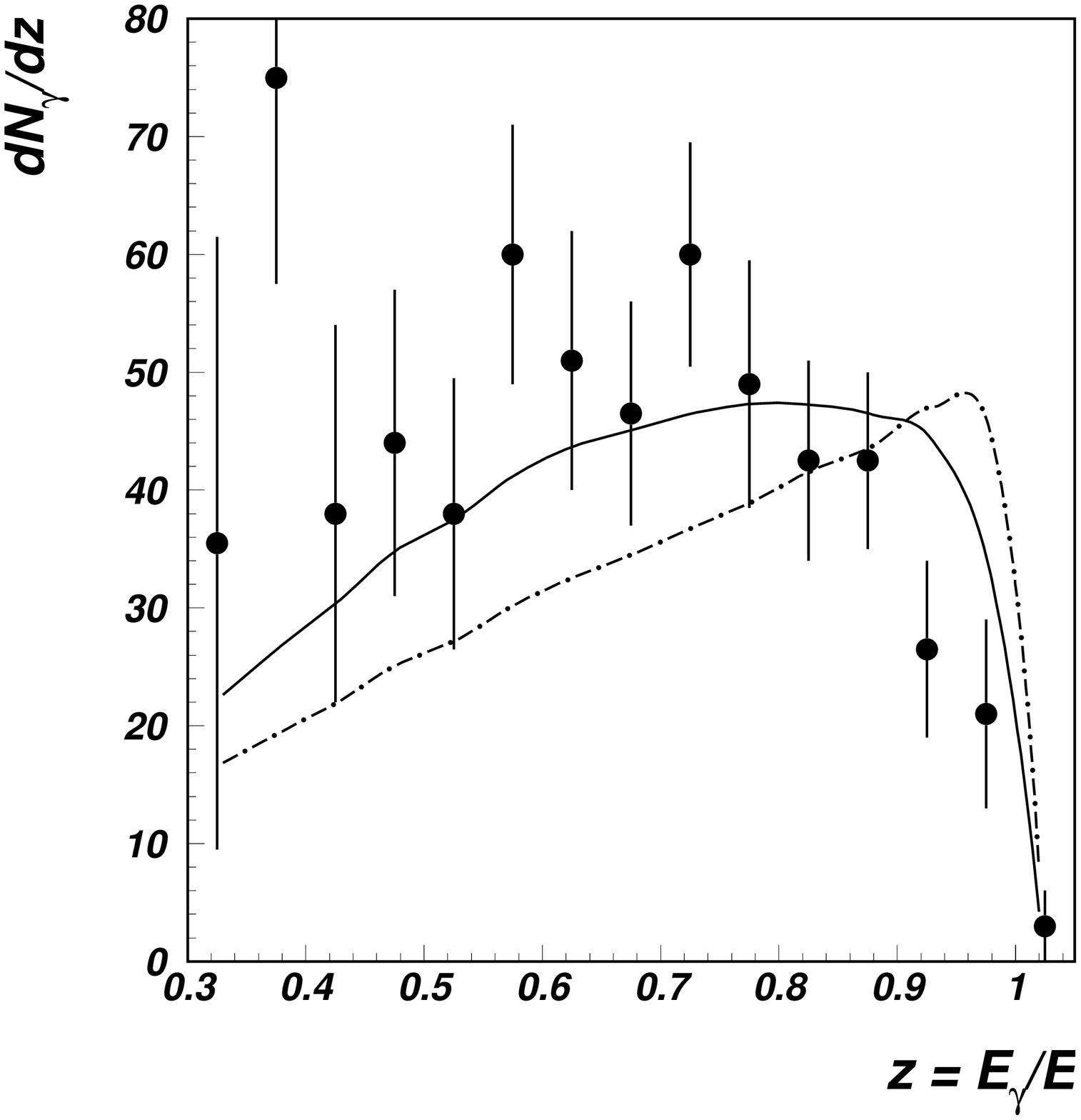}
   \includegraphics*[width=7cm]{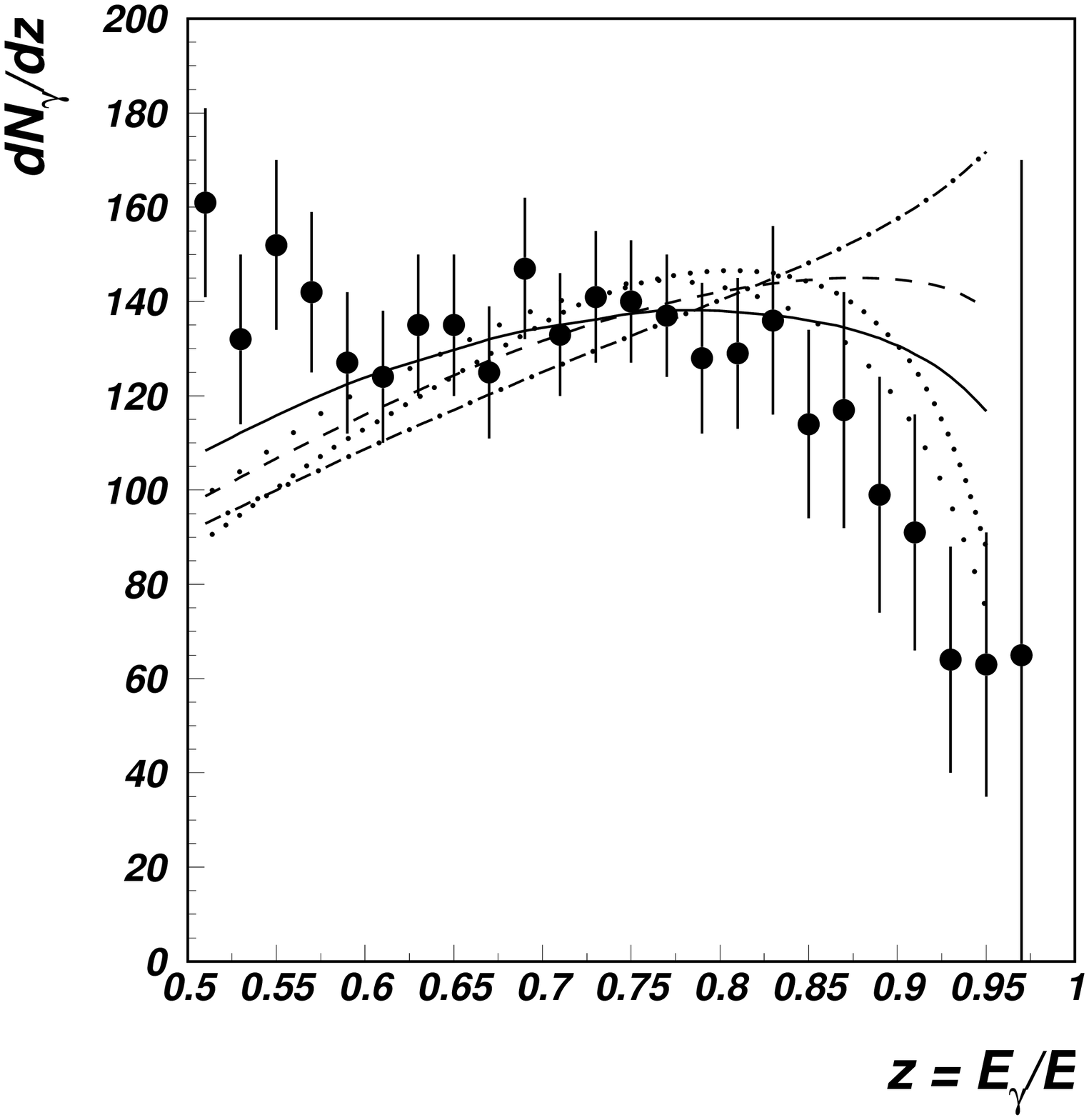}
 \end{center}
\caption{{\it Fits, assuming $\mg=0$,
 to inclusive photon spectra in $\Up$ decays: CUSB (top left), ARGUS
 (top right),
 Crystal Ball (bottom left),  CLEO2 (bottom right). Fit curves are
 defined as in Fig.3.}}
\label{Upsmg0}
\end{figure}
\begin{figure}
 \begin{center}
   \includegraphics*[width=7cm,]{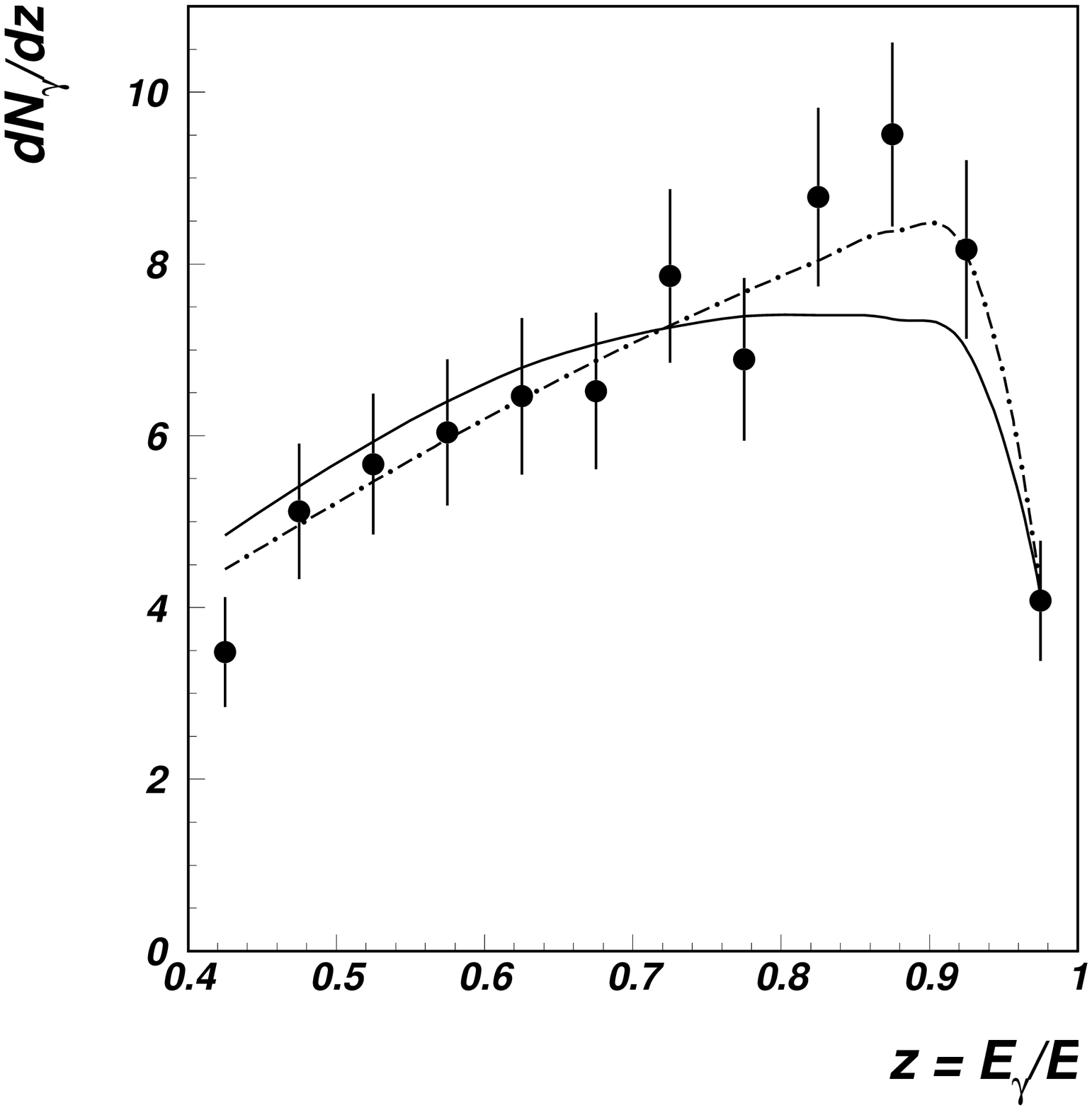} 
   \includegraphics*[width=7cm]{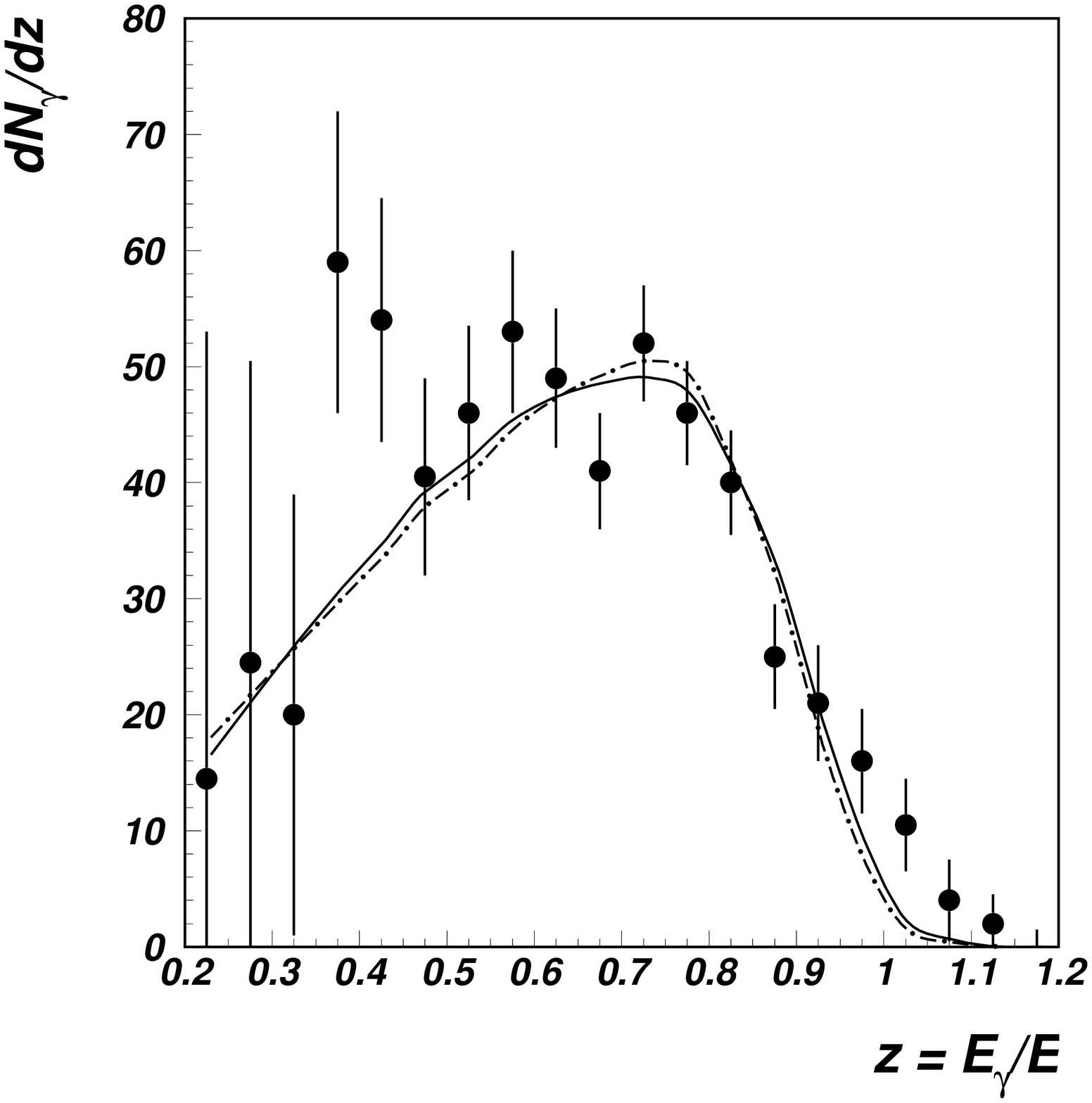}\\
\vspace{1cm}  
   \includegraphics*[width=7cm]{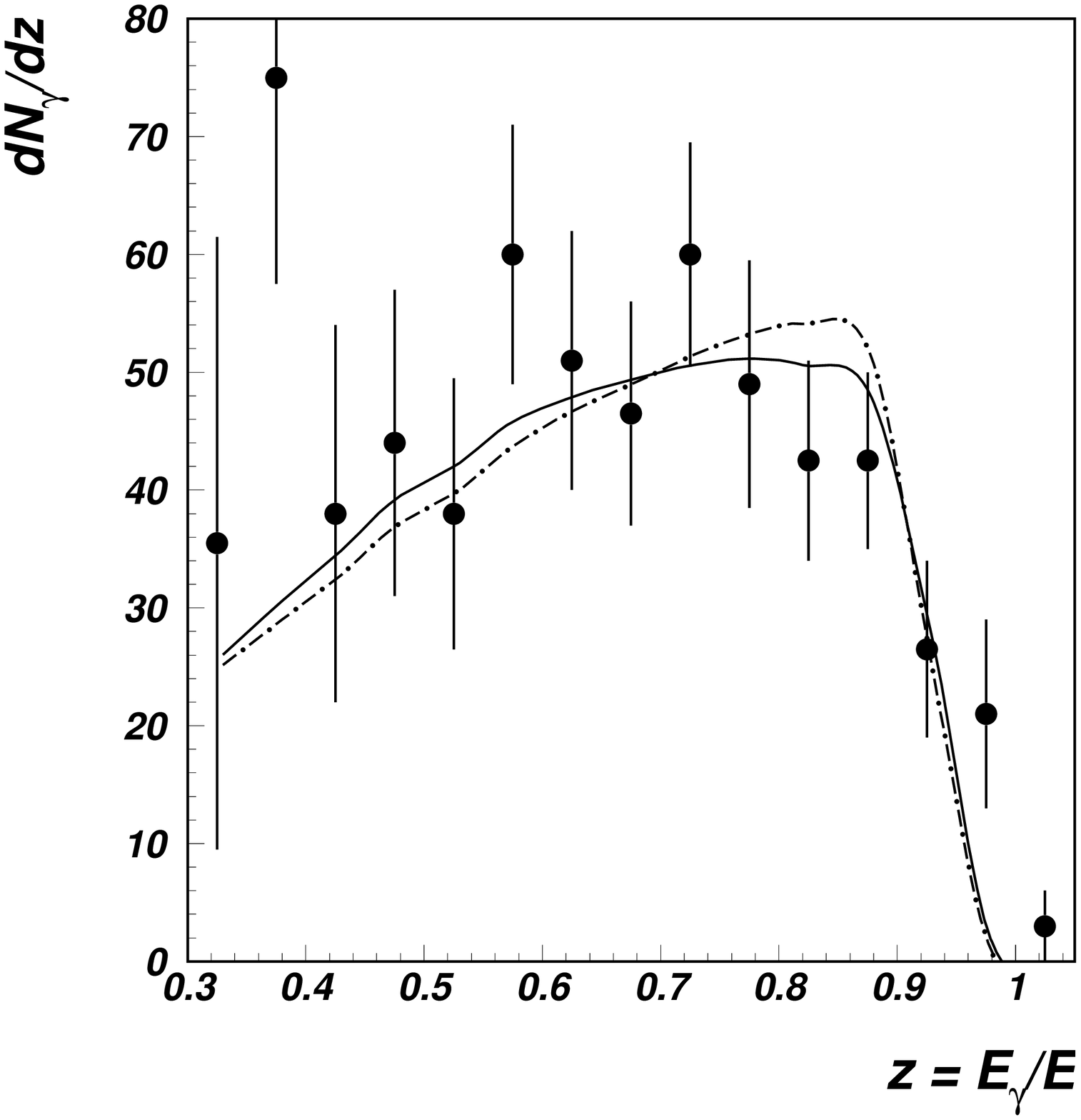}
   \includegraphics*[width=7cm]{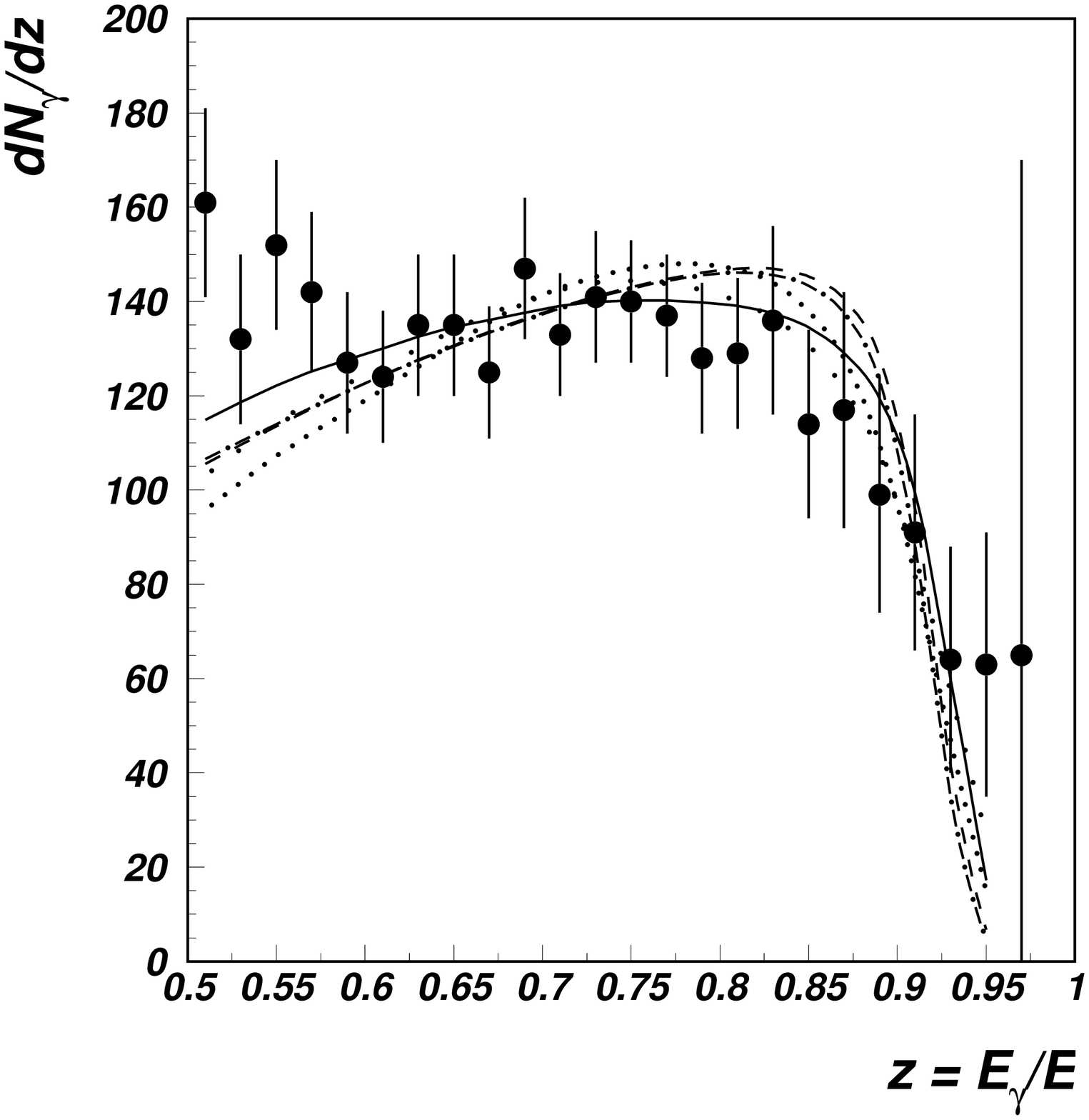}
 \end{center}
\caption{{\it Fits, for the effective gluon mass $\mg$,
 to inclusive photon spectra in $\Up$ decays: CUSB (top left), ARGUS
 (top right),
 Crystal Ball (bottom left),  CLEO2 (bottom right).
 Fit curves are defined as in Fig.3.}}
\label{Upsmg}
\end{figure}
\begin{table}
\begin{center}
\begin{tabular}{|c|c|c|} \hline 
 Fitted Model  & $\chi^2_{min}$ & CL \\
\hline
\hline
\multicolumn{3}{|c|}{CUSB (11 dof)}\\
\hline
 LO & 30.2 & $<$ 1.5$\times$10$^{-3}$ \\
\hline
Rel. Corr$^n$ & 10.9 &  0.45 \\
\hline
Rel. Corr$^n$, QCD(P) & 14.7 & 0.2 \\
\hline
Rel. Corr$^n$, QCD(K) & 13.0 & 0.29 \\
\hline
Rel. Corr$^n$, QCD(P$\times$K) & 21.3 & 0.03 \\
\hline
\multicolumn{3}{|c|}{ARGUS (19 dof)}\\
\hline
 LO & 95.3 & 3.8$\times$10$^{-12}$ \\
\hline
Rel. Corr$^n$ & 62.6 & 1.5$\times$10$^{-6}$ \\
\hline
Rel. Corr$^n$, QCD(P) & 44.7 & 7.5$\times$10$^{-4}$ \\
\hline
Rel. Corr$^n$, QCD(K) & 49.8 & 1.4$\times$10$^{-4}$ \\
\hline
Rel. Corr$^n$, QCD(P$\times$K) & 37.6 & 6.7$\times$10$^{-3}$ \\
\hline
\multicolumn{3}{|c|}{Crystal Ball (14 dof)}\\
\hline
 LO & 51.9 &  2.9 10$^{-6}$ \\
\hline
Rel. Corr$^n$ & 31.6 & 4.6$\times$10$^{-3}$ \\
\hline
Rel. Corr$^n$, QCD(P) & 22.4 & 0.071 \\
\hline
Rel. Corr$^n$, QCD(K) & 24.6 & 0.039 \\
\hline
Rel. Corr$^n$, QCD(P$\times$K) & 18.9 & 0.17 \\
\hline
\multicolumn{3}{|c|}{CLEO2 (22 dof)}\\
\hline
 LO & 90.7 &  2.6$\times$10$^{-10}$ \\
\hline
Rel. Corr$^n$ & 56.9 & 6.3$\times$10$^{-5}$ \\
\hline
Rel. Corr$^n$, QCD(P) & 32.1 & 0.076 \\
\hline
Rel. Corr$^n$, QCD(K) & 48.8 & 8.5$\times$10$^{-4}$ \\
\hline
Rel. Corr$^n$, QCD(P$\times$K) & 29.5 & 0.131 \\
\hline
\end{tabular}
\caption[]{{\it  Results of fits with $\mg = 0$ to $\Up$ inclusive photon
spectra. See Table 5 for the definitions of the different fits.}}
\end{center}
\end{table}

\begin{table}
\begin{center}
\begin{tabular}{|c|c|c|c|} \hline 
 Fitted Model & $\mg$ (GeV)  & $\chi^2_{min}$ & CL \\
\hline
\hline
\multicolumn{4}{|c|}{CUSB (10 dof)}\\
\hline
 LW & 0.66 $\pm$ 0.08 & 5.2 & 0.88 \\
\hline
 LW, Rel. Corr$^n$  & 0.64 $\pm$ 0.09 & 6.1 & 0.81 \\
\hline
 LW, Rel. Corr$^n$ & & & \\
 QCD(P)   & 0.54 $\pm$ 0.12 & 13.1 & 0.22 \\
\hline
 LW, Rel. Corr$^n$ & & & \\
 QCD(K)   & 0.16$_{-0.16}^{+0.17}$ & 12.6 & 0.25 \\
\hline
 LW, Rel. Corr$^n$ & & & \\
 QCD(P$\times$K)   & 0.15$_{-0.15}^{+0.18}$ & 21.0 & 0.021 \\
\hline
\multicolumn{4}{|c|}{ARGUS (18 dof)}\\
\hline
 LW & 1.39$_{-0.10}^{+0.08}$  & 27.3 & 0.074 \\
\hline
 LW, Rel. Corr$^n$  & 1.39$\pm$ 0.10  & 28.8 & 0.051 \\
\hline
 LW, Rel. Corr$^n$ & & & \\
 QCD(P)   & 1.27$_{-0.12}^{+0.11}$   & 23.3 & 0.18 \\
\hline
 LW, Rel. Corr$^n$ & & & \\
 QCD(K)   & 1.19$_{-0.12}^{+0.10}$ & 34.4 & 0.011 \\
\hline
 LW, Rel. Corr$^n$ & & & \\
 QCD(P$\times$K)   & 1.06$_{-0.16}^{+0.13}$ & 28.8 & 0.05 \\
\hline
\multicolumn{4}{|c|}{Crystal Ball (13 dof)}\\
\hline
 LW & 1.21$_{-0.09}^{+0.10}$  & 21.5 & 0.064 \\
\hline
 LW, Rel. Corr$^n$  & 1.21$_{-0.11}^{+0.10}$   & 22.2 & 0.052 \\
\hline
 LW, Rel. Corr$^n$ & & & \\
 QCD(P)   & 1.14$_{-0.16}^{+0.12}$ & 18.0 & 0.16 \\
\hline
 LW, Rel. Corr$^n$ & & & \\
 QCD(K)   & 0.90$_{-0.40}^{+0.19}$ & 23.5 & 0.036 \\
\hline
 LW, Rel. Corr$^n$ & & & \\
 QCD(P$\times$K)   & No $\chi^2_{min}$ & --- & --- \\
\hline
\multicolumn{4}{|c|}{CLEO2 (21 dof)}\\
\hline
 LW & 1.27 $\pm$ 0.07  & 29.5 & 0.103 \\
\hline
 LW, Rel. Corr$^n$  & 1.25$_{-0.07}^{+0.08}$   & 30.4 & 0.08 \\
\hline
 LW, Rel. Corr$^n$ & & & \\
 QCD(P)   & 1.15$_{-0.09}^{+0.08}$ & 16.9 & 0.72 \\
\hline
 LW, Rel. Corr$^n$ & & & \\
 QCD(K)   & 1.03$_{-0.12}^{+0.09}$ & 34.5 & 0.032 \\
\hline
 LW, Rel. Corr$^n$ & & & \\
 QCD(P$\times$K)   & 0.90$_{-0.13}^{+0.11}$ & 22.5 & 0.37 \\ 
\hline
\end{tabular}
\caption[]{{\it Results of fits with  variable $\mg$ to $\Up$ inclusive photon
spectra. The descriptions of the different fitted models are the same as in 
 Table 6.}}
\end{center}
\end{table}

Five different experiments have measured the inclusive photon spectrum 
in $\Up$ decays: CUSB~\cite{CUSB}, CLEO~\cite{CLEO}, ARGUS~\cite{ARGUS},
Crystal Ball~\cite{CB} and CLEO2~\cite{CLEO2}. The CLEO measurement, which,
like Crystal Ball, ARGUS and CLEO2, but unlike CUSB, is in good agreement 
with the RDF prediction, is not analysed in the present paper as no efficiency
corrected spectrum was provided.
\par Unlike for the case of the $\Jp$, no positive evidence has been found for
 the exclusive production of single resonances in the radiative decays of the
 $\Up$~\cite{PDG}. The efficiency corrected inclusive photon spectra 
measured by CUSB, ARGUS, Crystal Ball, and CLEO2 have therefore been
directly fitted to the relativistically corrected spectrum 
 Eqn(2.8), possibly also including HO QCD corrections as
 discussed in Section 3 and gluon mass effects as described in
 Section 4. Experimental resolution effects are included in the same way
 as for the fits to the $\Jp$ decays described above. The photon
 energy resolutions of the different experiments are given in Table 2.
 Results of fits assuming a vanishing effective gluon mass are presented
 in Table 7 and Fig.5, while fits to $\mg$ and an overall
 normalisation constant yield the results shown in Table 8 and Fig.6.
\par As can be seen in Figs.5 and 6, the CUSB spectrum differs markedly
 in shape from those of the later ARGUS, Crystal Ball, and CLEO2 
experiments. The suppression of the end-point
region, relative to the LO QCD prediction (the dash-dotted curves), is much reduced. The results
 of the fits to the CUSB data should be treated with caution as, unlike
 the other experiments, the published errors on the photon spectrum are 
 are purely statistical.
 A relatively large systematic error is expected (as found in the 
 other experiments) from the $\pi^0/\gamma$ separation procedure,
 especially at lower values of $z$.
\par Considering first the fits with $\mg = 0$ in Table 7, it can be
 seen that the LO spectrum is ruled out, individually, by all four
 experiments. Inclusion of the relativistic correction with
$<v^2> = 0.09$ gives a good description of the CUSB data but is
ruled out with a confidence level of less than 0.5$\%$ by each of 
the other experiments. The best overall description is given
by combining the relativistic and Photiadis HO QCD corrections.
Low, but acceptable, confidence levels of 0.2, 0.071 and 0.076
are found for the fits to CUSB, Crystal Ball and CLEO2 data, 
respectively. Only the ARGUS spectrum (CL = 7.5$\times$10$^{-4}$)
is inconsistent with this hypothesis. However, combining the fits to
 all four experiments gives $\chi^2/dof = 114/66$ (CL=2.3$\times$10$^{-4}$).

For ARGUS, Crystal Ball and CLEO2 the best fits are given by including
 both the Photiadis and Kr\"{a}mer HO QCD corrections. The combined fit
 gives, however, $\chi^2/dof = 86/66$ (CL=4.7$\times$10$^{-3}$).
 Thus no consistent
overall description of the  data is found for $\mg = 0$.
\par When the effective gluon mass $\mg$ is included as a fit parameter 
 it can be seen (Table 8) that fits with confidence levels $> 1\%$ are
 found for
 all fit hypotheses and all experimental spectra. However, the ARGUS,
 Crystal Ball and CLEO2 data give consistent values of $\mg$ in the
 range 0.9-1.4 GeV, whereas significantly lower values 0.15-0.66 GeV
 are found in the fits to the CUSB spectrum. Because of this discrepancy
 and the neglect of (potentially large) systematic errors in the latter
 experiment, only the results from the three most recent experiments are
 used to obtain the average value of $\mg$ quoted below. For these 
 experiments the best overall fit is given (as in the case $\mg = 0$)
 by including both the relativistic and the Photiadis HO QCD
 correction. This yields, for the weighted average value of the
 effective gluon mass: 
\[ \mg =1.18 \pm 0.06~\rm{GeV} \]
 where the error quoted is derived from fit errors of the different
 experiments.
Performing a fit to the ARGUS, Crystal Ball and CLEO2 spectra,
with $\mg$ fixed at the above value, and varying only the
normalisation constants of the fitted curves gives a good overall
fit with $\chi^2/dof = 59.0/55$ (CL$=$0.33). Making the same
type of fit, but including CUSB leads to:
$\chi^2/dof = 121.0/66$ (CL=4.3$\times$10$^{-5}$). The published
CUSB data is therefore clearly inconsistent with value of
$\mg$ favoured by ARGUS, Crystal Ball and CLEO2. This apparent 
inconsistency of the CUSB measurement, is, however, very sensitive
to the error assignement of the data. Increasing the quoted
(purely statistical) errors by a constant factor of 1.5 to account
 for systematic effects, modifies the last fit result quoted above to:
$\chi^2/dof = 89.0/66$ (CL$=$0.031). The CUSB measurement is now
 marginally consistent with the average of the three other experiments.
 \par The theoretical systematic error on $\mg$ is estimated in the same
  way as for $\Jp$ decays. This gives:
\[ \mg =1.18 \pm 0.06^{+0.07}_{-0.28}~\rm{GeV}~~(\Up)  \]
  where the first error is experimental and the second is a conservatively
 estimated theoretical error that that includes the full range of relativistic and
 HO QCD corrections in the fits to the CLEO2 data in Table 8.
\par To investigate the importance of the effects of longitudinal gluon
polarisation states for the case of the $\Up$, a fit is made to the CLEO2 data
including relativistic and the Photiadis HO QCD corrections,  
with $\mg = 0$ in the function $f_0$ of Eqn(4.7). The fit gives 
$\mg = 1.10+0.08-0.09$~GeV with $\chi^2/dof = 13.1/21$ (CL$=$0.91) to be
compared with $\mg = 1.15+0.08-0.09$~GeV with $\chi^2/dof = 16.9/21$
 (CL$=$0.70) when longitudinal gluon contributions are included
 in $f_0$. As in the case of the $\Jp$, longitudinal gluon states
 give only a small effect; they increase the fitted value of $\mg$
 by only 4.5$\%$.

\par Finally in this Section a few remarks on the CLEO measurement~\cite{CLEO},
 which is not included in the present analysis.
 Relativistic corrections were not taken into account in the theoretical predictions,
 which were suitably modified to account for detector acceptance and resolution
 effects before comparison with the (uncorrected) experimental data. Fitting the
LO QCD, Photiadis and RDF spectra to the data yields $\chi^2$ of 14.2, 10.0, and 8.1,
 respectively, for eleven degrees of freedom~\cite{CLEO}. The corresponding
 respective confidence
levels are 0.22, 0.46 and 0.70. Although the RDF model is slightly favoured, all
 the fits have acceptable confidence levels and no distinction between the different
 theoretical models is possible from this measurement.

\newpage
\SECTION{\bf{Determinations of $\alpha_s(\mQ)$}}

\begin{figure}[htbp]
\begin{center}\hspace*{-0.5cm}\mbox{
\epsfysize10.0cm\epsffile{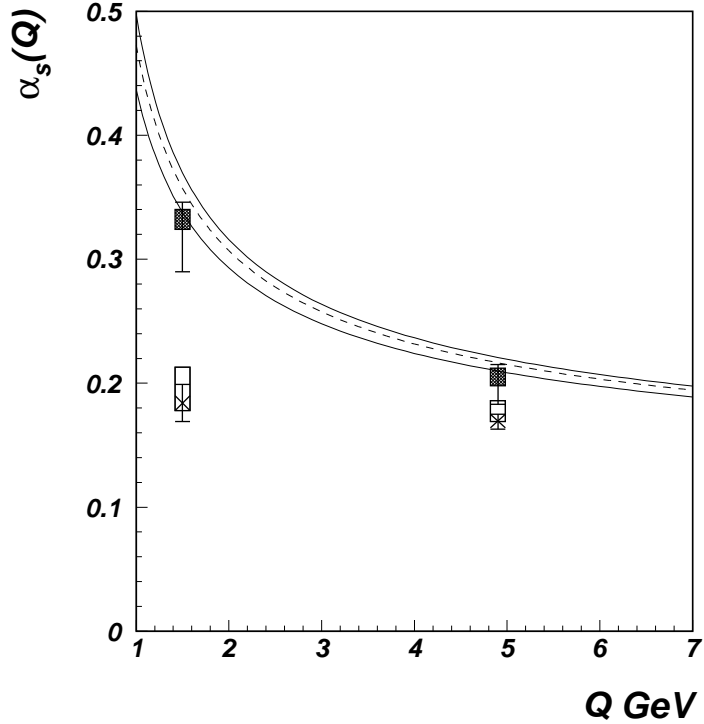}}
\caption{{\it Comparison of $\alsc$ and $\alsb$ values obtained from
 measurements of $R'_{\Jp}$ and $R'_{\Up}$ compared to the world average value
 (dashed curve with $\pm 1\sigma$ limits indicated by the solid curves)
 from Reference~\cite{PDG}. The crosses with error bars show values obtained assuming 
 $\mg = 0$ and $\mu/\mQ = 1.0$. The effect of varying the renormalisation scale 
 in the range $0.6 < \mu/\mQ < 2.0$ is indicated by the vertical boxes. The
 square points with error bars are obtained by applying full
 gluon mass corrections with $\mg = 0.721$ GeV for the $\Jp$ and $\mg = 1.18$ GeV
 for the $\Up$ with  $\mu/\mQ = 1.0$. NLO QCD corrections are applied in all cases.}}
\label{fig-fig7}
\end{center}
 \end{figure}
\begin{table}
\begin{center}
\begin{tabular}{|c|c|c|} \hline 
 Branching ratio  & Experimental value  & Reference \\
\hline
\hline
$\Gamma(\Jp \rightarrow
 {\rm hadrons})/
\Gamma_{\Jp}$ & 0.632 $\pm$ 0.022 & ~\cite{PDG} \\
\hline
$\Gamma(\Jp \rightarrow
\gamma+{\rm hadrons})/
\Gamma_{\Jp}$  & 0.0624 $\pm$ 0.0067 & This paper  \\
\hline
$R'_{\Jp}$ & 10.13 $\pm$ 1.14 & Ratio of above \\
\hline
\hline
$R'_{\Up}$ & 33.33 $\pm$ 2.44 & ARGUS~\cite{ARGUS} \\
\hline
$R'_{\Up}$ & 37.04 $\pm$ 6.17 & Crystal Ball~\cite{CB} \\
\hline
$R'_{\Up}$ & 36.36 $\pm$ 2.11 & CLEO2~\cite{CLEO2} \\
\hline
$R'_{\Up}$

 & 35.46 $\pm$ 1.51 & Weighted average \\
\hline
\end{tabular}
\caption[]{{\it  Experimental branching ratios used to determine 
$\alpha_s(\mQ)$}}
\end{center}
\end{table}

\begin{table}
\begin{center}
\begin{tabular}{|c|c|c|c|} \hline 
$\mu/\mQ$  & 0.6  & 1.0 & 2.0 \\
\hline
\hline
$\alsc$ & 0.178 $\pm$ 0.011 & 0.184 $\pm$ 0.015  & 0.213 $\pm$ 0.023 \\
$\Delta(PDG)$  & -0.179 $\pm$ 0.022 & -0.173 $\pm$ 0.024 & -0.144 $\pm$ 0.03 \\
Deviation($\sigma$) & -8.1 & -7.2 & -4.8   \\
\hline
$\alsb$ & 0.169 $\pm$ 0.005 & 0.169 $\pm$ 0.006  & 0.186 $\pm$ 0.008 \\
$\Delta(PDG)$  & -0.048 $\pm$ 0.008 & -0.048 $\pm$ 0.009 & -0.031 $\pm$ 0.01 \\
Deviation($\sigma$) & -6.0  & -5.3  & -3.0   \\
\hline
\end{tabular}
\caption[]{{\it $\alpha_s(\mQ)$ values obtained neglecting
 gluon mass corrections for different choices of the 
 renormalisation scale $\mu$. $\Delta(PDG)$ is the difference from the PDG~\cite{PDG}
 average values: $\alsc = 0.357_{-0.019}^{+0.013}$,
  $\alsb = 0.217_{-0.007}^{+0.004}$. `Deviation' is $\Delta(PDG)$ divided
 by its error}}
\end{center}
\end{table}

\newpage 
\begin{table}
\begin{center}
\begin{tabular}{|c|c|c|c|c|} \hline
   &   & $f_{\gamma \rm{gg}}$ &
$f_{ \rm{ggg}}$ &
$f_{ \rm{ggg}}/ f_{\gamma  \rm{gg}}$ \\ \cline{2-5} 
 \cline{2-5}
$\Jp$ & Phase Space & $0.40_{-0.02}^{+0.07}$ & $0.18_{-0.02}^{+0.08}$ &
 $0.45_{-0.02}^{+0.10}$ \\  \cline{2-5}
   & LW & $0.52_{-0.01}^{+0.06}$ & $0.18_{-0.01}^{+0.07}$ &
 $0.35_{-0.02}^{+0.09}$ \\  \hline 
$\Up$ & Phase Space & $0.74_{-0.04}^{+0.11}$ & $0.61_{-0.06}^{+0.15}$ &
 $0.83_{-0.04}^{+0.08}$ \\  \cline{2-5}
   & LW & $0.80_{-0.03}^{+0.08}$ & $0.61_{-0.06}^{+0.16}$ &
 $0.77_{-0.05}^{+0.11}$ \\  \hline  
\end{tabular}
\caption[]{{\it Gluon mass correction factors. `Phase Space' indicates that
 gluon mass effects are taken into account only in the kinematic limits. `LW'
 means that the complete calculation of Reference~\cite{LW} is used
The errors quoted are derived from the total uncertainties in the 
 fitted values of $\mg$. Note that $f_{\gamma \rm{gg}} = f_{ \rm{ggg}} = 1$ for 
$\mg = 0$.}}  
\end{center}
\end{table}

\begin{table}
\begin{center}
\begin{tabular}{|c|c|c|c|c|} \hline 
$\mu/\mQ$  & 0.6  & 1.0 & 2.0 & No O($\alpha_s$) Corr. \\
\hline
\hline
$\alsc$ & 0.221$^{+0.002}_{-0.008}$ & 0.298$^{+0.016}_{-0.041}$  & 0.467$^{+0.062}
_{-0.119}$ & 0.525$^{+0.066}_{-0.132}$ \\
$\Delta(PDG)$  & -0.136 $\pm$ 0.027 & -0.059 $\pm$ 0.025 & 0.110 $\pm$ 0.120
 & 0.168 $\pm$ 0.133 \\
Deviation($\sigma$) & -5.0 & -2.4 & 0.9 & 1.3  \\
\hline
$\alsb$ & 0.189$^{+0.005}_{-0.012}$ & 0.193$^{+0.008}_{-0.015}$ &
 0.221$^{+0.035}_{-0.015}$ &  0.248$^{+0.015}_{-0.025}$ \\
$\Delta(PDG)$  & -0.028 $\pm$ 0.009 & -0.024 $\pm$ 0.011  & 0.004 $\pm$ 0.016
& 0.032 $\pm$ 0.025 \\
Deviation($\sigma$) & -3.1 & -2.2 & 0.25 & 1.2 \\
\hline
\end{tabular}
\caption[]{{\it $\alpha_s(\mQ)$ values obtained including phase space
 gluon mass corrections for different choices of the 
 renormalisation scale $\mu$. See Table 10 for the definitions of $\Delta(PDG)$
 and `Deviation'.
 The quoted errors include the effect of the uncertainties in the
 gluon mass correction factors.}}
\end{center}
\end{table}

\begin{table}
\begin{center}
\begin{tabular}{|c|c|c|c|c|} \hline 
$\mu/\mQ$  & 0.6  & 1.0 & 2.0 & No O($\alpha_s$) Corr. \\
\hline
\hline
$\alsc$ & 0.224$^{+0.001}_{-0.005}$ & 0.332$^{+0.014}_{-0.042}$  & 0.617$^{+0.088}
_{-0.178}$ & 0.681$^{+0.086}_{-0.187}$ \\
$\Delta(PDG)$  & -0.133 $\pm$ 0.019 & -0.025 $\pm$ 0.024 & 0.260 $\pm$ 0.178
 & 0.324 $\pm$ 0.187 \\
Deviation($\sigma$) & -7.0 & -1.0 & 1.5 & 1.7  \\
\hline
$\alsb$ & 0.197$^{+0.005}_{-0.016}$ & 0.205$^{+0.010}_{-0.022}$ &
 0.239$^{+0.017}_{-0.033}$ &  0.271$^{+0.021}_{-0.040}$ \\
$\Delta(PDG)$  & -0.020 $\pm$ 0.009 & -0.012 $\pm$ 0.012  & 0.022 $\pm$ 0.033
& 0.054 $\pm$ 0.040 \\
Deviation($\sigma$) & -2.2 & -1.0 & 0.67 & 1.4 \\
\hline
\end{tabular}
\caption[]{{\it $\alpha_s(\mQ)$ values obtained, including full
 gluon mass corrections from Reference~\cite{LW}, for different choices of the 
 renormalisation scale $\mu$. See Table 10 for the definitions of $\Delta(PDG)$
 and `Deviation'.
 The quoted errors include the effect of the uncertainties in the
 gluon mass correction factors.}}
\end{center}
\end{table}

\begin{table}
\begin{center}
\begin{tabular}{|c|c|} \hline 
 Experiment  & $\alpha_s(\mu_{BLM}) = \alpha_s(1.5{\rm GeV})$  \\
\hline
\hline
 CUSB & $0.226^{+0.067}_{-0.042}$ \\
 ARGUS & $0.225 \pm 0.011 \pm 0.019$ \\
Crystal Ball & $0.25 \pm 0.02 \pm 0.04$ \\
 CLEO & $0.27^{+0.03~+0.03}_{-0.02~-0.02}$ \\
\hline
 Weighted Mean & $0.228 \pm 0.016 \pm 0.011$ \\
\hline
\end{tabular}
\caption[]{{\it Published values of $\alpha_s$ at the BLM scale 
 determined from $\Up$ radiative decays. For
 CLEO, the measured spectrum is extrapolated using the RDF prediction.
 The first error quoted in each case is statistical, the second 
 systematic.}}
\end{center}
\end{table}

The strong coupling constant $\alpha_s(\mQ)$ may be determined from the 
experimental measurements of the branching ratio:
\begin{equation}
 R'_V \equiv \frac{\Gamma(V \rightarrow  {\rm hadrons})}
{\Gamma(V \rightarrow \gamma + {\rm hadrons})}
\end{equation}
where $V$ denotes a vector heavy quarkonium ground state ($\Jp$ or $\Up$).
 Use of $R'_V$ has the advantage, as  compared with other branching
 ratios sensitive to  $\alpha_s$ (for example
$R_V \equiv \Gamma(V \rightarrow  {\rm hadrons})/
\Gamma(V \rightarrow \ell^+ \ell^-)$, that relativistic corrections
 cancel in the approximation used in the present paper.

In Eqn(7.1) the process `$V \rightarrow \gamma + {\rm hadrons}$' is understood
 to be the heavy quark annihilation process into light hadrons, for which the lowest 
 order QCD process is $V \rightarrow \gamma gg$. Thus the non-annihilation process
 $V \rightarrow \gamma \eta_Q$, where $\eta_Q$ is the lowest-lying pseudoscalar
 heavy quarkonium ground state, is not included. Similarly, `$V \rightarrow
  {\rm hadrons}$' is understood to be the direct (strong interaction) process
 for which
the lowest order QCD process is $V \rightarrow \rm{ggg}$. The contribution of the
radiative process: $V \rightarrow \gamma^{\star}  \rightarrow \rm{q}\rm{\overline{q}}
 \rightarrow {\rm hadrons}$
(branching ratio 17$\%$ for the $\Jp$) is, therefore, not included.

\par The experimental branching ratios used here to extract $\alpha_s(\mQ)$ are
summarised in Table 9. For the $\Jp$ the branching ratio $\Gamma(\Jp \rightarrow
\gamma + \rm{hadrons})/
\Gamma_{\Jp}$, where $\Gamma_
{\Jp}$ denotes the total width of the ${\Jp}$, is obtained from 
the fits to the branching ratio $R_{\eta'}$ presented in Table 4. The measured exclusive 
 branching ratio into $\gamma \eta'$~\cite{PDG}, is used to derive, from $R_{\eta'}$,
 the branching
 fraction for `$\gamma~{\rm continuum}$' (see Section 5 above). The measured exclusive
$\gamma \eta$ and $\gamma \eta$' fractions are then added to the `$\gamma~{\rm continuum}$'
fraction to give the full branching fraction into  $\gamma+{\rm hadrons}$ quoted in Table 9.
 The errors on this quantity are derived from that on the weighted average value
 of $R_{\eta'}$ and the experimental $\gamma \eta$ and $\gamma \eta$' fractions~\cite{PDG}.
 The branching fraction $\Gamma(\Jp \rightarrow {\rm hadrons})/
\Gamma_{\Jp}$ is derived from the total hadronic width of the $\Jp$ given in Ref~\cite{PDG}
 by subtracting the contribution~\cite{PDG}  of the process $\Jp \rightarrow
 \gamma^{\star}  \rightarrow \rm{q}\rm{\overline{q}} \rightarrow {\rm hadrons}$.

\par In the case of the $\Up$ measurements, the branching ratio $R'_{\Up}$ was directly
 measured by CUSB, CLEO, ARGUS, Crystal Ball and CLEO2.
 The values of  $R'_{\Up}$ obtained by the last three of these experiments are reported
 in Table 9. In all cases the extrapolation of the measured photon spectrum to $z=0$
 was done using the RDF theoretical spectrum. As will be shown below, the shape of this
 spectrum is in good agreement with the fit curves obtained in the present paper,
 which take into account relativistic corrections and explicit gluon mass effects at
 the born level. This shape is little affected by including HO QCD corrections
 according to Ref~\cite{Photiadis} and/or Ref~\cite{Kramer}. To extract $\alpha_s(m_b)$,
 the weighted average (also reported in 
 Table 9) of the $R'_{\Up}$ measurements of ARGUS, Crystal Ball and CLEO2 is used.
\par Consistent results for $R'_{\Up}$ were found by CUSB: 33.4 $\pm$ 6.6 and CLEO
 (using the RDF spectrum): 39.4 $\pm$ 3.6. Since, however, the shape of the inclusive
 photon spectrum measured by CUSB is inconsistent with those measured by the other four
 experiments, and the analysis of the CLEO photon spectrum could not be performed,
 these two measurements are omitted from the $R'_{\Up}$ average used here to determine
 $\alpha_s(m_b)$. Thus, the  $\alpha_s$ analysis is performed using only data
 from experiments for which a consistent determination of the effective gluon
 mass was possible. Including also the CUSB and CLEO measurements in the weighted
 average of $R'_{\Up}$ gives the value 35.94 $\pm$ 1.36 which differs by only
 0.35$\sigma$ from the weighted average of ARGUS, Crystal Ball and CLEO2 quoted
 in the last row of Table~9.

\begin{figure} 
 \begin{center}
   \includegraphics[width=10cm]{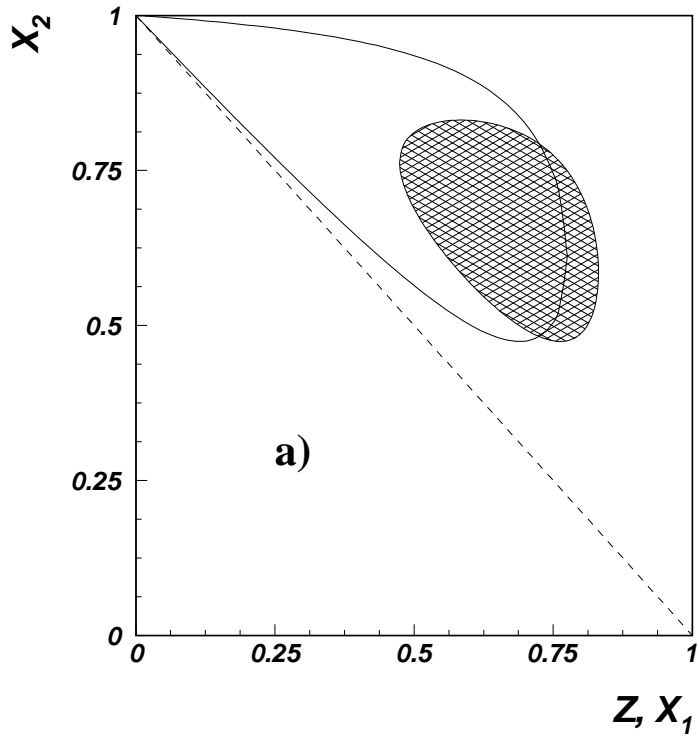}
   \includegraphics[width=10cm]{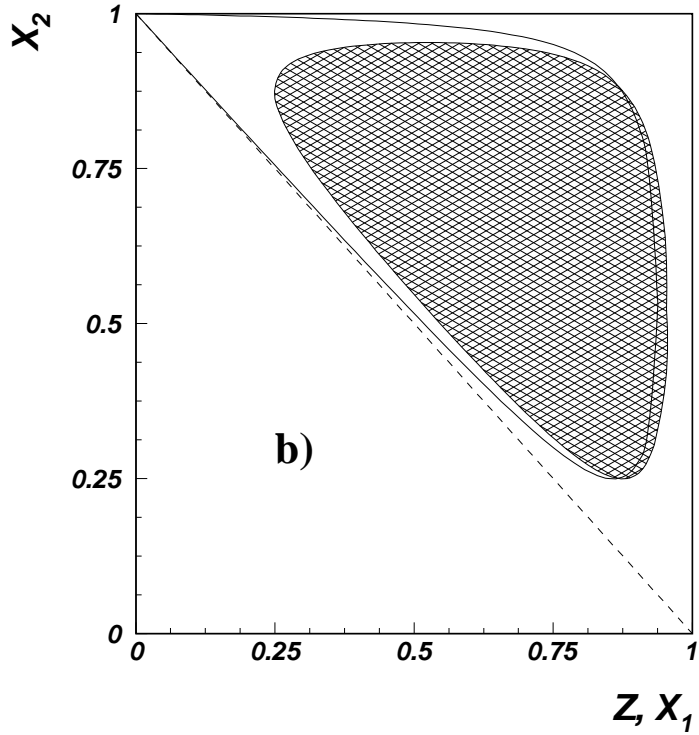} 
 \caption{{\it Allowed regions of the Dalitz plots for $V \rightarrow \gamma \rm{gg}$ 
 (open areas) 
 and $V \rightarrow  \rm{ggg}$ (cross-hatched areas), corresponding to the effective gluon mass
 values 0.721 GeV ($\Jp$), a), and 1.18 GeV ($\Up$). b). For vanishing effective gluon mass,
 the full areas above the dotted lines are kinematically allowed}}
\label{fig-fig8}
 \end{center}
\end{figure}
\newpage
\begin{figure} 
 \begin{center}
   \vspace*{-5cm}
   \includegraphics[width=11cm,bb=25 149 568 693]{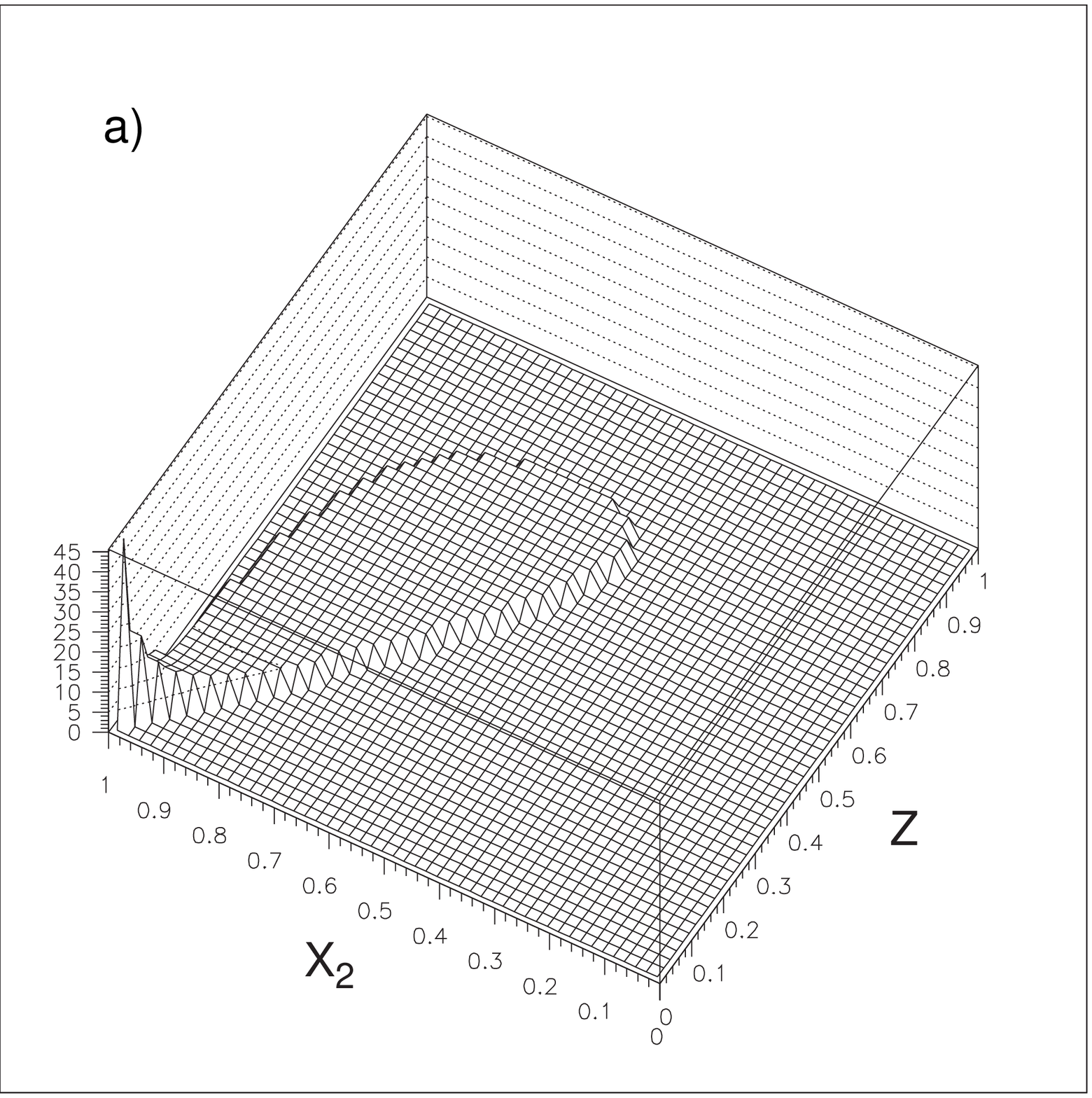}\\
   \vspace*{0.3cm}
   \includegraphics[width=11cm,bb=25 149 568 693]{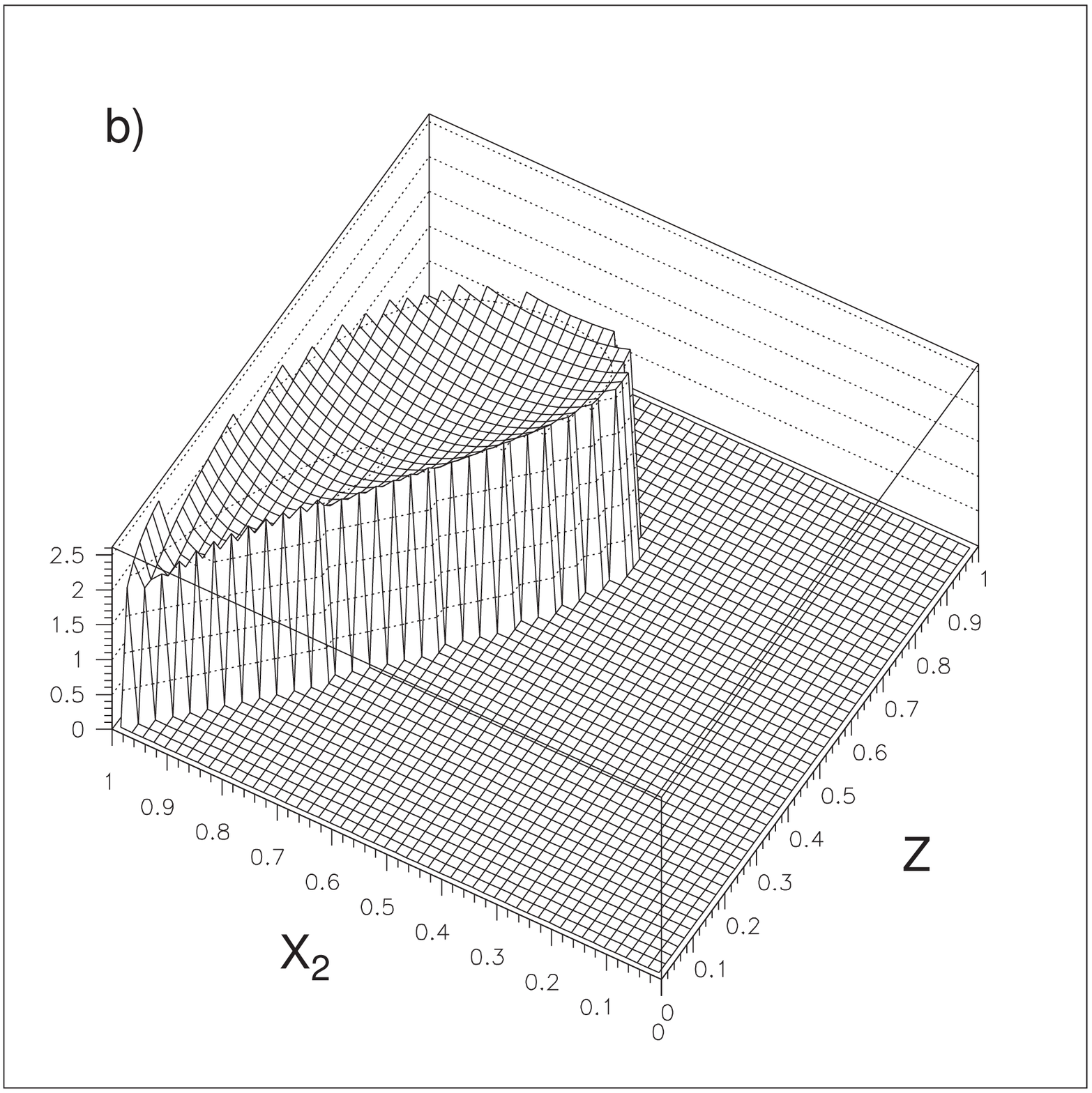}
   \vspace*{3cm} 
 \caption{{\it The effect of gluon mass corrections ($\mg = 0.721$ GeV)
  on the differential decay rate $d\Gamma/dzdx_2$ (Eqn.(4.1) for $\Jp \rightarrow \gamma \rm{gg}$ decays.  
 a) phase space effects only, b) inclusion also of longitudinal gluon polarisation states.}}
\label{fig-fig9}
 \end{center}
\end{figure}
\newpage
\begin{figure} 
 \begin{center}
   \includegraphics[width=11cm,bb=25 149 568 693]{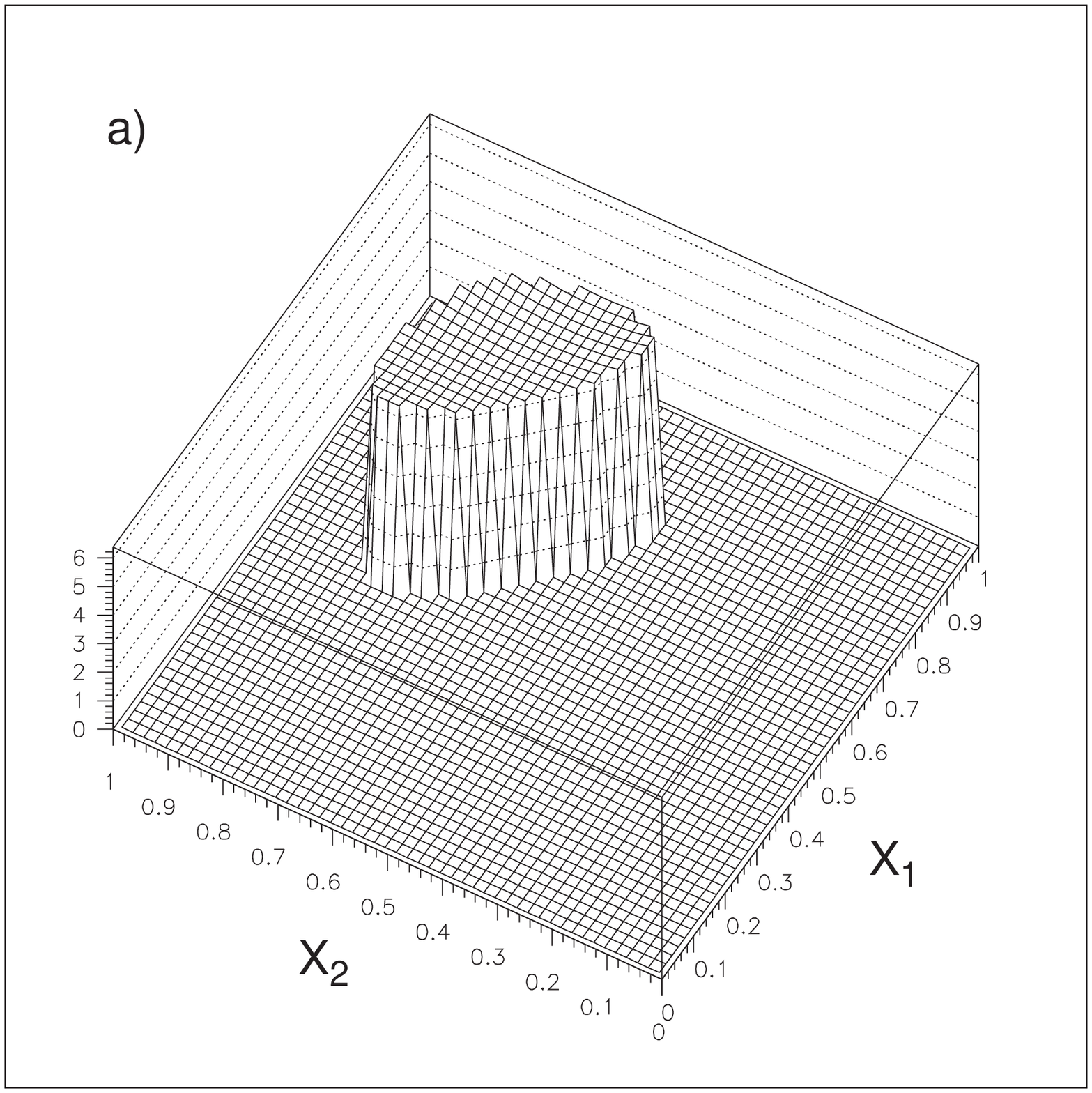}\\
  \vspace*{0.3cm}
   \includegraphics[width=11cm,bb=25 149 568 693]{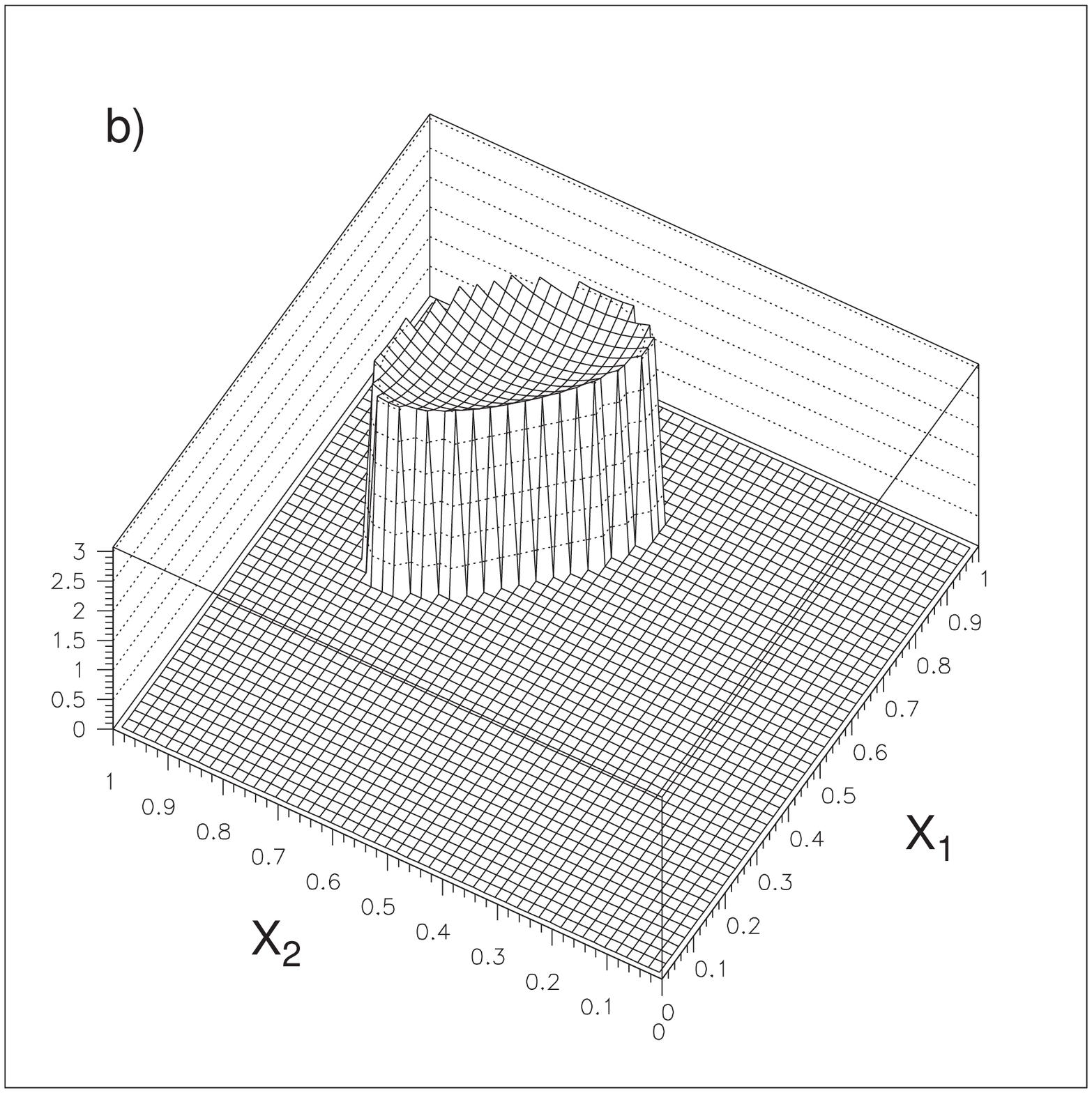} 
 \caption{{\it The effect of gluon mass corrections ($\mg = 0.721$ GeV)
  on the differential decay rate $d\Gamma/dx_1dx_2$ (Eqn.(7.6)
  for $\Jp \rightarrow  \rm{ggg}$ decays.  
 a) phase space effects only, b) inclusion also of longitudinal gluon polarisation states.}}
\label{fig-fig10}
 \end{center}
\end{figure}

\par Taking into account NLO QCD radiative corrections~\cite{KWMaC} as well as gluon
 mass corrections, 
$\alpha_s(\mQ)$ may be derived from the formulae:
\begin{eqnarray}
 R'_{\Jp} & = & \frac{5 \alsm}{16 \alpha}\frac{f_{\rm{ggg}}}{f_{\gamma \rm{gg}}}
 \frac{[1+\frac{\alsm}{\pi}(\frac{3}{2}\beta_0\ln \frac{\mu}{m_c}-3.74)]}
 {[1+\frac{\alsm}{\pi}(\beta_0\ln \frac{\mu}{m_c}-6.68)]}, \\
 R'_{\Up} & = & \frac{5 \alpha_s(\mu)}{4 \alpha}\frac{f_{\rm{ggg}}}{f_{\gamma \rm{gg}}}
 \frac{[1+\frac{\alsm}{\pi}(\frac{3}{2}\beta_0\ln \frac{\mu}{m_b}-4.90)]}
 {[1+\frac{\alsm}{\pi}(\beta_0\ln \frac{\mu}{m_b}-7.45)]}.
\end{eqnarray}
 Here $\beta_0$ is the one-loop QCD beta function coefficient: 
\[ \beta_0 = 11-\frac{2 n_f}{3} \]
 where the number of active quark flavours, $n_f$, is taken to be 3 for the $J/\psi$ and 
 4 for the $\Up$. The values of the heavy quark masses are assumed to be
 $m_c = 1.5$ GeV and $m_b = 4.9$ GeV. The parameter $\mu$ is an arbitary renormalisation 
 scale, and $f_{\rm{ggg}}$,  $f_{\gamma \rm{gg}}$ are correction factors taking into
 account gluon mass effects. As previously pointed out, relativistic corrections cancel
 in the ratio $R'_V$. For any given value of the renormalisation scale $\mu$, Eqns.(7.2) or
 (7.3) are solved for $\alsm$. The corresponding value of $\alsQ$ is then found by use
 of the one-loop QCD evolution formula:
\begin{equation}
\frac{1}{\alsQ} = \frac{1}{\alsm}-\frac{\beta_0}{2 \pi}\ln \frac{\mu}{m_Q} 
\end{equation}
\par Setting $f_{\rm{ggg}} = f_{\gamma \rm{gg}} = 1$ ( i.e. neglecting gluon mass
 corrections) and choosing the values\footnote{Eqns.(7.2) and (7.3)
 have no real solution for  $\alsm$ when $\mu/m_Q =$ 0.5} $\mu/m_Q =$ 0.6, 1.0 and 2.0\, yields
 the values of $\alsc$ and $\alsb$
 reported in Table~10. It can be seen that there is poor agreement
 with the world average values~\cite{PDG}:
\[ \alpha_s(1.5\rm{GeV}) = 0.357^{+0.013}_{-0.019},~~~~  \alpha_s(4.9\rm{GeV}) =  0.217^{+0.004}_{-0.007}. \] 
 These are calculated using Eqn.(9.5) of reference~\cite{PDG98} and correspond to four active
 quark flavours. Matching to the five flavour region where $\Lambda^{(5)}= 219^{+25}_{-23}$ MeV
 (corresponding
 to the world average value $\alpha_s(M_Z)=0.119 \pm 0.002$) is done using Eqn.(9.7) of
 reference~\cite{PDG98} at a matching scale of 4.3 GeV. For both the $J/\psi$ and the $\Up$ the
 best agreement is found for $\mu/\mQ =2.0$, but the respective deviations are still 
 4.8$\sigma$ and 3.0$\sigma$. Fig.7 shows a comparison of the $\alsc$ and $\alsb$ values
 quoted in Table 10 for $\mu/\mQ =1.0$  with the present world average value of
 $\alpha_s(Q)$~\cite{PDG}.
\newpage
 \par The gluon mass correction factors are calculated by integrating
the differential distributions of gluon and photon energies of the 
decay processes $V \rightarrow \gamma \rm{gg}$  or $V \rightarrow \rm{ggg}$
 over the kinematically allowed regions
 of their respective Dalitz plots. For $V \rightarrow \gamma \rm{gg}$ it
 is convenient to use the photon spectrum given by integrating Eqn.(4.1)
 over the gluon energies~\cite{LW}:
\begin{eqnarray}
\frac{1}{\Gamma_0}\frac{d\Gamma}{dz}& = &  
\frac{1}{\pi^2-9}\left[   \right.
\frac{x_+-x_-}{z^2}  \nonumber \\
 & &+\frac{\ln(x_+/x_-)}{z^2(-2+4\eta+z)^3}
 [8(2\eta-1)^2(2-4\eta+7\eta^2)  \nonumber \\
 & &+8(2\eta-1)(5-12\eta+10\eta^2+2\eta^3)z  \nonumber \\
 & &+2(2\eta-1)(-17+10\eta+6\eta^2)z^2
 +2(-5+2\eta+2\eta^2)z^3] \nonumber \\ 
& &+\frac{(1/x_--1/x_+)}{z^2(-2+4\eta+z)^2}[4(2\eta-1)^2(1+3\eta^2)
 +4(2\eta-1)(3-4\eta+2\eta^2+2\eta^3)z  \nonumber \\
& &+2(7-18\eta+10\eta^2+10\eta^3)z^2
 +4(2+\eta)(2\eta-1)z^3+(2+\eta)z^4] \left], \right. 
\end{eqnarray}  
 where:
\[ x_{\pm}=1-2\eta-\frac{z}{2}\left[ 1\mp\sqrt{1-\frac{4\eta}{1-z}}\right]. \]
  
For the decays $V \rightarrow \rm{ggg}$ a two dimensional integration 
is performed over the distribution~\cite{LW}:

\begin{eqnarray}
\frac{1}{\Gamma_0}\frac{d\Gamma}{dx_1dx_2dx_3} = &  & 
\frac{1}{(\pi^2-9)}\frac{1}{(x'_1)^2(x'_2)^2(x'_3)^2}\left[   \right. 
 \frac{16}{3} \eta (1-3\eta)^2(1-\frac{51}{8}\eta-\frac{15}{4}\eta^2) \nonumber \\
   &+&((x'_1)^2+(x'_2)^2+(x'_3)^2)(1-14\eta+48\eta^2+25\eta^3) \nonumber \\
   &-&2((x'_1)^3+(x'_2)^3+(x'_3)^3)(1-\frac{17}{3}\eta-3\eta^2) \nonumber \\
   &+&((x'_1)^4+(x'_2)^4+(x'_3)^4)(1+\frac{\eta}{2})  \left] \right.
\end{eqnarray} 
 The allowed phase space region is:

\begin{eqnarray}
2 & = & x_1+x_2+x_3    \\
  2\sqrt{\eta} & \le & x_1  \le ~ 1-3\eta    \\ 
 x_2^{min} & \le & x_2~\le~ x_2^{max}  \\
 x_2^{max}  & = & 1-\frac{x_1}{2}\left[1-D(x_1,\eta)\right] \\
 x_2^{min}  & = & 1-\frac{x_1}{2}\left[1+D(x_1,\eta)\right]  \\
D(x_1,\eta) & = & \sqrt{(1-\frac{4\eta}{x_1^2})
(1-\frac{4\eta}{1-x_1+\eta})}
\end{eqnarray}

\par The allowed regions of the Dalitz plots for $\Jp \rightarrow \gamma \rm{gg}$ 
 decays (open contour) and  $\Jp \rightarrow \rm{ggg}$ (cross-hatched contour)
 for $\mg = 0.721$ GeV are shown in Fig.8a. Similar contours for the corresponding 
$\Up$ decays and $\mg = 1.18$ GeV, are shown in Fig.8b. The phase space suppression
 factors due to gluon mass effects are the ratios of the areas inside the contours
 to the area above the dotted line, corresponding to $\mg =0$. It is seen that the
 phase space suppression is considerable for $\Up$ and very large for $\Jp$ decays.
 \par The effect of the inclusion of longitudinal polarisation states for the
 gluons is illustrated in Figs. 9 and 10, which show decay rates as a function of
 photon and gluon energies as given by Eqn.(4.1) ( $\Jp \rightarrow \gamma \rm{gg}$)
 and Eqn.(7.6) ($\Jp \rightarrow \rm{ggg}$), respectively, for $\mg = 0.721$ GeV. In 
 Fig.9a,10a the longitudinal gluon contributions are suppressed by setting 
 $\eta =0$ except in the equations defining the Dalitz plot boundary, i.e. only
 the phase space limitations due to the non-vanishing value of $\mg$ are taken into
 account. In Fig.9b and 10b the complete formulae (4.1) and (7.6),
 respectively, are
 used. The most dramatic effect of the longitudinal contributions is a strong 
 suppression of the decay rate for $\Jp \rightarrow \gamma \rm{gg}$
 in the region $z \simeq 0.0$, $x_2 \simeq 0.0$.
 As the experimental data analysed here has $z \geq 0.3$, this has no practical
 consequences for the present work. Indeed, in the region of small $z$, the dominant
 mechanism of direct photon production is expected to be the fragmentation
 of light hadrons into photons~\cite{Hautmn},
 \cite{HandC}, which is not taken into account in
 the NLO QCD calculation of Reference~\cite{Kramer}. As can be seen in Figs.9b,10b
 the other effect of the longitudinal contributions is a modest suppression
 of the decay rate, near the centre of the allowed region of the Dalitz plot,
 relative to the boundaries. The gluon mass correction factors given by 
 integrating Eqn.(7.5) over $z$ or Eqn.(7.6) over $x_1$ and $x_2$, are presented in
 Table 11. The rows labelled `LW' use the complete formulae, those labelled
 `Phase Space' have $\eta =0$ except in the equations defining the kinematic 
 limits. It can be seen that that longitudinal gluon effects are negligible 
 in the correction factors for $\rm{V} \rightarrow \rm{ggg}$ decays for both the
 $\Jp$ and the $\Up$. For  $\rm{V} \rightarrow \gamma \rm{gg}$ decays these
 effects increase $f_{\gamma \rm{gg}}$ by 30$\%$ and 8$\%$, respectively,
 for the $\Jp$ and the $\Up$.
 The errors quoted on the correction factors are derived from the total
 errors on $\mg$ given in Sections 5 and 6 above.

 \par The values of $\alsc$ and $\alsb$ derived from Eqns.(7.2) and (7.3), taking into
 account gluon mass effects according to the values of
 $f_{\rm{ggg}}/f_{\gamma \rm{gg}}$ given in Table 11 are presented in Table 12
 (phase space corrections only) and Table 13 (full gluon mass corrections).
 In each case the values for $\mu/\mQ =$ 0.6, 1.0 and 2.0 are presented as well as those
 given by neglecting the $O(\alpha_s)$ corrections in Eqns.(7.2) and (7.3). For all
 choices of renormalisation scale, the agreement with the world average values is
 improved as compared to the values presented in Table 10, where gluon mass effects
 are neglected. The best agreement (at the 0.3-1.5$\sigma$ level) is found for
 $\mu/\mQ =$ 2.0, though almost equally consistent results (deviations of 1.2-1.7$\sigma$)
 are found when the $O(\alpha_s)$ corrections are neglected. The inclusion of
 longitudinal gluon effects, in the latter case, increases the value of $\alsc$ by 30$\%$ and $\alsb$
 by 9$\%$. These shifts are comparable to the uncertainties on $\alpha_s$ due to the
 experimental errors on $R'_V$ and the gluon mass correction factors. Similar shifts are found
 for $\mu/\mQ =$ 2.0 and somewhat smaller ones for $\mu/\mQ =$ 1.0. The values of $\alsc$
 and $\alsb$ given by using the full
 gluon mass correction with $\mu/\mQ =$ 1.0 are compared, in Fig.7,
  with the world average value of $\alpha_s(\mu)$, and the values obtained for the
 same renormalisation scale, but without gluon mass corrections.

 \par Comparison of Tables 12 and 13 with Table 10 and inspection of Fig.7 shows
 that that the inclusion of effective gluon corrections are essential in order to obtain
 values of  $\alsc$ and $\alsb$, derived from measurements of $R'_V$, 
 that are consistent with the current world average determination
 of $\alpha_s(Q)$.
\par Each of the published experiments extracted a value of $\alpha_s$ from the measured
 values of $R'_{\Up}$. In all cases, good consistency was found with other available 
 measurements of $\alpha_s$. This is due to the relatively large errors, at the time, both on the
 individual measurements and on the average value of the other measurements with which they were
 compared. It does not at all contradict the results shown in Table 10, which show instead a 
 poor consistency of values derived from
 the weighted average value of $R'_{\Up}$ of the three latest experiments
 with the current world average value of $\alpha_s$.
 The experiments CUSB, ARGUS, Crystal Ball and CLEO all used the Brodsky-Lepage-Mackenzie (BLM)
 scale setting procedure~\cite{BLM} to determine $\alpha_s$ at a scale of $0.157M_{\Up}$, i.e.
 1.5 GeV. These results, together with their weighted mean, are presented in Table 14. The mean
 value of $\alpha_s(1.5{\rm GeV})$ of
 $0.228 \pm 0.019$
 differs from the current world average value of
 $0.357+0.013-0.019$ by 4.8 standard deviations, and is consistent with the results for $\alsb$
 given in Table 10. The CLEO2 experiment used the Principle of Minimal Sensitivity (PMS)
 ~\cite{PMS} to determine:
\[ \alpha_s(M_{\Up}) = 0.163 \pm 0.002 \pm 0.009 \pm 0.010 \]
 where the first error is statistical, the second systematic, and the third due to the 
 estimated uncertainty of the PMS scale setting procedure. Evolving to the scale $m_b = 4.9$
 GeV using Eqn.(7.4) gives
\[ \alpha_s(m_b) = 0.190^{+0.020}_{-0.019}  \]
 This differs from the current world average value cited above by only 1.3$\sigma$, but also
 agrees within 0.19-1.1$\sigma$ with the values quoted in Table 10. The latter, however, differ from
 the world average by 3.0-5.7$\sigma$. The results of the present analysis, and the combined
 average of those already published in the literature are thus in agreement, and lie 3$\sigma$
 or more below the world average. No consistent description is obtained unless the effective gluon
 mass effects are taken into account.

\SECTION{\bf{Discussion}}

\begin{figure}[htbp]
\begin{center}\hspace*{-0.5cm}\mbox{
\epsfysize10.0cm\epsffile{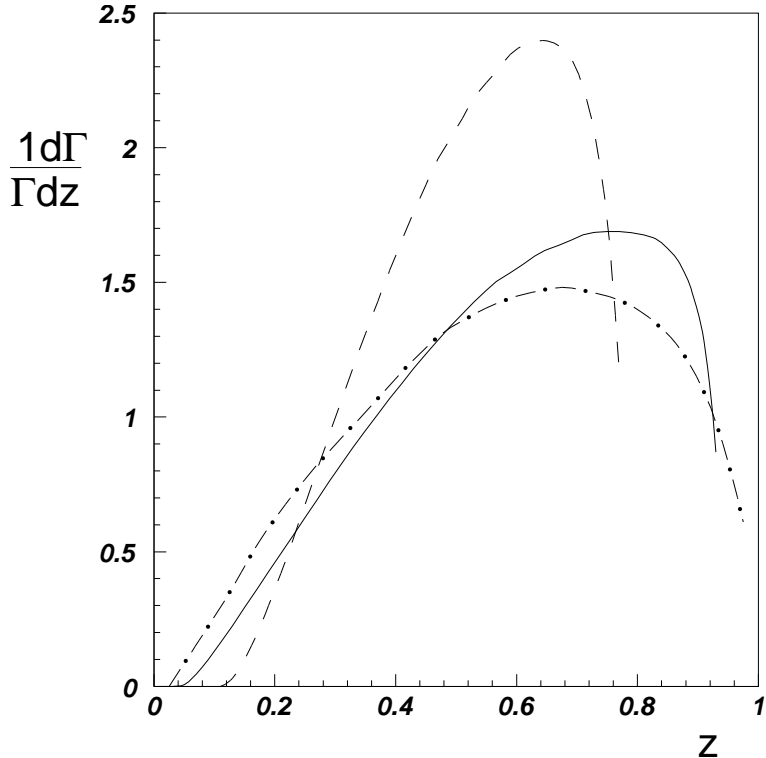}}
\caption{{\it Inclusive photon spectra including
 gluon mass effects. Dashed curve: $\Jp \rightarrow \gamma X$, $\mg = 0.721$ GeV ;
 solid curve: $\Up \rightarrow \gamma X$, $\mg = 1.18$ GeV. The dot-dashed curve 
 shows the RDF~\cite{RDF} prediction for $\Up \rightarrow \gamma X$.}}
\label{fig-fig11}
\end{center}
 \end{figure}

The inclusive photon spectrum for $\Jp$ decays (dashed curve) and $\Up$ decays (solid curve) 
obtained from the fits performed here to all available experimental data
are shown in Fig.11. In both cases, the relativistic corrections and the Photiadis HO QCD 
correction are included. The values 0.721 GeV and 1.18 GeV of $\mg$ found in Sections 5 and 6 
above, are 
used for the $\Jp$ and $\Up$ respectively. It is clearly seen that the end-point suppression
is much more severe for the $\Jp$ than the $\Up$. Also shown in Fig.11 is the RDF spectrum,
which has been found to describe well all the measurements of the photon spectrum in
 $\Up \rightarrow \gamma X$ except that of CUSB. It is seen to be in good qualitative
 agreement with the $\Up$ fit curve, but to predict a much harder spectrum than the
 fit curve for the $\Jp$. In fact, the RDF model, where the gluon mass is perturbatively
 generated using a low cut-off value of 0.45 GeV, predicts that the spectra
 are very similar in shape for the $\Jp$ and the $\Up$. This is clearly not the case.
\par The very different shapes of the spectra for the $\Jp$ and the $\Up$ can only be
 understood  if the scale introduced into the kinematics of the process
 by the effective gluon mass is not small in
comparison with the mass of the $\Jp$. This condition is very well satisfied, since
 the rest mass of the two effective gluons of $2\times0.721 = 1.44$ GeV is 47$\%$
 of the $\Jp$ mass. As previously discussed in MCJHF1
 and MCJHF2, the stronger suppression of the end point in $\Jp$ decays
 can be understood as a propagator effect acting on off-shell gluons if
 the genuine gluon mass is $\simeq 1.0-1.5$ GeV, i.e. larger than the fitted
 value of $\mg$ for the $\Jp$, and similar to that found for the $\Up$.
 Corroborative evidence for this picture is provided by the structure of
 the hadronic final state.
 The dominance, for massive gluons,
 of the process $\rm{g}\rm{g} \rightarrow q \overline{q}$
 over gluon splitting
 $\rm{g}\rm{g} \rightarrow  \rm{gggg}, \rm{gg} q \overline{q}, q \overline{q} q \overline{q}$
 leads to a similar hadronic final state $X$ in radiative $\Jp$
 decays, to the annihilation process $e^+e^- \rightarrow \rm{q} \overline{q}$ at the
 same energy~\cite{Mark II}~\cite{mcjhf2},
 consistent with the experimental observations.
\par The huge difference observed in the shape of the $\Jp$ and $\Up$ spectra in
 Fig.11 is clearly at variance with the principle of Local Parton Hadron Duality
 (LPHD) where parton level
 pQCD calculations are used (as in the RDF model) down to scales of a few
 hundred MeV~\cite{LPHD}. Indeed, in Monte Carlo models that give a good detailed 
 description of hadronisation effects~\cite{JETSET}~\cite{HERWIG} the cut-off
 scale of perturbative QCD effects is in the range 1-2 GeV, comparable to the
 effective gluon mass in $\Up$ decays, and much larger than $\Lambda_{QCD}$.
 This point will be further discussed below.
\par Some remarks are now made on the related work presented in MCJHF1 and MCJHF2.
In this case fits were performed only to the MARKII data for the $\Jp$, and to those
 of ARGUS and Crystal Ball for the $\Up$. In the fit to the $\Jp$, only phase space
gluon mass corrections were included, with no HO QCD or relativistic corrections.
The $\Up$ fits used phase space gluon mass corrections and the Photiadis HO QCD
 correction, but no relativistic correction. The values obtained for $\mg$ of 
 $0.66 \pm 0.01$\footnote{There is a misprint in MCJHF1, propagated also to MCJHF2, 
 where this error is wrongly quoted as 0.08} GeV and $1.17 \pm 0.08$ GeV,
 respectively, are similar to those $0.721+0.016-0.068$ GeV, $1.18+0.09-0.29$ GeV, found
 in the present paper. The larger errors quoted here result from a study
 of theoretical systematics (relativistic corrections,
 different HO QCD corrections) not done in MCJHF1 and MCJHF2. Due to a
 programming error, the resolution functions used in these papers had a 
 width that was too large by a factor $\sqrt{2}$. This had the effect of
 destroying the sensitivity of the high $z$ part of the $\Jp$ spectrum to the 
 process $\Jp \rightarrow \gamma \eta'$. A fit with an acceptable Confidence
 Level was then obtained without explictly taking into account this decay channel
 as described in Section 5 above. An important difference between the present 
 work and MCJHF2 is an improved understanding of the effect of relativistic
 corrections. As is clear from the discussion at the beginning of Section 2 above,
 the `binding energy' $E(p)-\mQ$ introduced by KM must be a positive definite
 quantity. The same conclusion can be reached from simple physical reasoning. In the
 presence of the relativistic correction the heavy quark-antiquark annihilation
 process occurs over a finite spatial region around the origin of the
 radial wave function, instead of at the origin as in the static limit.
 As the ground state wave function peaks at the origin, relativistic
 corrections must always reduce the decay rate, not increase it.
 In MCJHF2, following the NRQCD~\cite{NRQCD} approach, the corresponding parameter
 $r$ was taken to be free, to be determined from experiment,
 and was allowed to take positive or negative values. In the present paper the
 relativistic correction parameter $<v^2>$ is set to the fixed values
 0.28 and 0.09, respectively, for the $\Jp$ and $\Up$ on the basis of
 potential model calculations.
 Finally, the NLO QCD correction to the inclusive photon spectrum
 ~\cite{Kramer} was not available when MCJHF2 was written.
\par The analysis presented in this paper has neglected possible Colour
Octet contributions to the radiative decay rates. These have been calculated for $\Up$
decays at NLO in the NRQCD formalism by Maltoni and Petrelli~\cite{COCTET}. In the
 region of interest for the fits performed in the present paper, $z > 0.3$,
 the corrections to the LO spectrum were found to be modest, $\simeq 10-15 \%$.
 In a more recent study~\cite{Wolf} in which octet operators were resummed to
  yield the so-called shape functions~\cite{Shapf} a much larger
 contribution was predicted in the near end-point region. However, comparison with the 
 CLEO2 data showed that the colour octet contribution, with normalisation fixed 
 by the velocity counting rules of NRQCD, exceeds the experimental data
 by between one and two orders of magnitude. It may also be remarked that the
 result of the shape function calculation, in which clusters of `non-perturbative'
 soft gluons are summed, is expected to be
 drastically affected (reduced) if the phase space suppression 
 associated with a gluon mass of $\simeq 1$GeV is taken
 into account.
 In view of the small correction to the shape of the spectrum due even to the 
 full colour singlet NLO correction~\cite{Kramer} as compared to that resulting
 from gluon mass effects (see Figs.3-6), the neglect of possible colour octet
 contributions is not expected to modify, in any essential way, the conclusions
 of this paper. In fact, according to Reference~\cite{Wolf} the colour octet 
 contributions are expected to strongly enhance the rate in the end-point 
 region, whereas what is observed is actually a strong suppression.  
 Indeed, both colour octet and colour singlet NLO QCD corrections should be
 redone, taking into account gluon mass effects. Phase space restrictions
 are expected, in this case, to significantly reduce the NLO corrections
 to both colour singlet and colour octet contributions, 
 especially for $\Jp$ decays.
\par The values of the effective gluon mass determined, in the present 
 paper, from radiative $\Jp$ and $\Up$ decays are now compared with the
 results of other studies in the literature of gluon mass effects.
 A number of representative estimates of the gluon mass are presented in 
 Table 15. In the following the generic symbol $\Mg$ will be used for 
 the genuine gluon mass, reserving the symbol $\mg$ for the `effective
 mass' in the sense described in Section 4 above, determined at tree level
 in the radiative decays of heavy quarkonia to light hadrons.
\begin{table}
\begin{center}
\begin{tabular}{|c|c|c|c|} \hline 
 Author  & Reference & Estimation Method & Gluon Mass  \\
\hline
\hline
Parisi, Petronzio & \cite{PP} & $\Jp \rightarrow \gamma X$ & 800 MeV  \\
\hline
Cornwall & \cite{Cornwall1} & Various & 500 $\pm$ 200 MeV \\
\hline
Donnachie, Landshoff & \cite{DL} & Pomeron parameters & 687-985 MeV \\
\hline
Hancock, Ross & \cite{HanRos} & Pomeron slope & 800 MeV  \\
\hline
Nikolaev {\it et al.} & \cite{Nickol}  & Pomeron parameters & 750 MeV \\
\hline
Spiridonov, Chetyrkin & \cite{SpirChet} & $\Pi^{em}_{\mu \nu}$,
 $\langle Tr G^2_{\mu \nu}\rangle$
& 750 MeV \\
\hline
Lavelle & \cite{Lavelle} & $qq \rightarrow qq$, $\langle Tr G^2_{\mu \nu}\rangle$ & 
640 MeV$^2/Q(\rm{MeV})$ \\ 
\hline
Kogan, Kovner & \cite{KK} & QCD vacuum energy, $\langle Tr G^2_{\mu \nu}\rangle$ &
 1.46 GeV \\
\hline
Field & \cite{JHFGM} & pQCD at low scales (various) & 1.5$_{-0.6}^{+1.2}$ GeV \\
\hline 
Liu, Wetzel & \cite{LW} & $\Pi^{em}_{\mu \nu}$, $\langle Tr G^2_{\mu \nu}\rangle$ &
 570 MeV \\
       &     & Glue ball current, $\langle Tr G^2_{\mu \nu}\rangle$ &
 470 MeV \\
\hline 
Yndur\'{a}in & \cite{Yndurain} & QCD potential & 10$^{-10}$-20 MeV \\
\hline
Leinweber {\it et al.} & \cite{LSWP}  & Lattice Gauge & 1.02 $\pm$ 0.10 GeV \\
\hline 
Field & This paper & $\Jp \rightarrow \gamma X$ & 0.721$_{-0.068}^{+0.016}$ GeV \\
   &      & $\Up \rightarrow \gamma X$ & 1.18$_{-0.29}^{+0.09}$ GeV \\ 
\hline      
\end{tabular}
\caption[]{{\it Estimates of the value of the gluon mass from the literature.
 For Donnachie and Landshoff, the inverse of the correlation length $a$ is
 quoted. }}
\end{center}
\end{table}
 
\par Pioneering work in this field was done by PP ~\cite{PP}.
 An estimate of $\simeq 800$ MeV for the gluon mass was made
 from the observed softening of the end-point of the inclusive photon
 spectrum in radiative $\Jp$ decays. The work presented in this paper
 is, in essence, a more refined version of the analysis of PP, taking
 into account experimental resolution effects, and the including
 $\Up$ decays as well as the best current knowledge on relativistic
 and HO QCD corrections.
\par The first extended theoretical discussion of gluon mass effects
 within QCD was made by Cornwall~\cite{Cornwall1}. Estimates were made
 of the possible value of a dynamically generated gluon mass by several
 different methods: phenomenological glueball regularisation, considerations
 based on the gluon condensate, the glueball spectrum and lattice gauge
 calculations. In fact, many of the gluon mass estimates shown in Table 15
 are based on the use of the gluon condensate:
 \begin{equation}
\langle Tr G^2_{\mu \nu}\rangle = \langle 0 |\alpha_s \rm{T}\left[ G_{\mu \nu} 
(x)G^{\mu \nu}(0)\right]|0\rangle = M^4_cf(\frac{x^2}{a^2}) 
\end{equation}
 introduced by Shiftman, Vainstein and Zakharov (SVZ)~\cite{SVZ}.
 Here $G_{\mu \nu}$ is the gluon field tensor. The gluon mass is
 related to the inverse of the correlation length, $a$, of the 
 gluonic vacuum fields. In Reference~\cite{SVZ} the phenomenological
 determination of $\langle Tr G^2_{\mu \nu}\rangle$ by the use of
 QCD Sum Rules is described.
\par Donnachie and Landshoff~\cite{DL}, identified the Pomeron
trajectory, used to describe diffractive scattering, with the QCD
 two gluon exchange process~\cite{2gEP}. They modified the perturbative
 gluon propagator in the long distance region by introducing a finite
 correlation length. The value of the latter was derived from 
 phenomenological Pomeron exchange parameters. In Table 15 the
 reciprocal of the correlation length is equated to the gluon mass.
 Other studies of the sensitivity of the Pomeron parameters to 
 non-perturbative modifications of the gluon propagator were made by
 Hancock and Ross~\cite{HanRos} and Nickolaev {\it et al.}~\cite{Nickol}.
 Again, an effective gluon mass of about 800 MeV was found.
\par Spiridonov
 and Chetyrkin ~\cite{SpirChet} estimated the gluon mass 
 by calculating power corrections to the polarisation tensor 
 $\Pi_{\mu \nu}^{em}(q)$ of the electromagnetic current of light quarks, 
 and identifying them with the gluon condensate term in the
 Operator Product Expansion (OPE) for this quantity derived by
 SVZ. The calculation was later repeated by Liu and Wetzel~\cite{LW},
 who obtained a very similar, though not identical, result (see below).
 \par Lavelle~\cite{Lavelle}, by considering a long-distance 
 modification of the gluon propagator in the amplitude for
 quark-quark scattering, established a relation between a running,
 dynamically generated, gluon mass and the gluon condensate of
 SVZ. It was emphasised by Lavelle that the derived formula,
 based on the OPE is only valid in the deep Euclidean region.
 Even so, in a recent paper Mihara and Natale (MN)~\cite{MN} applied the 
 running gluon mass formula of Lavelle to decays of the $\Jp$ and
 $\Up$ to ggg of $\gamma$gg where the gluons are either on shell
 or have time-like virtualities. MN concluded that the average
 effective gluon mass should be smaller for $\Up$ than for $\Jp$
 decays, at variance with the results of the fits presented in the
 present paper. The correction factors due to gluon mass effects:
 $f_{\rm{ggg}}$ were calculated for the $\Jp$ and
 $\Up$ and found to be 0.47 $\pm$ 0.30 and 0.94 $\pm$ 0.03 respectively,
 to be compared with the values found here (see Table 11) of
 0.18+0.08-0.02 and 0.61+0.16-0.06. The smaller gluon mass corrections
 found by MN are unable to explain the large differences in 
 the values of $\alpha_s(\mQ)$ determined from $R'_V$ and the
 the world average value of $\alpha_s$ (see Table 10 and Fig.7).

 \par The gluon mass estimates presented in Table 15 are, with one 
 exception, in the range from a few hundred MeV to about 1.5 GeV.
 The exception is the paper of Yndur\'{a}in~\cite{Yndurain} which
 claims that experimental upper bounds in the range from
 20 MeV to 10$^{-10}$ MeV may be set. These limits are based on
 considerations of the quantum mechanical potential between a quark
 and an anti-quark. This is assumed to be Coulombic for short 
 distances $r$ ($r \ll (\Lambda_{QCD})^{-1} $) where $\Lambda_{QCD}$
 is the QCD scale parameter, linear (~$\simeq Kr,~K\simeq(0.5)~\rm{GeV}^2$)
 for $(\Lambda_{QCD})^{-1} < r < \mg^{-1}$ and, finally, for $r > \mg^{-1}$,
 to exhibit a Yukawa form: $\simeq e^{r\mg}/r$. The linear portion
 of the potential gives rise to a barrier of height $E_{crit}
 \simeq K \mg^{-1}$. Yndur\'{a}in argues that $E_{crit}$ may
  be identified with the highest energy at which unsuccessful
  searches for liberated quarks have been performed.
  For example, $E_{crit} \simeq 200$ GeV leads to the limit:
\[ \mg \simeq \frac{K}{200 \rm{GeV}} = 2.5~\rm{MeV}. \]
 Other limits are given by applying similar arguments to the
 absence of proton decay into free quarks ( $\mg < 20$ MeV),
 and the non-observation of free quarks on cosmological
 scales ( $\mg < 2 \times 10^{-10}$ MeV). The arguments 
 leading to these limits are clearly untenable because of
 the neglect of the Quantum Field Theory (QFT) aspects of the
 problem\footnote{Interestingly enough, Yndur\'{a}in
 mentions, near the end of his paper, the screening of the
 potential by quark pair production, but does not draw the
 conclusion that the existence of such effects invalidate
 his limits on $\mg$ derived from a pure quantum mechanical
 potential. In the real world it is impossible to
 deconfine a quark since the production
 of low mass $\rm{q} \overline{q}$ pairs (mesons) is 
 always energetically favoured.}. In fact, when a colour
 field is stretched between a quark and an anti-quark,
 the number of colour charges is not conserved. The 
 vacuum energy materialises as quark-antiquark pairs,
 which form bound states of light mesons ( $\pi$, $\rho$,...)
 This mechanism, as implemented in the JETSET Monte Carlo
 Program~\cite{JETSET} is found to give a good description 
 of the observed hadron multiplicity in $e^+e^- \rightarrow q
 \overline{q} \rightarrow hadrons$, where the final state
 is just a colour singlet $q \overline{q}$ pair of the
 type discussed by Yndur\'{a}in. This neglect of the
 QFT aspects of the problem renders it unecessary to discuss
 further the contradiction between Yndur\'{a}in's upper
 limits and the other gluon mass estimates in Table 15.
 It is interesting to note, however, that the paper of Yndur\'{a}in is
 the only one cited on the subject of experimental limits on
 the gluon mass in the current Review of Particle Properties
 ~\cite{PDG}.
 \par The estimations of the gluon mass based on the
 phenomenologically determined value of the gluon condensate
 ~\cite{SVZ} cited in Table 15 are, with one exception,
 in the range from 500-700 MeV. As discussed above, such
 a value for the gluon mass could not explain the
 much stronger suppression of the end point of the inclusive
 photon spectrum in $\Jp$ decays, as compared to $\Up$ decays.
 The exception, that is much more consistent with the gluon
 mass value suggested by the radiative decays, is the calculation
 of Kogan and Kovner~\cite{KK}. This uses an analytical approach
 in which the vacuum energy of a gauge-invariant QCD wave functional
 is minimised. It leads to the relation, for a pure SU(3) Yang-Mills
 theory\footnote{i.e. effects of quark fields are neglected}:
\begin{equation}
\langle Tr G^2_{\mu \nu}\rangle = \frac{1}{40 \pi} M^4_g
\end{equation}  
Substituting the value of $\langle Tr G^2_{\mu \nu}\rangle$ of 
 0.038 GeV$^4$ from Reference~\cite{SVZ} gives $\Mg = 1.48$ GeV.
 This value is quite consistent with the effective gluon
 masses derived earlier~\cite{mcjhf1,mcjhf2} and confirmed in
 the present study. It also agrees well with a previous, independent,
 estimate of the present author, based on a pQCD analysis
 of several processes with low physical scales~\cite{JHFGM},
 and a recent Lattice Gauge estimate~\cite{LSWP}.
\par An effective gluon mass has also been introduced in the 
 context of the estimation of power-corrections to various hard QCD
 processes~\cite{Powcor}.
 This was already briefly discussed in MCJHF2. The leading correction to the
 mean thrust in $e^+e^-$ annihilation into hadrons, for example, 
 is found to be $\simeq \Mg/\sqrt{s}$. The gluon mass appears
 as an intermediate parameter in these calculations, but no explicit 
 values are quoted, and to date, no comparisons have been made with other
 estimates of the effective gluon mass, e.g. lattice calculations.
 Power corrections are an example
 of `higher twist' effects in the language of the OPE formalism.
 In fact, it was pointed out in Reference~\cite{SVZ}
that, as a consequence of the dimensionality of the relevant
 operators, physical systems described by an OPE (as in the case of
 QCD Sum Rules) are expected to have a leading power correction 
 $\simeq 1/Q^4$. A OPE description, as used in QCD Sum Rules, is not
 appropriate to describe higher twist effects in the quarkonium
 radiative decay spectrum discussed in the present paper. As can 
 be seen by inspection of Eqns.(7.5) and (7.6) the leading higher
 twist effects are $\simeq (\mg/Q)^2$ where $Q = \mv$.

 \par The gluon mass estimate labelled `pQCD at low scales' in Table 15,
  is based on phenomenological arguments (similar to those used later
  in  Reference~\cite{KK}), proposed in Reference~\cite{JHFOPT}
  and further developed in References~\cite{JHFGM,JHFKLN,JHFMp96}.
  In this approach, effective gluon and quark masses are related to the
   QCD $\Lambda$ parameter, that plays the role of an infra-red cut-off
   of the theory.

 \par It is interesting to note that the approach just mentioned, which relates the
 $\Lambda$ parameter of pQCD to effective quark and gluon
 masses, is complementary to the SVZ approach of describing the perturbative/non-perturbative
 interface in terms of QCD Sum Rules. In this context it is interesting to consider
 the relation between $\alpha_s$ , $\langle Tr G^2_{\mu \nu}\rangle$ and $\Mg$~\cite{SpirChet}:
 \begin{equation}
\langle Tr G^2_{\mu \nu}\rangle = \frac{\alpha_s(Q^2)}{\pi^2} \Mg^4
 \left(3 \ln\frac{Q^2}{\Mg^2}+\frac{23}{2}-12\zeta(3)\right) 
\end{equation}
 The value $\Mg = 750$ MeV quoted in Table 15, was obtained by setting $Q = 10$ GeV in 
 Eqn(8.3), but
 the equation is expected to be valid for any value of $Q$ in the perturbative region.
 Substituting the values $Q_0 = 2.88$ GeV, $\alpha_s(Q_0^2) = \alpha_s(0) = 0.27$
 and $\Mg = 1.0$ GeV, which reproduce well the measured value of
 $\alpha_s(M_Z^2)$~\cite{JHFGM}, into Eqn(8.3) gives, for the gluon condensate,
 $\langle Tr G^2_{\mu \nu}\rangle = 0.094$ (GeV)$^4$, which may be compared with 
 the average value derived by Narison~\cite{Narison} of  
 $\langle Tr G^2_{\mu \nu}\rangle = 0.071 \pm 0.009$ (GeV)$^4$. Liu and Wetzel~\cite{LW}
 derived a formula identical to Eqn(8.3) except that the term 23/2 is replaced by 10. 
 With the same parameters as quoted above, the value
 $\langle Tr G^2_{\mu \nu}\rangle = 0.053$ (GeV)$^4$ is obtained. In view of the 
 $\simeq$ factor of two difference between the original SVZ estimation of
 the numerical value of $\langle Tr G^2_{\mu \nu}\rangle$, as compared to that of Narison,
 the overall  consistency of the value of $\langle Tr G^2_{\mu \nu}\rangle$
 derived from the QCD Sum Rule approach, and the pQCD phenomenology of
 Reference~\cite{JHFGM}, is very satisfactory. 
  
 \par Thus, although gluon mass effects are only {\it directly} observable in processes with
  pure gluonic parton-level final states, such as the radiative heavy quarkonia decays
 analysed in detail in this paper, the existence of a gluon mass of order one GeV is
 already implicit in all pQCD analyses that use, as an infra-red cut-off, the conventional
 $\Lambda$ parameter. It is important to stress, however, that as previously pointed 
 out~\cite{JHFMp96,CDL}, the scale at which pQCD is expected to break down is 
 $\simeq$ gluon mass $\simeq 1$ GeV, {\it not} $\Lambda_{QCD}$, which is typically a
 factor of five smaller. It is interesting  to note
 that the same conclusion has recently been reached in a study of the five-loop
 QCD $\beta$-function, using Pad\'{e}-approximant methods~\cite{CES}.  
  From this point-of-view the success of pQCD, in association
 with the LPHD hypothesis, in describing observed particle multiplicity distributions,
 as well as their energy dependence, using cut-off scales $Q_0$ as low as 270 or
 even 150 MeV~\cite{LPHD}, seems somewhat mysterious and may be accidental. As previously
 mentioned, Monte-Carlo
 generators that actually simulate in detail both the partonic and hadronic phases of the
 space-time evolution of the final state use infra-red 
 cut-off parameters of about 1-2 GeV, of the same order as the observed gluon mass.

 \par Since the main effect of the inclusion of a phenomenological
 gluon mass on the QCD predictions is phase spaee suppression,
 it is clear that the associated mass must be time-like: $m_g^2 > 0$.
 This behaviour is also consistent with an analytical parameterisation
 of the lattice results of Reference~\cite{LSWP}. Another recent lattice
 study using a Coulomb gauge gluon propagator~\cite{CuZwan} has suggested 
 rather a {\it pure imaginary} pole mass: $i(575 \pm 124)$ MeV. This
 work is related to the suggestion of Gribov~\cite{Gribov} that colour
 confinement is due to a long range Coulomb force. In this model
 physical (transverse) gluons disappear form the physical spectrum in the
 infra-red region. This would appear to be at variance with the essential
 role of massive, time-like, and dominantly transverse, physical gluons
 in the description of
 inclusive photon spectra that is demonstrated in the present paper.   
 Actually, renormalisability remains to be proved for the Coulomb
 gauge and, as pointed out by the authors of Reference~\cite{CuZwan},
 except possibly in a confined phase, in quantum field theory the 
 square of a particle mass identified with the pole of a propagator
 must be real and positive. The author's 
 opinion is that pure quantum field theory studies of the 
 type carried out by Gribov and Zwanziger~\cite{Zwan} are no better
 adapted than the quantum mechanical potential model of Yndur\'{a}in
 ~\cite{Yndurain}, discussed above, to understand the physical
 mechanism of confinement. This appears to occur via a
 transition beween partonic and hadronic phases of matter, after which
 colour charges are all confined within hadrons, which are mostly
 bound states of quarks. Clearly quenched lattice calulations of the
 type done in Reference~\cite{CuZwan} are unable to describe such 
 a mechanism.

 \par It has recently been pointed out that the introduction of
 a term containing 
 a tachyonic gluon mass $\lambda$ with $\lambda^2 \simeq -0.5$ GeV$^2$
 in the OPE of QCD Sum Rules is able to explain
 some long standing puzzles in the related phenomenology~\cite{ChetNarZak}.
 This approach was justified by evidence from lattice gauge calculations
 for non-perturbative contributions leading to a linearly increasing term
 in the static QCD potential at short distances~\cite{Bali}. The connection
 of this gluon mass parameter, which gives an economical description,
 within the QCD Sum Rule formalism, of a non perturbative {\it
 short distance} effect with the time-like effective mass discussed
 in the present paper (and all references cited in Table 15), is unclear.
 This latter mass describes rather the {\it long distance} behaviour of the
 gluon propagator. It is quite possible that both the time-like and
 tachyonic gluon masses may be appropriate and 
 consistent phenomenological parameters within their different domains
 of applicablity.

 \SECTION{\bf{Summary and Outlook}}
 In this paper a phenomenological QCD analysis has been performed using all 
 available data on the inclusive photon spectrum in the decays of the $\Jp$
 (Mark II Collaboration) and the $\Up$ (CUSB, ARGUS, Crystal Ball and
 CLEO2 Collaborations). The fits performed to the shape of the spectra
 included, for the first time, the combined contributions of relativistic
 corrections, NLO pQCD corrections, and corrections due to the
 non-vanishing gluon mass. For the relativistic correction, fixed values
 of $<v^2>$ of 0.28, and 0.09 were assumed for the $\Jp$ and $\Up$
 repectively, in accordance with recent potential model calculations.
 Both the new, complete, NLO pQCD calculation by Kr\"{a}mer~\cite{Kramer} of the  $\Up$
 photon spectrum and the old resummed calculation of Photiadis~\cite{Photiadis}
 (applicable
 only in the end-point region $z \simeq 1$) were used in the fits. Gluon 
 mass effects were estimated using the complete tree-level calculation
 of $V \rightarrow \gamma \rm{gg}$ of Liu and Wetzel~\cite{LW}. 
\par For neither the $\Jp$ nor the $\Up$ was any consistent description
 of the experimental data possible in the absence of gluon mass corrections.
 In this case, for the Mark II data, no fit was
 obtained with a confidence level greater
 than 10$^{-30}$ (Table 5). For the $\Up$, acceptable confidence levels
 of 0.17 and 0.13 were found for some fits to the Crystal Ball and CLEO2
 data, but not for ARGUS where the best confidence level obtained
 was $6.7 \times 10^{-3}$ (Table 7). However the best confidence level
 for the combined fit to these three experiments was only
 $4.7 \times 10^{-3}$.
\par Including gluon mass corrections yielded confidence levels of
 greater than 1$\%$ for all the fit hypotheses tried (see Tables 6 and 8)
 and values of the effective gluon mass, $\mg$, of:
\begin{eqnarray}
 \mg & = & 0.721^{+0.010~+0.013}_{-0.009~-0.068} \rm{GeV}~~(\Jp) 
 \nonumber \\
     &   &~~~~~~~(\rm{Mark~II})
 \nonumber \\
 \mg & = & 1.18^{+0.06~+0.07}_{-0.06~-0.28} \rm{GeV}~~(\Up) 
 \nonumber \\
     &   &\rm{ (Mean~of~ARGUS,~ Crystal~Ball~and~CLEO2)}
 \nonumber 
\end{eqnarray}
 The first errors quoted are experimental, and correspond to the
 68$\%$ error contour of the fit, while the second errors are theoretical,
 reflecting the spread in the best fit values of $\mg$ resulting from
 different treatments of relativistic and HO pQCD corrections.
 It is clear from the fit results shown in Tables 6 and 8 that the shape
 of the photon spectrum for both the $\Jp$ and the $\Up$ is completely
 dominated by gluon mass effects. Introducing the relativistic correction
 leaves the fitted values of $\mg$ almost unchanged, while the shifts
 produced by different HO pQCD corrections are less than, or comparable
 to, the fit errors. The gluon mass corrections are due essentially to
 phase space restrictions. Including or excluding contributions from
 longitudinal gluon polarisation states changes the fitted values of 
 $\mg$ by only $\simeq~5 \%$.
 \par It can be seen, from the theoretical photon spectra calculated
 including the effects given by the experimentally determined values
 of $\mg$ (Fig. 11), that the suppresssion of the end point of the
 spectrum is much stronger in the case of the $\Jp$, than that of
 the $\Up$. This is in contradiction with the prediction of the 
 QCD parton shower model of R.D.Field~\cite{RDF} that agrees well with the
 measured $\Up$ spectrum, but predicts, for the $\Jp$, a much 
 harder spectrum than that actually observed. Such a strong
 suppression of the $\Jp$ end-point, in comparison with that of the $\Up$
 is only possible if the gluon mass is $\simeq$ 1 GeV. All these
 conclusions are in agreement with those of two earlier, closely
 related, papers~\cite{mcjhf1,mcjhf2}, in which it was conjectured,
 following the work of Parisi and Petronzio~\cite{PP}, that gluon mass 
 corrections must be much more important than either relativistic
 of HO pQCD corrections in determining the shapes of the inclusive
 photon spectra. 
\par The QCD coupling constants $\alpha_s(1.5{\rm GeV})$ and 
$\alpha_s(4.9{\rm GeV})$ were determined from the experimental
 measurements of the branching ratio $R'_V$ (Eqn. 7.1) for the
 $\Jp$ and $\Up$ respectively. Use of $R'_V$ has the advantage
 of being insensitive to relativistic correction effects. As 
 shown in Fig. 7, even allowing for a variation of the 
 renormalisation scale in the range from $0.6\mQ$ to $2.0\mQ$,
 poor agreement (deviations of $> 4.8\sigma,~3.0\sigma$ for
 the $\Jp$, $\Up$ respectively) is found with the current
 world average value of $\alpha_s(Q)$. Using the measured 
 $\mg$ values to calculate the gluon mass correction 
 factors in the theoretical expressions for $R'_V$
 (Eqns. (7.2) and (7.3)) leads to values of $\alpha_s$ 
 that are consistent, albeit within much larger errors,
 (due mainly to the large theoretical error on $\mg$), with the
 expected value of $\alpha_s(Q)$ (see Fig. 7).
\par The results obtained in this paper for the effective gluon
 mass in radiative $\Jp$ and $\Up$ decays are compared, in Table 15,
  with some other
 estimates of the gluon mass that have appeared in the literature
 over the last 20 years. Apart from the upper limits of
 Yndur\'{a}in~\cite{Yndurain}, that have been critically discussed in the previous
 Section of this paper, all the estimates lie in the range from
 500 MeV to 1.5 GeV. The values of $\mg$ obtained here favour
 somewhat higher values of the genuine gluon masss $\Mg$ of $\ge$ 1 GeV.
 In particular, there is good agreement with the estimates of
 Kogan and Kovner~\cite{KK}, (minimisation of the energy of a pure 
 Yang-Mills QCD wave functional), a previous estimate of
 the present author, using a pQCD analysis of several 
 processes with low physical scales,  and the recent Lattice
 Gauge estimate of Leinweber {\it et al.}~\cite{LSWP} with improved
 lattice sampling in the far infra-red region. It is also
 mentioned in the previous Section that the conventional
 value for the phenomenological scale parameter $\Lambda$ of 
 pQCD of several hundred MeV actually implies a gluon
 mass some five times larger. Since for scales less than
 $\Mg$, $\Lambda$ is actually a (calculable) scale dependent
 parameter~\cite{JHFOPT}, there is no Landau pole in QCD at the scale
 $Q=\Lambda$, but rather a break-down of pQCD at a much
 larger infra-red cut-off scale $Q_0 \ge 1 $ GeV. 
 
 \par A model of confinement in which a purely imaginary gluon mass 
 is introduced~\cite{CuZwan} predicts the decoupling of physical
 gluon states at low physical scales. Such behaviour would seem to
 be inconsistent with the radiative decay data which can only be
 described by the contribution, in the same infra-red region, of
 physical gluons with a time-like effective mass.
 \par A contribution in the OPE of QCD
 Sum Rules corresponding to a tachyonic gluon mass  was recently
 proposed~\cite{ChetNarZak}
 to describe some short distance properties of the QCD potential
 suggested by lattice studies~\cite{Bali}. This work has no
 obvious connection with the present one where it is shown
 instead that
 a gluon with a time-like effective mass apparently plays an
 important role in the long-distance region. 

 \par Since the theoretical uncertainties on the values
of the effective gluon mass, $\mg$, determined here are
 much larger than the experimental ones ( the former
 are $9\%$ ($\Jp$), $23\%$ ($\Up$) as compared to 
$1.3\%$ ($\Jp$), $5\%$ ($\Up$) for the latter) the most urgent
 need is for improved theory predictions rather than more precise
 experimental data\footnote{ It is interesting to note that the effective
 gluon mass is determined with a much better relative precision
 for the $\Jp$ than for the $\Up$. This is a consequence of the
 larger size of the gluon mass corrections for the $\Jp$ resulting
 in a greater sensitivity to $\mg$.}. As a first step, NLO pQCD
 calculations should be repeated including gluon mass effects, as
 has been done at tree level by Parisi and Petronzio and Liu
 and Wetzel. Of even more interest may be the calculation of:
\[ \Jp \rightarrow \gamma {\rm q \overline{q}} \]
 via the exchange of two massive virtual gluons (actually a
 NNLO process) to test the conjectured dominance of this 
 process as already indicated by the structure of the 
 hadronic final state.
 \par As was already evident from previously presented results~\cite{mcjhf2},
 the usefulness of the NRQCD approach of Reference~\cite{NRQCD}
 will be very limited unless some means is found of incorporating
 the numerically very important gluon mass effects within the
 formalism. Another important problem that this approach
  must address is the possible double counting of pQCD
 and relativistic corrections~\cite{JHFMp96}.
\par Finally it should not be forgotten that the most dramatic 
 effects discussed in this paper (in the process $\Jp
 \rightarrow \gamma X$) are based on the results of a single 
 experiment~\cite{Mark II} performed more than 20 years ago now. It is clearly
 important that this remarkable experimental result, strongly
 suggesting that the gluon mass is~ $\simeq 1$ GeV should be
 confirmed. In a world where half dozen or so b-factories 
 (or their equivalents) exist, or are under construction, it
 is high time that the $\tau$ charm-quark energy region 
 be revisited with modern general-purpose detectors and
 much higher integrated luminosity than
 that which yielded the data from the Mark II collaboration
 that has been analysed in this paper\footnote{It seems, at the time of writing,
  that there is now a very good chance that this will soon be done~\cite{CLEOC}}.
 In view of the less than perfect consistency of the different measurements
 of $\Up \rightarrow \gamma X$, an improved measurement of this process,
 with at least an order of magnitude greater statistics, and reduced systematic
 errors, would be of great interest. This is perhaps possible at existing
 b-factories.
\pagebreak
\par {\bf Acknowledgements}
\par
\ I specially thank M.Consoli for a collaboration that is at the 
 inception of the work described in this paper. I thank
 S.Catani, M.Kr\"{a}mer, M.Mangano and D.Ross for their interesting critical comments.
 Finally, I am indebted to D.Duchesneau for a careful reading of the manuscript, and
  a discussion that helped to improve the clarity of the presentation.
\pagebreak
\par {\Large Appendix}
\par In the previously published analysis of the $\Jp$ and $\Up$ inclusive
 photon spectra~\cite{mcjhf2}, resolution effects were simulated by smearing
 the theoretical distributions with a gaussian random number. In order to obtain
 stable fit results, an unsmeared histogram with a very large number of entries
 was necessary, and the random number throwing had to be repeated at each fit 
 iteration, which was both time consuming and resulted in different statistical
 errors in the fitted function at each iteration. Even with $\simeq 10^6$ events in
 the histogram, the determination of the exact position of the $\chi^2$ minimum
 was quite difficult. To avoid these problems, a new, purely analytical smearing
 technique was used for the fits presented in this paper. The method was faster
 and eliminated all problems related to the statistical errors inherent in
 Monte Carlo methods.
\par A theoretical histogram of the inclusive photon spectrum,
 with bin index $J$, $H0(J)$, is
 generated with a fine binning, typically $NBIN = 1000$. 
 The corresponding resolution smeared histogram with bin index $JS$, $HS(JS)$,
 is the generated according to the following algorithm:
 \[ HS(JS)= \sum_{J=1}^{NBIN} H0(J) F(J,JS) \]
 where
\begin{eqnarray}
  F(J,JS) &=& \frac{\exp[-\frac{1}{2} \Delta(J,JS)^2]}{\sigma[z(J)]}
 ~~~~~~~~-6 \le \Delta(J,JS) \le 6 \nonumber \\
          &=&  0.0~~~~~~~~~~~~~~~~~~~~~~~~\Delta(J,JS) > 6,~~\Delta(J,JS) < -6  \nonumber 
\end{eqnarray}
 and 
\[ \Delta(J,JS) = \frac{z(J)-z(JS)}{\sigma[z(J)]} \]
 Here $z(J)$ is the central value of the scaled photon energy in 
 bin $J$ and $\sigma[z(J)]$ is the photon energy resolution at 
 energy $E_{\gamma}=z(J)M_V/2$. Values of $\sigma[E_{\gamma}]/E_{\gamma}$
 for the different experiments analysed here are reported in Table 2. 
 In this way, a gaussian resolution smearing of the photon energy by
 up to $\pm 6 \sigma$
 around the true value is performed.
 Precise results were obtained by normalising the unsmeared histogram to a
 million events:
 \[\sum_{J=1}^{NBIN} H0(J) = 10^6 \]
 but, unlike in the case of a Monte-Carlo simulation, the execution time of
 the program does not depend on this number. The bins of the smeared histogram
 are grouped to correspond to those of the experimental histogram before
 fitting~\cite{MINUIT}. In all the fits the relative normalisation of the
 experimental and the smeared theoretical histogram was allowed to float.

\pagebreak

\end{document}